\documentclass[fleqn,usenatbib]{mnras}

\usepackage{newtxtext,newtxmath}

\usepackage[T1]{fontenc}
\usepackage{ae,aecompl}

\usepackage{graphicx}	
\usepackage{amsmath}	
\usepackage{amssymb}	
\usepackage{textcomp}
\usepackage[section]{placeins}
\usepackage{adjustbox}
\usepackage{array}

\usepackage[normalem]{ulem}

\DeclareMathAlphabet{\mathsc}{OT1}{cmr}{m}{sc}
\def\testbx{bx}%
\DeclareRobustCommand{\ion}[2]{%
\relax\ifmmode
\ifx\testbx\f@series
{\mathbf{#1\,\mathsc{#2}}}\else
{\mathrm{#1\,\mathsc{#2}}}\fi
\else\textup{#1\,{\mdseries\textsc{#2}}}%
\fi}
\newcommand{\Hei}{\ion{He}{i}}
\newcommand{\Heii}{\ion{He}{ii}}
\newcommand{\Feii}{\ion{Fe}{ii}}
\newcommand{\Oi}{\ion{O}{i}}
\newcommand{\Oiii}{\ion{O}{iii}}
\newcommand{\Cii}{\ion{C}{ii}}
\newcommand{\Caii}{\ion{Ca}{ii}}
\newcommand{\Mgii}{\ion{Mg}{ii}}
\newcommand{\Ciii}{\ion{C}{iii}}
\newcommand{\Coii}{\ion{Co}{ii}}

\newcommand{\Nii}{\ion{N}{ii}}
\newcommand{\Niii}{\ion{N}{iii}}
\newcommand{\Sii}{\ion{S}{ii}}
\newcommand{\Siii}{\ion{Si}{ii}}

\newcommand{\msun}{\mbox{M$_{\odot}$}}
\newcommand{\msol}{\mbox{M$_{\odot}$}}

\newcommand{\rsun}{\mbox{R$_{\odot}$}}
\newcommand{\kms}{\mbox{$\rm{km}\,s^{-1}$}}
\newcommand{\ms}{\mbox{$\rm{ms}$}}
\newcommand{\G}{\mbox{$\rm{G}$}}

\newcommand{\nick}{\mbox{$^{56}$Ni}}

\title[LSQ13ddu]{LSQ13ddu: A rapidly-evolving stripped-envelope supernova with early circumstellar interaction signatures}

\author[P. Clark et al.]{Peter Clark$^{1}$,
	Kate Maguire$^{1,2}$,
	Cosimo Inserra$^{3}$,
	Simon Prentice$^{1,2}$,
	Stephen J. Smartt$^{1}$,
	\newauthor
	Carlos Contreras$^{4}$,
	Griffin Hossenizadeh$^{5}$,
	Eric Y. Hsiao$^{6}$,
	Erkki Kankare$^{7}$,
	\newauthor
	Mansi Kasliwal$^{8}$,
	Peter Nugent$^{9}$,
	Melissa Shahbandeh$^{6}$,
	Charles Baltay$^{10}$,
	\newauthor
	David Rabinowitz$^{10}$,
	Iair Arcavi $^{11}$,
	Chris Ashall$^{6}$,
	Christopher R. Burns$^{12}$,
	\newauthor
	Emma Callis$^{13}$,
	Ting-Wan Chen$^{14}$,
	Tiara Diamond$^{15}$,
	Morgan Fraser$^{13}$,
	\newauthor
	D. Andrew Howell$^{16,17}$,
	Emir Karamehmetoglu$^{18,19}$,
	Rubina Kotak$^{7}$,
	Joseph Lyman$^{20}$,
	\newauthor
	Nidia Morrell$^{4}$,
	Mark Phillips$^{4}$,
	Giuliano Pignata$^{21,22}$,
	Miika Pursiainen$^{23}$,
	\newauthor
	Jesper Sollerman$^{19}$,
	Maximilian Stritzinger$^{18}$,
	Mark Sullivan$^{23}$,
	David Young$^{1}$
	\\
$^{1}$Astrophysics Research Centre, School of Mathematics and Physics, Queen's University Belfast, Belfast BT7 1NN, UK.\\
$^{2}$School of Physics, Trinity College Dublin, Dublin 2, Ireland.\\
$^{3}$School of Physics and Astronomy, Cardiff University, The Parade, Cardiff, CF24 3AA, UK\\
$^{4}$Las Campanas Observatory, Carnegie Observatories, Casilla 601, La Serena, Chile\\
$^{5}$Center for Astrophysics \textbar{} Harvard \& Smithsonian, 60 Garden Street, Cambridge, MA 02138-1516, USA \\
$^{6}$Department of Physics, Florida State University, Tallahassee 32306, USA \\
$^{7}$Tuorla Observatory, Department of Physics and Astronomy, University of Turku, FI-20014 Turku, Finland \\
$^{8}$Division of Physics, Mathematics and Astronomy, California Institute of Technology, Pasadena, CA 91125, USA \\
$^{9}$Lawrence Berkeley National Laboratory, Berkeley, CA 94720, USA \\
$^{10}$Physics Department Yale University 210 Prospect Street, New Haven CT\\
$^{11}$School of Physics and Astronomy, Tel Aviv University, Tel Aviv 69978, Israel \\
$^{12}$Observatories of the Carnegie Institution for Science, 813 Santa Barbara St., Pasadena, CA 91101, USA\\
$^{13}$School of Physics, O'Brien Centre for Science North, University College Dublin, Belfield, Dublin 4, Ireland \\
$^{14}$Max-Planck-Institut f{\"u}r Extraterrestrische Physik, Giessenbachstra\ss e 1, 85748, Garching, Germany \\
$^{15}$Laboratory of Observational Cosmology, Code 665, NASA Goddard Space Flight Center, Greenbelt, MD 20771, USA \\
$^{16}$Las Cumbres Observatory, 6740 Cortona Drive, Suite 102, Goleta, CA 93117-5575, USA \\
$^{17}$Department of Physics, University of California, Santa Barbara, CA 93106-9530, USA \\
$^{18}$Department of Physics and Astronomy, Aarhus University, Ny Munkegade 120, DK-8000 Aarhus C, Denmark\\
$^{19}$The Oskar Klein Centre \& Department of Astronomy, Stockholm University, AlbaNova, SE-106 91 Stockholm, Sweden\\
$^{20}$Department of Physics, University of Warwick, Coventry, CV4 7AL, UK \\
$^{21}$Departamento de Ciencias F\'isicas, Universidad Andres Bello, Avda. Fern\'andez Concha 700, 7591538, Santiago, Chile\\
$^{22}$Millennium Institute of Astrophysics, Nuncio Monsenor S\'otero Sanz 100, Providencia, Santiago, 8320000 Chile\\
$^{23}$School of Physics and Astronomy, University of Southampton, Southampton, SO17 1BJ, UK \\
}

\date{Accepted 2019 December 20. Received 2019 December 20; in original form 2019 March 29}

\pubyear{2019}

\begin{document}
\label{firstpage}
\pagerange{\pageref{firstpage}--\pageref{lastpage}}
\maketitle

\begin{abstract}
This paper describes the rapidly evolving and unusual supernova LSQ13ddu, discovered by the La Silla-QUEST survey. LSQ13ddu displayed a rapid rise of just 4.8$\pm$0.9~d to reach a peak brightness of $-$19.70$\pm$0.02 mag in the \textit{LSQgr} band. Early spectra of LSQ13ddu showed the presence of weak and narrow $\Hei$ features arising from interaction with circumstellar material (CSM). These interaction signatures weakened quickly, with broad features consistent with those seen in stripped-envelope SNe becoming dominant around two weeks after maximum. The narrow $\Hei$ velocities are consistent with the wind velocities of luminous blue variables but its spectra lack the typically seen hydrogen features. The fast and bright early light curve is inconsistent with radioactive \nick\ powering but can be explained through a combination of CSM interaction and an underlying \nick\ decay component that dominates the later time behaviour of LSQ13ddu. Based on the strength of the underlying broad features, LSQ13ddu appears deficient in He compared to standard SNe Ib. 

\end{abstract}




\section{Introduction}
\label{sec:Introduction}
	
In recent years transient surveys such as the Asteroid Terrestrial-impact Last Alert System \citep[ATLAS;][]{Tonry_2018}, the La-Silla QUEST survey \citep[LSQ;][]{Baltay2013_LaSillaQuest}, the All-Sky Automated Survey for Supernovae \citep[ASAS-SN;][]{Shappee2014}, the Dark Energy Survey \citep[DES;][]{Pursiainen2018}, and the Zwicky Transient Facility \citep[ZTF;][]{ZTF_2019} have used high-cadence observations (repeat visits every $\sim$1 -- 3~d) to reveal and study new classes of luminous and rapidly evolving transients \citep{Drout2014, Arcavi2016}. Numerous other objects showing exotic and unusual behaviour have been observed such as iPTF16asu \citep{Whitesides2017a} and AT~2018cow \citep{Prentice2018_cow}, with these transients presenting challenges to existing explosion models and progenitor scenarios. Transitional objects whose properties span more than one class of transients have also been observed, such as SN~2016coi that had properties consistent with SNe Ic but with evidence of residual He within its ejecta \citep{Prentice2018_coi}. 

There are also known classes of luminous, and sometimes fast-evolving, transients showing clear signs of interaction with circumstellar material (CSM). Type IIn SNe show features associated with H-rich Type II SNe, along with narrow (`n') H emission lines superimposed on their spectra \citep{Schlegel1990}. Similarly, SNe Ibn \citep{Pastorello2008a} are a rare subclass of helium-rich Type Ib SNe that display narrow He emission features within their spectra. These features are produced through interaction between the SN ejecta and surrounding CSM with SNe Ibn showing wide variation in their spectral evolution. Some develop weak H features at late times \citep{Smith2012,Pastorello2015e}, thought to be indicative of the SN ejecta interacting with a more distant shell of H-rich CSM, while others evolve to more closely resemble `normal' Type Ib SNe displaying broad He features at late times \citep{Pastorello2015e,Pastorello_15ed}.

Despite their varied spectral evolution, SNe Ibn display broadly similar photometric behaviour \citep{Hosseinzadeh2017}, with an initial rise to a peak \textit{R}-band absolute magnitude between $-$18.5~mag and $-$20~mag over the course of a few days up to two weeks. This rise is then usually followed by a decline over the course of $\sim$1--2 months. A small subset of SNe Ibn have been observed to have much slower post-maximum decline rates, with OGLE-2012-SN-006 being notable for entering an extended plateau-like phase \citep{Pastorello2015c}. Additionally one SN Ibn (OGLE-2014-SN-131) was observed to have an extended rise to peak \citep{Karamehmetoglu2017}, though such events form a small proportion of the overall class.

The most popular explanation for the production of SNe Ibn involves the core collapse of a H/He poor WCO-type Wolf-Rayet (WR) star embedded in a He-rich CSM \citep{Foley2007,Pastorello2007}. The CSM is expected to have velocities consistent with the wind speed of known WR stars of a few thousand \kms \citep{Rochowicz1999,Crowther2007}. Alternatively, the CSM present in these events could originate from material stripped from companion stars with evidence of a luminous blue variable-type (LBV-type) outburst from the system (either the progenitor or companion) of the prototype event SN~2006jc \citep{Pastorello2007}. It has also been suggested that stars within in a transitional phase between LBV and WR may be responsible for producing SNe Ibn. The origin of these SNe as core-collapse events is supported by their occurrence within spiral host galaxies suggesting a link to young stellar populations. PS1-12sk is an exception in that it occurred within a bright, cluster hosted elliptical galaxy \citep{Sanders2013, Pastorello2015e, Hosseinzadeh2019}. 

\begin{figure}
\centering
\includegraphics[width=7cm]{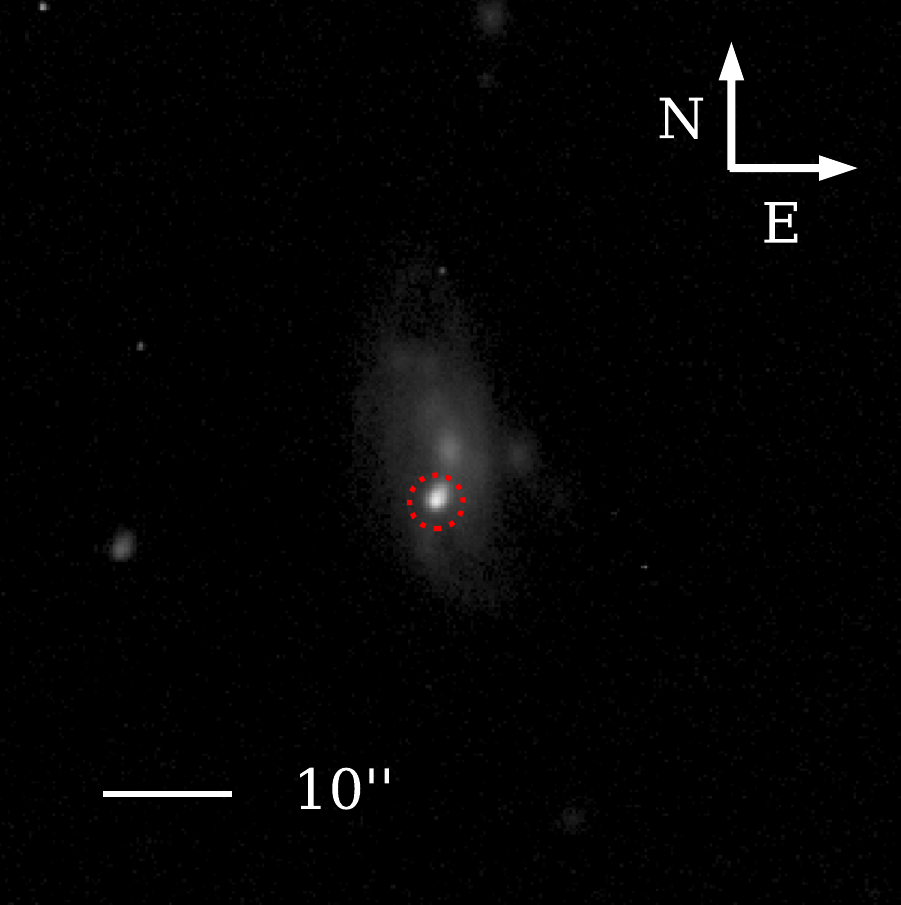}
\caption{NTT EFOSC2 \textit{V}-band image obtained on MJD~56626.1 with LSQ13ddu marked by the red circle.}
\label{2MASXJ035849232925086}
\end{figure}

We present photometric and spectroscopic observations and analysis of the unusual and fast evolving transient, LSQ13ddu. It was discovered by LSQ \citep{Baltay2013_LaSillaQuest} on MJD~56622.3 at R.A.~=~03:58:49.09 Decl.~=~$-$29:25:11.8 (J2000.0), which is offset by 4~kpc in projection from the centre of the galaxy, 2MASX~J03584923-2925086 \citep{Jones2009}, an Sc spiral \citep{Makarov2014} at a redshift of $z = 0.05845\pm0.00015$\footnote{Retrieved from the NASA/IPAC Extragalactic Database (NED)} (Fig.~\ref{2MASXJ035849232925086}). Using measurements of host galaxy emission lines in the spectra of LSQ13ddu at the position of SN, we find a slightly lower redshift of $z = 0.057787\pm0.000125$, which we use throughout this paper. LSQ13ddu was observed spectroscopically three days after the discovery by the Public ESO Spectroscopic Survey of Transient Objects \citep[PESSTO;][]{Smartt2014}. Due to its rapid rise in brightness and unusual early-time spectra, an extensive follow-up campaign was triggered. Throughout this paper, we assume a Hubble constant, H$_0$ of 70 \kms\ Mpc$^{-1}$ and adopt a standard cosmology with $\Omega_M$=0.27 and $\Omega_{\Lambda}$=0.73.

\section{Observations and data reduction}
Ultra-violet (UV), optical and near-infrared (NIR) photometric and spectroscopic observations of LSQ13ddu were obtained at a number of facilities detailed below. These observations span from very soon after explosion out to a few months after maximum light when LSQ13ddu had faded below the detection limit of available instrumentation. All reduced data will be available through the WISeREP archive \citep{Yaron2012a}.\footnote{The WISeREP archive is accessible at https://wiserep.weizmann.ac.il/} 

\begin{table*}
\centering
\caption{Summary of the optical photometry collected for LSQ13ddu, with \textit{BVgri} from LCOGT, \textit{LSQgr} from La Silla-QUEST, \textit{V} from NTT, \textit{BVri} from Swope and \textit{griz} photometry from the LT. Photometric detections at $<$3$\sigma$ are given as upper limits. Photometry is presented in the observer frame with no applied extinction corrections.}
\begin{adjustbox}{width=\textwidth}
\label{Optical_photom}
\begin{tabular}{ccccccccccc}
\hline
MJD & Date & Phase (d)$^a$ & \textit{B} (mag) & \textit{V} (mag) & \textit{g} (mag) & \textit{LSQgr} (mag) & \textit{r} (mag) & \textit{i} (mag) & \textit{z} (mag) & Source \\ \hline
56616.3 & 2013 Nov 19 & $-$9.7 & - & - & - & $>$21.79 & - & - & - & LSQ \\
56618.3 & 2013 Nov 21 & $-$7.7 & - & - & - & $>$21.72 & - & - & - & LSQ \\
56620.3 & 2013 Nov 23 & $-$5.7 & - & - & - & $>$21.84 & - & - & - & LSQ \\
56622.3 & 2013 Nov 25 & $-$3.7 & - & - & - & 18.45$\pm$0.03 & - & - & - & LSQ \\
56624.2 & 2013 Nov 27 & $-$1.8 & 17.47$\pm$0.01 & 17.58$\pm$0.01 & - & 17.56$\pm$0.02 & 17.74$\pm$0.01 & 17.99$\pm$0.02 & - & LSQ, Swope \\
56625.2 & 2013 Nov 28 & $-$0.8 & 17.30$\pm$0.02 & 17.36$\pm$0.02 & - & - & 17.38$\pm$0.01 & 17.74$\pm$0.02 & - & Swope \\
56626.1 & 2013 Nov 29 & $+$0.1 & - & 17.45$\pm$0.25 & - & - & - & - & - & NTT \\
56626.2 & 2013 Nov 29 & $+$0.2 & - & - & - & 17.35$\pm$0.01 & - & - & - & LSQ \\
56628.3 & 2013 Dec 01 & $+$2.3 & - & - & - & 17.61$\pm$0.02 & - & - & - & LSQ \\
56629.3 & 2013 Dec 02 & $+$3.3 & - & 17.74$\pm$0.27 & - & - & - & - & - & NTT \\
56632.0 & 2013 Dec 05 & $+$6.0 & 18.34$\pm$0.04 & 18.15$\pm$0.04 & 18.22$\pm$0.02 & - & - & - & - & LCOGT \\
56632.1 & 2013 Dec 05 & $+$6.1 & 18.47$\pm$0.06 & 18.27$\pm$0.03 & - & - & - & - & - & LCOGT \\
56632.2 & 2013 Dec 05 & $+$6.2 & - & - & - & 18.26$\pm$0.02 & 18.30$\pm$0.04 & 18.39$\pm$0.06 & - & LCOGT, LSQ \\
56632.3 & 2013 Dec 05 & $+$6.3 & 18.57$\pm$0.04 & 18.21$\pm$0.02 & 18.23$\pm$0.03 & - & - & - & - & LCOGT \\
56633.1 & 2013 Dec 06 & $+$7.1 & - & 18.43$\pm$0.02 & - & - & - & - & - & LCOGT \\
56633.6 & 2013 Dec 06 & $+$7.6 & 18.79$\pm$0.07 & 18.43$\pm$0.04 & 18.45$\pm$0.02 & - & - & - & - & LCOGT \\
56636.2 & 2013 Dec 09 & $+$10.2 & - & - & - & 18.97$\pm$0.04 & - & - & - & LSQ \\
56636.5 & 2013 Dec 09 & $+$10.5 & 19.43$\pm$0.11 & 18.88$\pm$0.09 & - & - & 18.85$\pm$0.07 & 18.58$\pm$0.06 & - & LCOGT \\
56636.6 & 2013 Dec 09 & $+$10.6 & - & - & 19.00$\pm$0.05 & - & - & 18.85$\pm$0.12 & - & LCOGT \\
56638.2 & 2013 Dec 11 & $+$12.2 & - & - & - & 19.26$\pm$0.07 & - & - & - & LSQ \\
56638.3 & 2013 Dec 11 & $+$12.3 & - & 19.23$\pm$0.07 & 19.31$\pm$0.05 & - & 19.43$\pm$0.08 & 19.53$\pm$0.14 & - & LCOGT \\
56639.2 & 2013 Dec 12 & $+$13.2 & - & - & - & 19.28$\pm$0.08 & - & - & - & LSQ \\
56639.3 & 2013 Dec 12 & $+$13.3 & 19.97$\pm$0.18 & 19.32$\pm$0.08 & 19.35$\pm$0.06 & - & 19.42$\pm$0.08 & - & - & LCOGT \\
56640.2 & 2013 Dec 13 & $+$14.2 & - & - & 19.65$\pm$0.06 & - & 19.44$\pm$0.12 & 19.78$\pm$0.23 & - & LCOGT \\
56641.2 & 2013 Dec 14 & $+$15.2 & - & - & - & 19.55$\pm$0.12 & - & - & - & LSQ \\
56642.9 & 2013 Dec 16 & $+$16.9 & - & - & 19.79$\pm$0.11 & - & 19.86$\pm$0.12 & 20.26$\pm$0.16 & 19.75 $\pm$0.12 & LT \\
56643.2 & 2013 Dec 16 & $+$17.2 & - & - & - & 19.85$\pm$0.17 & - & - & - & LSQ \\
56643.9 & 2013 Dec 17 & $+$17.9 & - & - & 19.93$\pm$0.10 & - & 19.90$\pm$0.08 & 20.01$\pm$0.09 & 19.60$\pm$0.06 & LT \\
56644.7 & 2013 Dec 18 & $+$18.7 & 20.02$\pm$0.30 & 19.39$\pm$0.30 & 20.02$\pm$0.16 & - & 20.12$\pm$0.15 & 20.31$\pm$0.21 & - & LCOGT \\
56644.9 & 2013 Dec 18 & $+$18.9 & - & - & - & - & 19.87$\pm$0.09 & 20.19$\pm$0.11 & 19.82$\pm$0.11 & LT \\
56645.1 & 2013 Dec 18 & $+$19.1 & - & - & - & 20.04$\pm$0.17 & - & - & - & LSQ \\
56645.9 & 2013 Dec 19 & $+$19.9 & - & - & 19.85$\pm$0.09 & - & 20.07$\pm$0.12 & 20.04$\pm$0.11 & 19.74$\pm$0.12 & LT \\
56646.7 & 2013 Dec 20 & $+$20.7 & - & - & - & - & 20.35$\pm$0.20 & 20.68$\pm$0.30 & - & LCOGT \\
56646.9 & 2013 Dec 20 & $+$20.9 & - & - & 20.02$\pm$0.06 & - & 20.09$\pm$0.08 & 20.44$\pm$0.09 & 19.80$\pm$0.07 & LT \\
56647.1 & 2013 Dec 20 & $+$21.1 & - & - & - & 20.23$\pm$0.15 & - & - & - & LSQ \\
56648.5 & 2013 Dec 21 & $+$22.5 & 20.64$\pm$0.13 & - & 20.27$\pm$0.16 & - & 20.75$\pm$0.27 & 21.02$\pm$0.33 & - & LCOGT \\
56648.9 & 2013 Dec 22 & $+$22.9 & - & - & 20.37$\pm$0.05 & - & - & 20.88$\pm$0.13 & 20.02$\pm$0.08 & LT \\
56649.1 & 2013 Dec 22 & $+$23.1 & - & - & - & 20.43$\pm$0.17 & - & - & - & LSQ \\
56649.9 & 2013 Dec 23 & $+$23.9 & - & - & - & - & - & - & 20.16$\pm$0.08 & LT \\
56650.9 & 2013 Dec 24 & $+$24.9 & - & - & 20.50$\pm$0.04 & - & 20.80$\pm$0.06 & 20.76$\pm$0.07 & - & LT \\
56651.1 & 2013 Dec 24 & $+$25.1 & - & - & - & 20.70$\pm$0.18 & - & - & - & LSQ \\
56651.9 & 2013 Dec 25 & $+$25.9 & - & - & 20.58$\pm$0.03 & - & 20.75$\pm$0.08 & 20.94$\pm$0.09 & 20.17$\pm$0.10 & LT \\
56652.5 & 2013 Dec 25 & $+$26.5 & 21.51$\pm$0.45 & 20.41$\pm$0.50 & - & - & 20.83$\pm$0.14 & - & - & LCOGT \\
56653.1 & 2013 Dec 26 & $+$27.1 & - & - & - & 20.89$\pm$0.13 & - & - & - & LSQ \\
56654.4 & 2013 Dec 27 & $+$28.4 & - & - & - & - & 21.20$\pm$0.37 & - & - & LCOGT \\
56654.5 & 2013 Dec 27 & $+$28.5 & - & 20.49$\pm$0.38 & - & - & - & 21.44$\pm$0.36 & - & LCOGT \\
56655.1 & 2013 Dec 28 & $+$29.1 & - & - & - & 21.26$\pm$0.19 & - & - & - & LSQ \\
56656.4 & 2013 Dec 29 & $+$30.4 & - & - & - & - & 21.18$\pm$0.50 & 21.63$\pm$0.46 & - & LCOGT \\
56656.5 & 2013 Dec 29 & $+$30.5 & - & 20.86$\pm$0.41 & - & - & - & - & - & LCOGT \\
56657.1 & 2013 Dec 30 & $+$31.1 & - & - & - & 21.07$\pm$0.23 & - & - & - & LSQ \\
56658.9 & 2014 Jan 01 & $+$32.9 & - & - & 20.94$\pm$0.06 & - & 21.30$\pm$0.09 & 21.49$\pm$0.13 & 20.81$\pm$0.29 & LT \\
56659.1 & 2014 Jan 01 & $+$33.1 & - & - & - & 20.96$\pm$0.15 & - & - & - & LSQ \\
56661.1 & 2014 Jan 03 & $+$35.1 & - & - & - & $>$21.76 & - & - & - & LSQ \\
56663.1 & 2014 Jan 05 & $+$37.1 & - & - & - & 21.03$\pm$0.17 & - & - & - & LSQ \\
56665.2 & 2014 Jan 07 & $+$39.2 & - & - & - & $>$21.88 & - & - & - & LSQ \\
56667.1 & 2014 Jan 09 & $+$41.1 & - & - & - & $>$22.25 & - & - & - & LSQ \\
56668.8 & 2014 Jan 11 & $+$42.8 & - & - & 21.71$\pm$0.31 & - & 21.57$\pm$0.60 & 21.69$\pm$0.32 & - & LT \\ \hline
\end{tabular}
\end{adjustbox}
\begin{flushleft}
$^a$Phase is relative to maximum light in the \textit{LSQgr} band: 17.35$\pm$0.01 on MJD~56626.2.\\
 \end{flushleft}
\end{table*}

\begin{table*}
\centering
\caption{Summary of the \textit{Swift} UVOT photometry, with detections at $<$3$\sigma$ given as upper limits.}
\begin{adjustbox}{width=1\textwidth}
\label{SwiftPhotomSummary}
\begin{tabular}{ccccccccc}
\hline
MJD     & Date        & Phase (d)$^a$& \textit{UVW2} (mag)     & \textit{UVM2} (mag)     & \textit{UVW1} (mag)     & \textit{U} (mag)        & \textit{B} (mag)        & \textit{V} (mag)        \\ \hline
56630.8 & 2013~Dec~04 & $+$4.8  & 18.43$\pm$0.20 & 18.09$\pm$0.20 & 17.82$\pm$0.15 & 17.25$\pm$0.11 & 18.15$\pm$0.15 & 17.98$\pm$0.22 \\
56632.7 & 2013~Dec~06 & $+$6.7  & -              & $>$18.94       & 18.55$\pm$0.24 & -              & -              & -              \\
56633.0 & 2013~Dec~06 & $+$7.0  & $>$18.82       & -              & -              & 17.73$\pm$0.16 & 18.66$\pm$0.24 & $>$18.43       \\
56634.5 & 2013~Dec~07 & $+$8.5  & 19.20$\pm$0.39 & $>$19.06       & 18.72$\pm$0.26 & -              & -              & 18.40$\pm$0.31 \\
56634.6 & 2013~Dec~07 & $+$8.6  & -              & -              & -              & 18.49$\pm$0.24 & 18.94$\pm$0.25 & -              \\
56635.4 & 2013~Dec~08 & $+$9.4  & $>$19.20       & $>$19.04       & 18.96$\pm$0.31 & 18.54$\pm$0.25 & 18.96$\pm$0.25 & 18.69$\pm$0.36 \\
56636.7 & 2013~Dec~10 & $+$10.7 & 19.06$\pm$0.34 & $>$18.93       & -              & -              & -              & $>$18.64       \\
56636.8 & 2013~Dec~10 & $+$10.8 & -              & -              & 18.84$\pm$0.35 & $>$18.82       & $>$19.30       & -              \\ \hline
\end{tabular}
\end{adjustbox}
\begin{flushleft}
$^a$Phase is relative to maximum light in the \textit{LSQgr} band in the observed frame.\\
 \end{flushleft}
\end{table*}

\subsection{Optical Photometry}\label{OpticalPhotom}
\begin{figure}
\includegraphics[width=8.3cm]{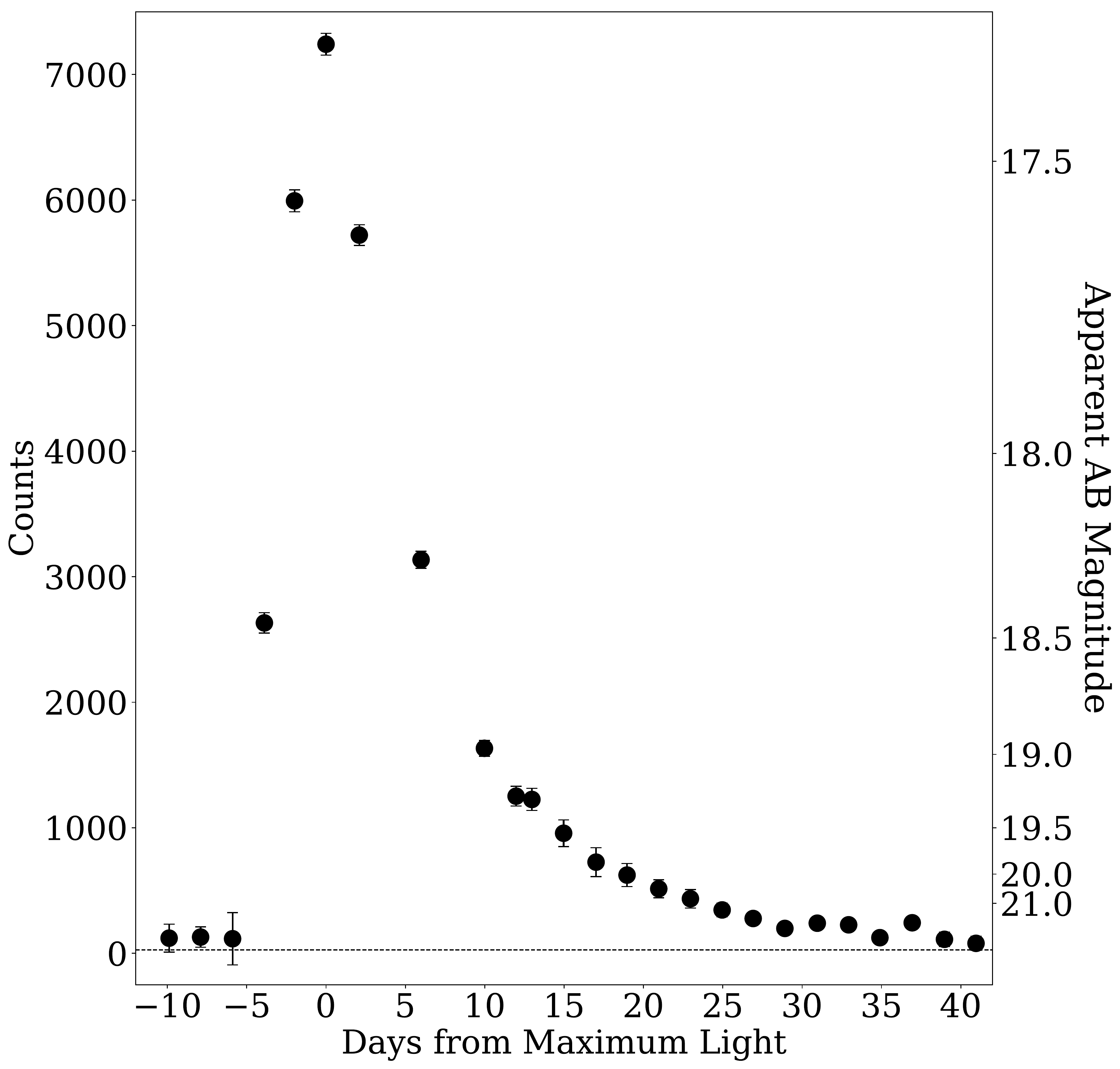}
\caption{\textit{LSQgr}-band observer frame light curve of LSQ13ddu, with the uncertainties generally smaller than the data point size.}
\label{FullLSQLC}
\end{figure}

LSQ13ddu was discovered by LSQ on 2013 November 25, Modified Julian Date (MJD) of 56623.3, in the broad \textit{g+r} search filter, \textit{LSQgr} \citep[e.g.][]{Baltay2013_LaSillaQuest}. The \textit{LSQgr} light curve of LSQ13ddu covers from the time of discovery until $\sim$40~d after peak. The \textit{LSQgr} data were reduced following the method described in \cite{Scalzo2014a}, making use of stars in the field of the SN calibrated to the AAVSO All-Sky Photometric Survey (APASS) catalogue \citep{Henden2009}. A non-detection at the position of LSQ13ddu on MJD~56620.3 (2~d prior to the first detection) gives a 3-$\sigma$ apparent limiting \textit{LSQgr}-band magnitude of 21.84. The \textit{LSQgr} light curve of LSQ13ddu is presented in Fig.~\ref{FullLSQLC}.

Multi-colour photometry was obtained using a number of facilities: the Las Cumbres Observatory (LCOGT) 1-m robotic telescope network with the SBIG instrument, the LCOGT 2-m robotic telescopes with the Spectral camera \citep{Brown2013}, the European Southern Observatory (ESO) New Technology Telescope (NTT) using the EFOSC2 instrument \citep{Buzzoni1984}, and the Liverpool Telescope (LT) using the IO:O instrument \citep{Steele2004}. Additional follow up was conducted as part of the Carnegie Supernova Project-II \citep[CSP-II;][]{Phillips2019} using the Henrietta Swope Telescope \citep[Swope;][]{Perez_2012}. The LSQ, LCOGT, LT, NTT and Swope optical photometry is presented in Table \ref{Optical_photom} and in Fig.~\ref{FullLC}.

The LCOGT \textit{BVgri} photometry was processed using the \textsc{orac-dr} pipeline \citep{Jenness2015}. After undergoing pre-processing using the \textsc{banzai} pipeline \citep{McCully2018} and image subtraction using the routine \textsc{hotpants}\footnote{This routine is available from: https://github.com/acbecker/hotpants} \citep{Becker2015}, which is an implementation of the image subtraction routine described in \cite{Alard1997}, the source was extracted using \textsc{lcogtsnpipe} \citep{Valenti2016} with the resulting magnitudes calibrated using the APASS catalogue. The \textit{BV} data and \textit{gri} data were calibrated to Vega and AB magnitudes, respectively. 

The LT~\textit{griz}-band photometry was measured on pre-processed imaging using a custom pipeline that includes subtraction of a reference template obtained at late times and calibration to the Pan-STARRS magnitude system, which is close to the AB system. 

The \textit{V}-band images obtained using the NTT were processed using the PESSTO photometric pipeline \citep{Smartt2014}, followed by image subtraction using \textsc{hotpants} and NTT images obtained of the field taken after the SN had faded with calibration using the APASS catalogue in Vega magnitudes. The photometry was measured using the \textsc{snoopy}\footnote{\textsc{snoopy} is a SN photometry package developed by E.~Cappellaro} software. 

Two epochs of Swope \textit{BVri} data were reduced in the manner outlined in \cite{Phillips2019} including image subtraction, with these observations calibrated to the Vega magnitude system. 

\subsection{UV and NIR Photometry}
Several epochs of \textit{Swift} UVOT data spanning from $+$4.8 to $+$10.8~d post \textit{LSQgr} maximum light were obtained. These data were retrieved from the Swift Optical/Ultraviolet Supernova Archive (SOUSA) and were processed using the method outlined in \cite{Brown2014}, which includes subtraction for the underlying host flux. 

A single epoch of pre-maximum \textit{YJH} photometry was obtained with the Ir\'en\'ee du Pont 2.5-m telescope using the RetroCam instrument \citep[du Pont+RetroCam;][]{CSP1_Lowz}. This photometry was reduced in the manner outlined in \cite{Phillips2019} including the subtraction of the underlying host galaxy using reference images.

\begin{figure}
\includegraphics[width=\columnwidth]{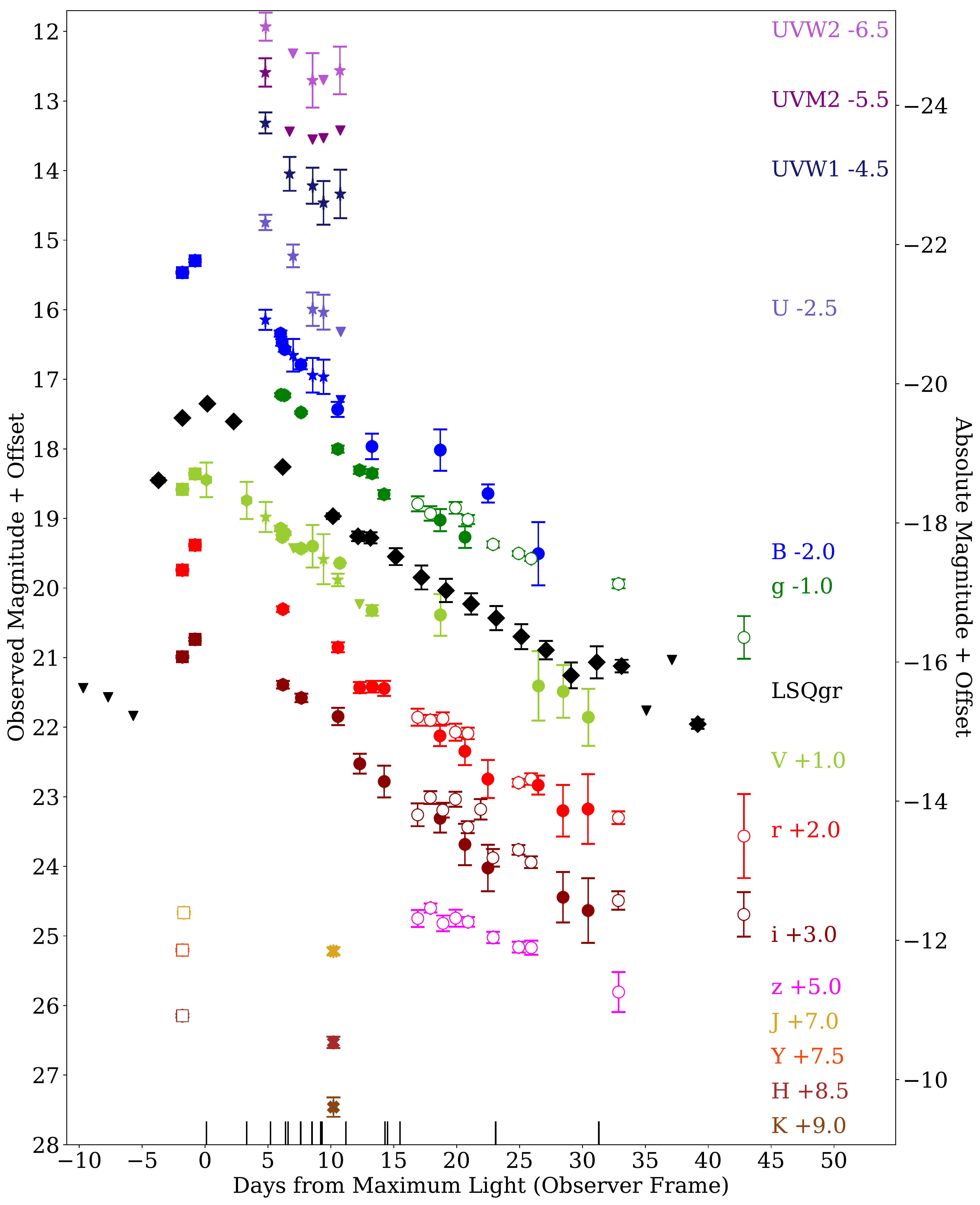}
\caption{UV, optical and NIR light curves of LSQ13ddu in observed and absolute magnitudes with the different instruments shown as follows: \textit{Swift} UV (stars), LCOGT optical (filled circles), LT optical (open circles), LSQ optical (filled diamonds), Swope optical (filled squares), NTT optical (hexagons), du Pont NIR (open squares), and NTT NIR (crosses). Data points have the offsets applied as specified in the legend. The late-time NTT NIR data are excluded for plotting purposes. Upper limits are displayed as downward facing triangles. The phases of the spectral observations are indicated with tick marks along the bottom axis.}
\label{FullLC}
\end{figure}

Two epochs of NIR \textit{JHK} photometry were obtained at the ESO NTT telescope with the SOFI instrument \citep{Moorwood1998}. We performed image subtraction using a late-time image of the field and \textsc{hotpants}. The reduction of the SOFI photometry was performed using \textsc{iraf} and \textsc{snoopy}. The UV and NIR photometry of LSQ13ddu are presented in Table \ref{SwiftPhotomSummary} and Table \ref{SOFIPhotomSummary}, respectively and are included with the optical light curves in Fig.~\ref{FullLC}. 

\begin{table}
\centering
\caption{Details of the NIR photometry of LSQ13ddu.}
\label{SOFIPhotomSummary}
\begin{adjustbox}{width=1\columnwidth}
\begin{tabular}{cccccccc}
\hline
MJD & Date & Phase (d)$^{a}$ & Magnitude & Band & Source\\  \hline
56624.2 &2013~Nov~27&$-$1.8 &17.70$\pm$0.02& \textit{Y} & (1) \\
56624.2 &2013~Nov~27&$-$1.8 &17.65$\pm$0.03& \textit{H} & (1)\\
56624.3 &2013~Nov~27&$-$1.7 &17.67$\pm$0.02& \textit{J} & (1) \\
56636.2 & 2013~Dec~09 & $+$10.2 &18.22$\pm$0.06& \textit{J} & (2) \\
56636.2 & 2013~Dec~09 & $+$10.2 &18.03$\pm$0.08& \textit{H} & (2) \\
56636.2 & 2013~Dec~09 & $+$10.2 &18.46$\pm$0.14& \textit{K} & (2)\\
56695.1 & 2014~Feb~06 & $+$69.1 &21.34$\pm$0.38& \textit{J} & (2)\\
56695.1 & 2014~Feb~06 & $+$69.1 &20.65$\pm$0.35& \textit{H} & (2)\\ 
\hline
\end{tabular}
\end{adjustbox}
\begin{flushleft}
$^a$Phase is relative to \textit{LSQgr} band maximum (observer frame).\\
(1) du Pont+RetroCam, (2) NTT+SOFI\\
\end{flushleft}
\end{table}

\subsection{Optical and NIR spectroscopy}
Optical spectroscopy was obtained using the NTT with EFOSC2, the VLT with XShooter \citep{Vernet2011a}, the LCOGT 2-m with FLOYDS\footnote{https://lco.global/observatory/instruments/floyds}, WiFeS at the ANU 2.3-m \citep{Dopita2007}, and with the IMACS f/4 camera mounted on the Magellan--Baade telescope \citep{Dressler2011}. The spectroscopic data spans from maximum light to $+$31~d post maximum. The details of these spectra are provided in Table \ref{SpectraLog}. 
The EFOSC2 spectra were reduced using a custom pipeline, applying bias-subtraction, flat-fielding, wavelength and flux calibration, and a telluric correction was applied as described in \cite{Smartt2014}. 
The XShooter spectrum was taken using 1.0$\arcsec$, 0.9$\arcsec$ and 0.9$\arcsec$ wide slits in the UVB, VIS, and NIR arms, respectively. It was reduced in the standard manner using the ESO Reflex pipeline \citep{Freudling2013}. Telluric corrections were applied to the XShooter VIS arm using the \textsc{molecfit} software \citep{Smette2015,Kausch2015}, but not to the NIR arm due to its lower signal to noise. 
The WiFeS spectra were obtained as part of the ANU WiFeS Supernova Programme (AWSNAP) with the reduction outlined in \cite{Childress2016}. 
The IMACS spectrum was reduced using the \textsc{cosmos} software package \citep{Dressler2011, Oemler2017}. 
The FLOYDS spectra were reduced using a custom pipeline. A telluric correction was applied to the IMACS and FLOYDS spectra following the method used for the EFOSC2 spectra. Two NIR spectra were obtained with Magellan+FIRE \citep{Simcoe2013} and were reduced following the method described in \cite{Hsiao2019} using the tailored pipeline \textsc{firehose} \citep{Simcoe2013}. An absolute flux calibration was applied using coeval photometric data.

The spectra have been corrected for Milky Way extinction using the \cite{Fitzpatrick1998} law and an $\textit{E(B-V)}$ value of 0.008 mag from \cite{Schlafly2011} retrieved from NED. The spectra were corrected to the rest frame using a value of $z=0.057787$, measured using host H$\alpha$, H$\beta$, \Nii\ 6548, 6583, \Oiii\ 5007 and \Sii\ 6717~\AA\ lines in the XShooter and WiFeS spectra. The optical and NIR spectral sequences are presented in Fig.~\ref{LSQ13dduSpectralSequenceVIS} and \ref{LSQ13dduSpectralSequenceNIR}, respectively. 

\section{Analysis}
\label{sec:Analysis}
In this section, we analyse the light curves of LSQ13ddu, construct a bolometric light curve, estimate the blackbody temperatures, and measure the evolution of the main spectral features. We also summarise the host galaxy properties. The use of \textsc{synapps} for spectral line identification is described, as well as potential light curve powering mechanisms for LSQ13ddu using photometric modelling.

\subsection{Light curve parameters}
\label{sec:Photometric Analysis}
Using a cubic spline fit to the peak data of the \textit{LSQgr} band light curve, we obtained a date of maximum light of 56626.0$\pm$0.2 MJD and apparent peak magnitude of 17.36$\pm$0.02 mag with the peak absolute \textit{LSQgr} magnitude found to be $-$19.70$\pm$0.02 mag. A cubic spline fit to the early light curve constrains the rise time to peak brightness to be 4.8$\pm$0.9~d with an explosion epoch of 56620.9$\pm$0.7 MJD, consistent with the deep \textit{LSQgr} non-detection at $-$5.7~d prior to peak. During its rise to peak, LSQ13ddu brightened at a rate of $\sim$0.75 mag~d$^{-1}$ with a $\Delta m_{15} (LSQgr)$ of 2.20$\pm$0.14 mag giving a decline rate of 0.15$\pm$0.01 mag~d$^{-1}$ in the \textit{LSQgr} band over the first 15~d post maximum light.

\subsubsection{Pseudo-bolometric light curve}
\label{sec:BolometricLightCurves}

\begin{figure*}
\includegraphics[width=12.75cm]{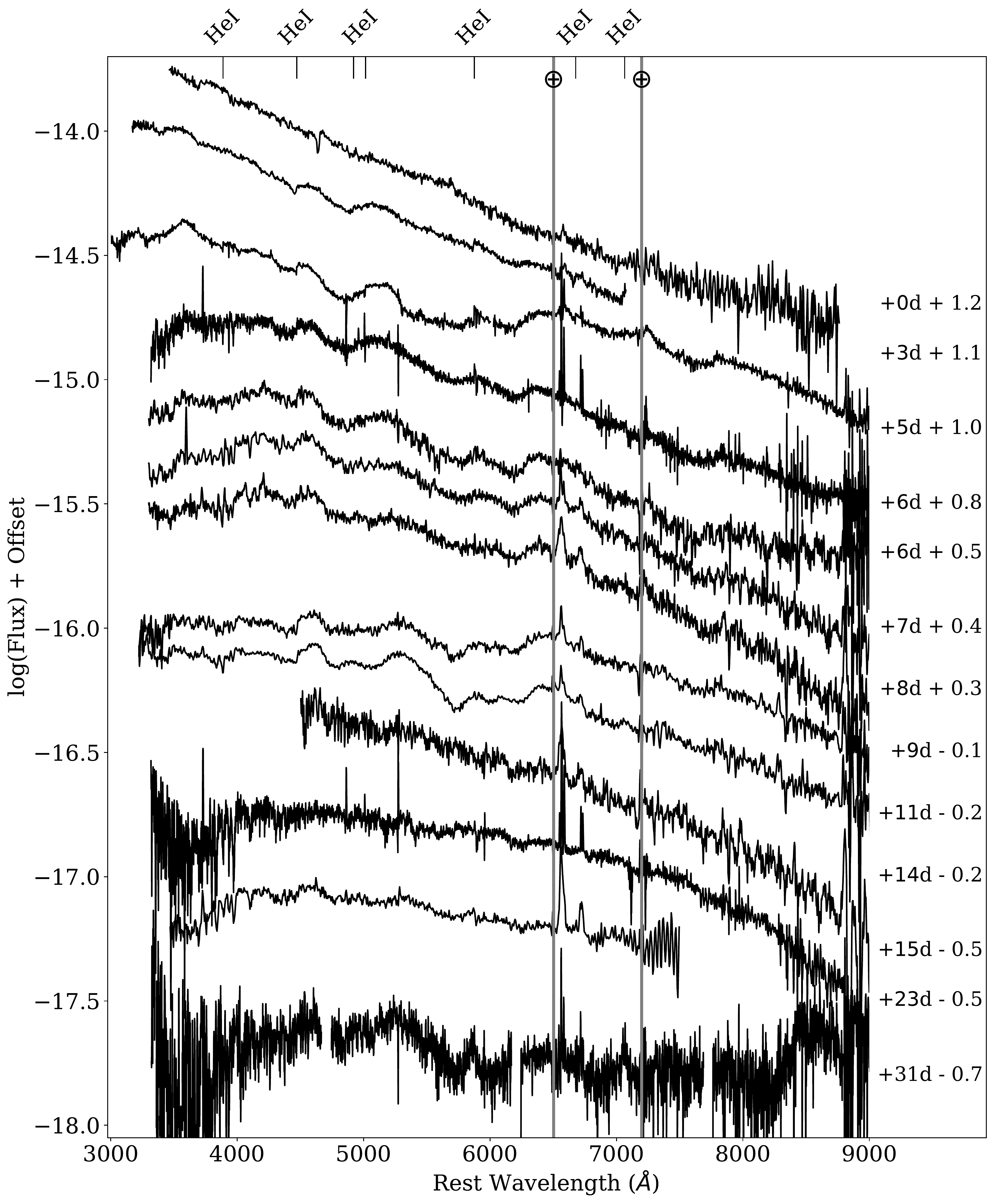}
\caption{Optical spectral sequence of LSQ13ddu. The location of the \Hei\ line sequence is shown by tick marks. The position of telluric regions are marked with cross hairs. For comparison with the lower resolution spectra, the XShooter spectrum at +5~d has undergone a 5-$\sigma$ clipping and been rebinned to 3~\AA\ in this plot.}
\label{LSQ13dduSpectralSequenceVIS}
\end{figure*}

\begin{figure*}
\includegraphics[width=12.cm]{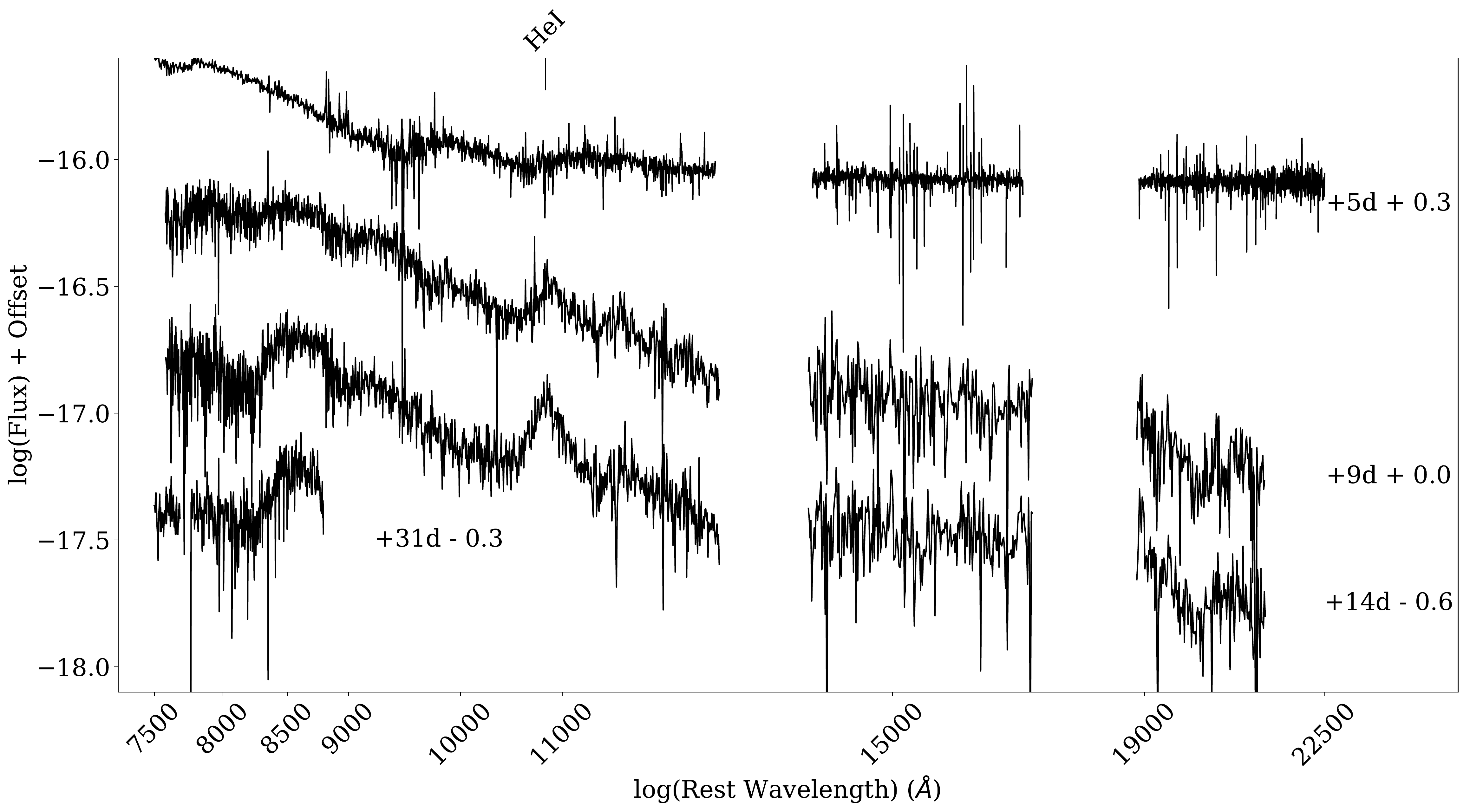}
\caption{NIR spectral sequence of LSQ13ddu. The $+$31~d Magellan Baade IMACS optical spectrum is also partially included to highlight the evolution of the \Caii\ triplet region. The XShooter spectrum at $+$5~d has undergone a 5$\sigma$ clipping procedure for display purposes. We note the low S/N of the XShooter NIR spectrum as a result of a short exposure time. Regions of significant telluric contamination across all spectra are not shown. The position of the strong \Hei\ 10830~\AA\ line is also indicated.}
\label{LSQ13dduSpectralSequenceNIR}
\end{figure*}

\begin{figure}
\includegraphics[width=7.7cm]{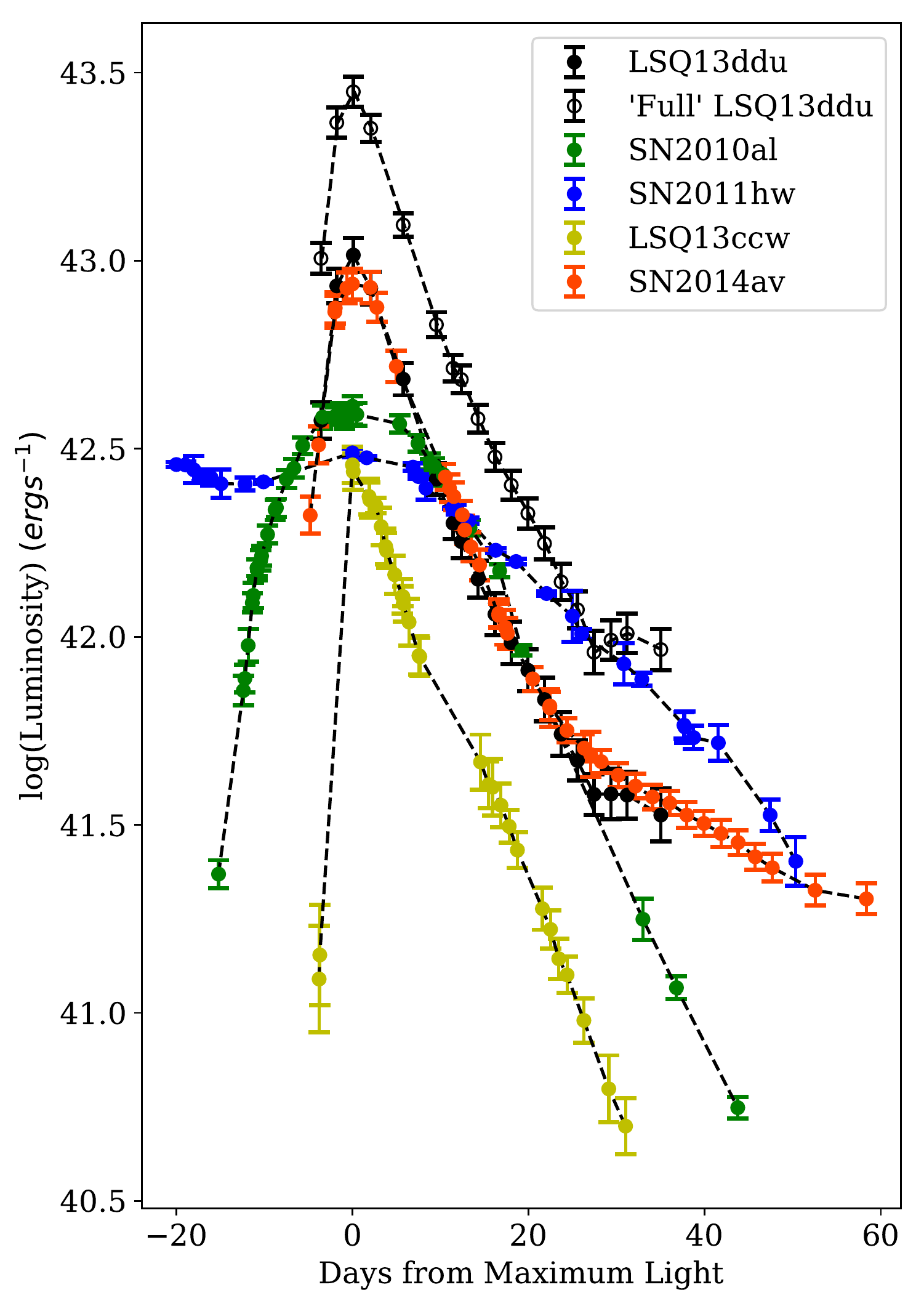}
\caption{Pseudo-bolometric (\textit{B} $\rightarrow$ \textit{i}) band light curve (solid black points) for LSQ13ddu compared to the light curves of a sample of SNe Ibn (detailed in the text). The `full' bolometric light curve of LSQ13ddu is also shown in open black points. A time dilation correction to account for the redshift of the objects has been applied to the light curves.}
\label{Bolometric_LC}
\end{figure}

\begin{table*}
\centering
\caption{Log of the spectroscopic observations of LSQ13ddu presented in this paper.}
\begin{adjustbox}{width=1\textwidth}
\label{SpectraLog}
\begin{tabular}{cccccccc}
\hline
MJD       & Date & Phase (d) & Telescope  & Instrument & Configuration       & Wavelength Range (\AA)$^a$ & Resolution ($\lambda$/$\Delta \lambda$)$^b$\\ \hline
56626.1   &  2013~Nov~29    & $+$0.1      & NTT        & EFOSC2     & GR13               & 3460 -- 8760     &  310   \\
56629.3   &  2013~Dec~02    & $+$3.3      & NTT        & EFOSC2     & GR11               & 3170 -- 7070    &  280      \\
56631.2   &  2013~Dec~04    & $+$5.2      & VLT        & XShooter  & UV, VIS, NIR$^c$   & 2970 -- 22500       &   5400,8820,5600        \\
56632.6   &   2013~Dec~05   & $+$6.4      & ANU 2.3m   & WiFeS      & B, R               & 3310 -- 9040    & 4430,2350  \\
56632.6   &   2013~Dec~05   & $+$6.6      & Las Cumbres        & FLOYDS     &-& 3060 -- 9450       &  470,380   \\
56633.6   &   2013~Dec~06   & $+$7.6      & Las Cumbres        & FLOYDS     &-& 3030 -- 10000    &    470,350        \\
56634.5   &   2013~Dec~07   & $+$8.5     & Las Cumbres        & FLOYDS     &-& 3030 -- 10000    &      280,230        \\
56635.2   &   2013~Dec~08   & $+$9.2      & NTT        & EFOSC2     & GR11, GR16 OG530 & 3180 -- 9450   &      390,440  \\
56635.3          &  2013~Dec~08    & $+$9.3      & Magellan Baade & FIRE       &  Longslit   & 7650 -- 22500  &  400  \\
56137.2   &  2013~Dec~10    & $+$11.2     & NTT        & EFOSC2     & GR11, GR16 OG530 & 3170 -- 9450   &  400,430    \\
56640.3  &  2013~Dec~13   & $+$14.3     & Magellan Baade & FIRE       & Longslit    & 7650 -- 22500   &  400 \\
56640.5 &   2013~Dec~13   & $+$14.5     & Las Cumbres        & FLOYDS     & Red           & 4500 -- 9600      & 200  \\
56641.5 &  2013~Dec~14   & $+$15.5     & ANU 2.3m   & WiFeS      & B, R               & 3310 -- 9040   & 4220,2380 \\
56649.1 &   2013~Dec~22   & $+$23.1     & NTT        & EFOSC2     & GR13               & 3470 -- 8760      &310 \\
56657.3       &   2013~Dec~30  & $+$31.3     & Magellan Baade & IMACS--f/4     & 300   & 3400 -- 9100  &   1300 \\ 
\hline
\end{tabular}
\end{adjustbox}
\begin{flushleft}
$^a$Wavelength ranges are given in the rest frame of the SN.\\
$^b$The resolution is calculated where possible from the night sky lines or taken from the instrument specifications.\\
$^c$UVB and VIS arms consisted of two 1300s exposures while due to an error the NIR arm was two 130s exposures resulting in low S/N in the NIR spectrum.\\
\end{flushleft}
\end{table*}

We computed two pseudo-bolometric light curves for LSQ13ddu using the python code, \textsc{superbol} \citep{Nicholl2018}. One using the data from the optical (\textit{B} $\rightarrow$ \textit{i}) bands to serve as a comparison with literature objects and as an input for photometric modelling using the light-curve fitting code \textsc{MOSFiT} \citep{Guillochon2018} (see Section~\ref{mostfitanalysis}). The second bolometric light curve calculated includes the early-time \textit{Swift} UV data and an extrapolation assuming a single temperature blackbody to account for non-observed bands. This `full' bolometric curve spans the range 1000--25000 \AA\ and is used to constrain the total peak luminosity. In both cases, we chose the band with the most consistent coverage, \textit{LSQgr}, as the reference filter. The other filters are interpolated to the epochs of \textit{LSQgr} data using a polynomial with the photometric uncertainties providing a weighting for the fits. At phases prior to or after the availability of multi-band photometry, constant colour approximations are used. 

The use of the optical pseudo-bolometric light curve is twofold. Firstly, literature objects do not typically have as extensive a photometric dataset as LSQ13ddu, with UV photometry not routinely obtained. Secondly, UV data is only available for LSQ13ddu at early times with the constant colour approximation significantly overestimating the UV contribution to the overall luminosity at late times. The LT \textit{z} band, RetroCam and SOFI NIR data have been excluded due to lack of consistent coverage and/or insufficient sampling.

The peak luminosity of the pseudo-bolometric light curves are after accounting for extinction 2.81$\pm$0.26$\times$10$^{43}$ erg s$^{-1}$ for the `full' version and 1.03$\pm$0.11$\times$10$^{43}$ erg s$^{-1}$ for the optical version. Optical pseudo-bolometric light curves of a sample of SNe Ibn (SN~2011hw; \citealt{Smith2012}, LSQ13ccw; \citealt{Pastorello2015}, SN~2014av; \citealt{Pastorello2016} and SN~2010al; \citealt{Pastorello2015e}) were computed in a similar manner and are compared to that of LSQ13ddu in Fig.~\ref{Bolometric_LC}. The behaviour of the optical bolometric light curve of LSQ13ddu is found to match that of SN~2014av, with SN~2014av having a peak luminosity of 8.49$\pm$0.82$\times$10$^{42}$ erg s$^{-1}$ along with very similar pre and post-maximum light curve evolution.

   \begin{figure}
    \includegraphics[width=8cm]{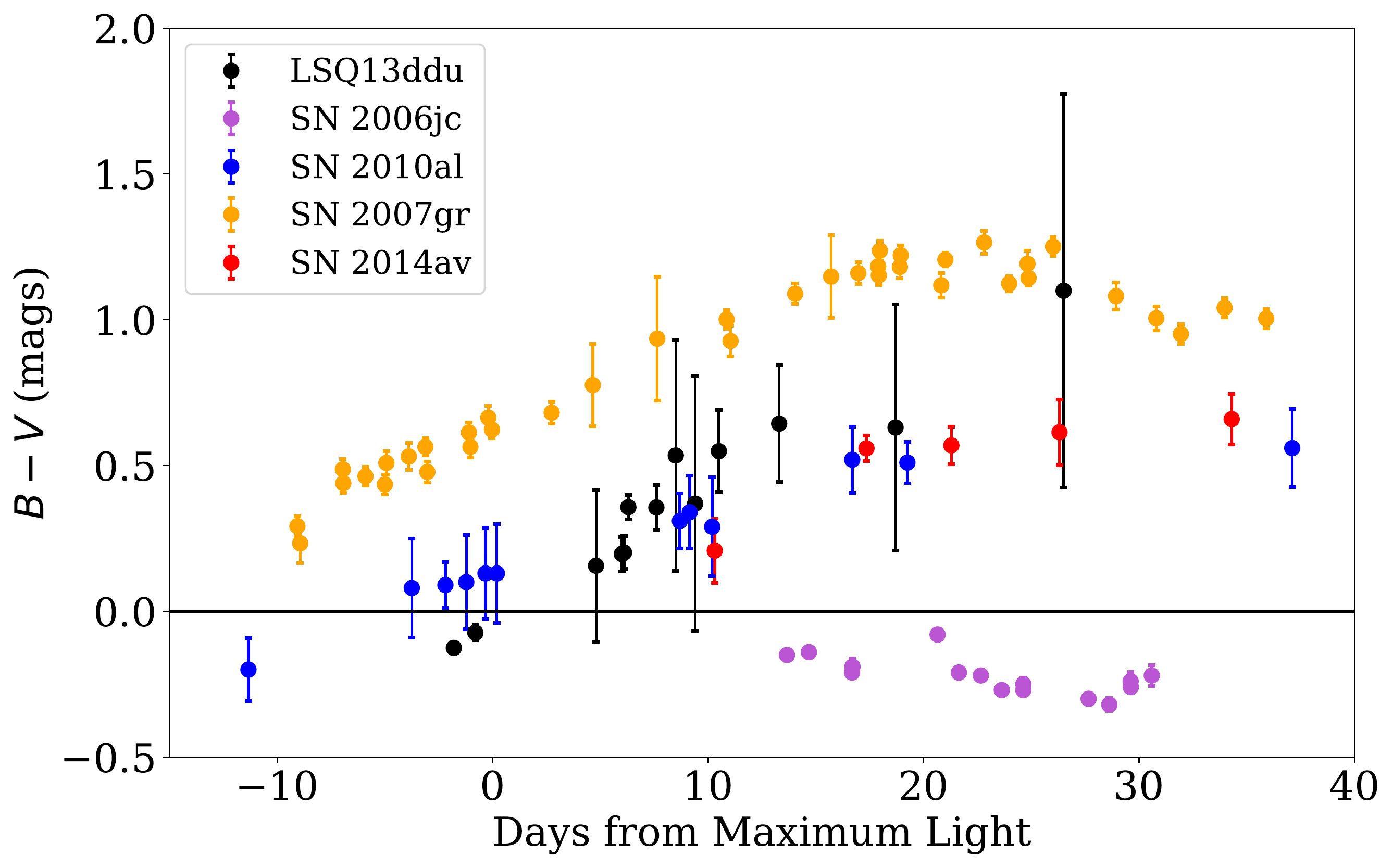}
    \caption{A comparison of the \textit{B-V} colour evolution of LSQ13ddu with SNe Ibn SN~2006jc \citep{Pastorello2007}, SN~2010al \citep{Pastorello2015e}, SN~2014av \citep{Pastorello2016}, and the Type SN Ic, SN~2007gr \citep{Hunter2009}.}	
    \label{ColourCurves}
    \end{figure}

\subsection{Blackbody temperature and colour evolution}
\label{sec:BBandColours}
The blackbody temperature of LSQ13ddu was estimated using a single temperature fit to both the spectroscopic and photometric data. The assumption is that the ejecta is a uniform temperature and lacks additional components from other effects such as dust re-radiating at redder wavelengths. In the spectral analysis, we measured the temperature using a least-squares fit to each spectrum using the \textsc{scipy} curve fitting routine \citep{SciPy}. The spectra have not undergone additional processing (e.g. no emission or absorption features have been removed prior to fitting) with the exception of the removal of the high noise regions at the edges of several spectra. As a consistency check, the blackbody temperature was also calculated using the available multi-band photometry at the same epochs as the spectra where there is sufficient multi-band observations ($\geq$4 bands) to produce a cubic spline interpolation to the photometry for fitting.

The earliest spectra ($+$0.1~d and $+$3.3~d) are relatively featureless and are well fitted by the blackbody model, with an initial temperature of 13000$\pm$500 K at $+$0.1~d, cooling to 11300$\pm$500 K at $+$3.3~d. At epochs after $\sim$3~d post maximum, the spectra of LSQ13ddu start to show stronger spectral features that deviate from the underlying blue continuum and make fitting a blackbody to the data more difficult. However, we measured a general cooling with time with the ejecta reaching a blackbody temperature of $\sim$7000 K at $+$6--8~d past maximum. The values obtained using both the spectra and multi-band photometry show good agreement. 

While the continued strengthening of broad spectral features prevents the determination of a reliable blackbody temperatures at later phases, colour measurements using photometric data can provide some insight into the continued temperature evolution of LSQ13ddu. Fig.~\ref{ColourCurves} is a comparison between the colour evolution of LSQ13ddu, the SNe Ibn, SN~2006jc \citep{Pastorello2007}, SN~2010al \citep{Pastorello2015e} and SN~2014av \citep{Pastorello2016}, along with the normal Type Ic SN~2007gr \citep{Hunter2009}. From this comparison the observed colour evolution of LSQ13ddu is found to be similar to that of SN~2010al and SN~2014av, with all three being redder than SN~2006jc but significantly bluer than the normal Type Ic SN~2007gr at similar phases.

\subsection{Spectroscopic analysis}
\label{sec:SpectroscopicAnalysis}

The features in the spectra of LSQ13ddu are analysed using the spectral fitting code \textsc{synapps} \citep{Thomas_Synapps}. Velocity measurements of the main features (both the narrow He and broader underlying features) are also analysed using Gaussian fits to the lines. The contamination of SN spectral features by narrow host galaxy emission lines, as well as the properties of the host, are also discussed.

    \begin{figure*}
    \includegraphics[width=14cm]{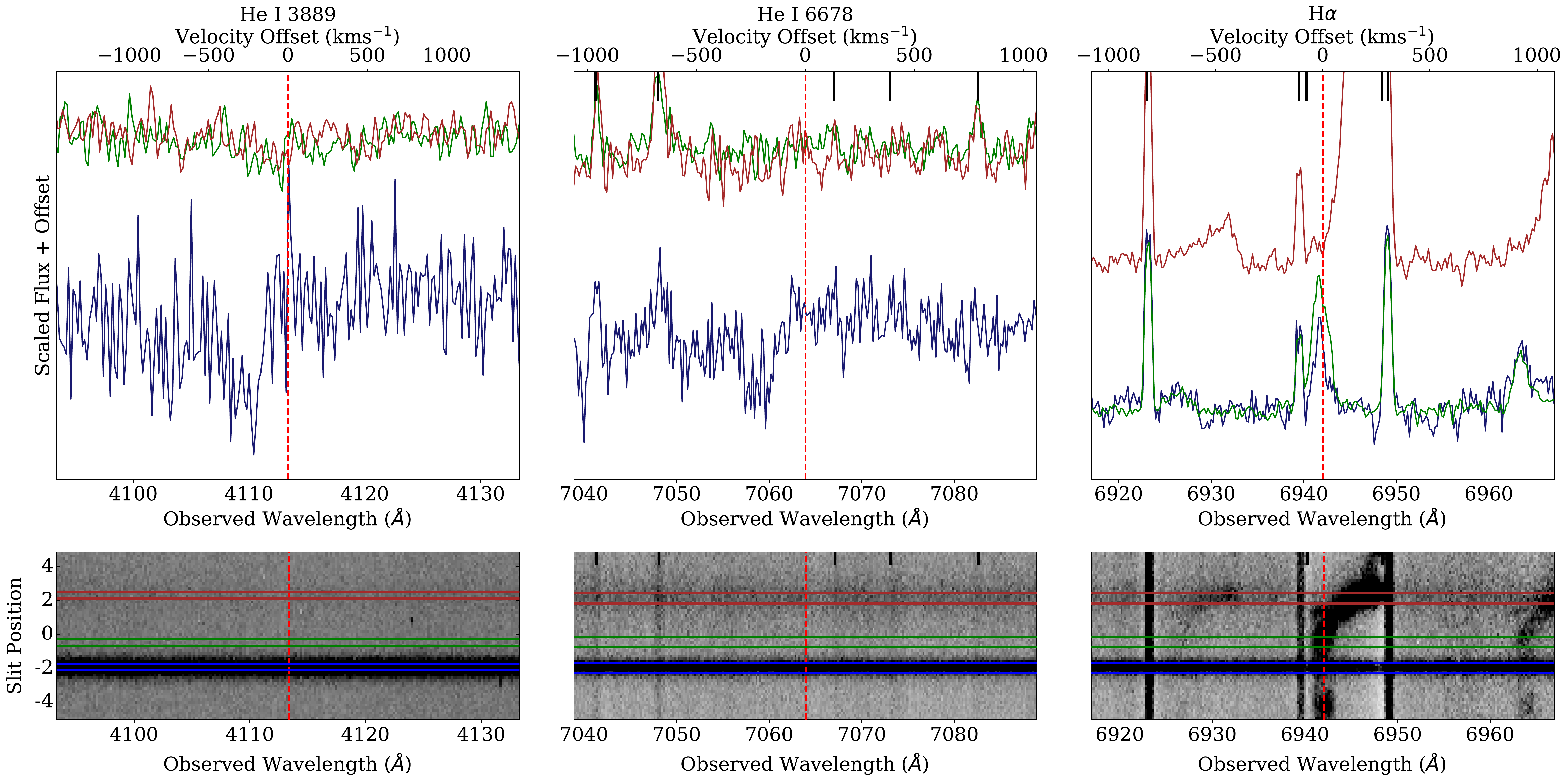}
    \caption{The upper panels show 1D spectral extractions at the position of the SN (blue), at a position close to the SN (green), and through the nucleus of the host galaxy (brown) in the observed wavelength range of \Hei\ 3889~\AA\ (left), \Hei\ 6678~\AA\ (middle) and H$\alpha$ (right). Zero velocity is based on the SN position and labelled as a red dashed line. The \Hei\ SN spectrum is the median combination of both stare exposures. Y-axis offsets have been made for clarity. The corresponding regions in the 2D frames are displayed in the bottom panels with the same colour coding. Strong sky lines are indicated with short black ticks at the top of the plots. In the right-hand plots of centred on the H$\alpha$ feature, the \Nii\ 6583~\AA\ line is seen observed wavelength of 6963~\AA\ for both the offset position close to the SN and at the SN position with both closely matched in their profile shape and strength.}
    \label{2DFrame_Comparison}
    \end{figure*}
    
\subsubsection{Host galaxy properties and line contamination}
\label{sec:HostGalaxy}

Due to its location within its host, the spectra of LSQ13ddu show contamination from narrow host emission lines. As the CSM signatures are narrow and weak, care must be taken to disentangle the host contamination from potential SN-related features. Their relative contributions were determined in three ways: i) a line velocity analysis and ii) a spectral comparison between the host and LSQ13ddu using multiple extractions of the 2D spectra, and iii) a comparison between the H$\alpha$ and \Nii\ 6583~\AA\ features. 

Firstly, we measured the full width half-maximum (FWHM) velocity of the host galaxy emission lines using Gaussian fits in one of our highest resolution spectra obtained with XShooter, which has a velocity resolution of $\sim$35~\kms\ (measured from sky lines). The FWHM velocity of the H$\alpha$ feature was measured to be 260$\pm$40 \kms, which was consistent with the widths of the nebular emission lines (not typically seen in SN spectra) of 220$\pm$50, 250$\pm$40 and 180$\pm$40 \kms\ for the \Nii\ 6548~\AA, \Nii\ 6583~\AA, and \Sii\ 6715~\AA\ lines, respectively.

Secondly, to determine if there is a contribution from CSM associated with the SN to the narrow H$\alpha$ signature, extractions from stare-mode 2D spectra were made at three locations along the slit as shown in Fig.~\ref{2DFrame_Comparison}: i) at the SN location, ii) at a position offset but close to the SN, and iii) at the nucleus of the host galaxy. As the extended emission from the host spans the full frame, clean sky subtraction was difficult and to remove the possibility of introducing artefacts into the spectra, was not attempted. The location of the sky lines have been marked in both the 1D and 2D spectra using black ticks. For the H$\alpha$ region, a velocity gradient is seen across the slit with the peak of the H$\alpha$ emission of the nucleus offset from the SN position by $\sim$200--250~\kms (no such feature is visible at the \Hei\ lines). The H$\alpha$ and \Nii~6583 \AA\ emission features at the offset position close to the SN and at the location of the SN are well matched in both width and strength.

Thirdly, we estimated the size of the potential SN contribution to the H$\alpha$ feature by subtracting a Gaussian with a width equivalent to the measured FWHM of the host galaxy \Nii~6583\AA\ line from the H$\alpha$ feature of the NOD mode +~5~d XShooter spectrum. When the Gaussian estimates of the contribution from the host galaxy are subtracted off, there is a residual flux of 14$\pm$14 per cent (using the uncertainties on the measured FWHM), which within the uncertainties is consistent with zero, suggesting little or no SN contribution to the H$\alpha$ feature.

Taken together, the consistent line widths and strengths at the location of LSQ13ddu compared to the offset positions shows that the narrow emission feature of H$\alpha$ is dominated by host light and does not appear to be intrinsic to the environment of LSQ13ddu, although a small contribution of up to $\sim$14 percent cannot be ruled out. 

Fig.~\ref{2DFrame_Comparison} also includes the corresponding 1D and 2D frames at the locations of the \Hei\ 3889~\AA\ and \Hei\ 6678~\AA\ lines (features from the UVB and VIS XShooter arms respectively). The \Hei\ features are seen as weak P-Cygni profiles at the SN position of both lines but are not seen at the positions of the host nucleus or the offset position from the SN. 

The local metallicity of the host galaxy at the location of LSQ13ddu was estimated from measurements of the line strength ratios of key galaxy emission lines in the $+$15~d WiFeS spectrum. The gas-phase metallicity, 12~+~log([OIII]/H), of the host galaxy was calculated to be 8.51$\pm$0.25 based on the O3N2 conversion ratio from \cite{Pettini2004}. This value is equivalent to $\sim$0.7 times the solar metallicity value of \cite{Asplund2009} and is consistent with previous metallicity studies of SNe Ibn that find sub-solar host metallicities for the class \citep{Pastorello2015e, Taddia2015}.
Additionally, an archival 6DF spectrum \citep{Jones2009} of the nuclear region of the host of LSQ13ddu was used to identify any active galactic nucleus (AGN) activity. From this spectrum a log([NIII/H$\alpha$]) value of $-$0.425 and a log([OIII/H$\beta$]) value of $-$0.177 were estimated, placing the host below both the Kewley theoretical dividing line \citep{Kewley2001} and the Kauffmann SDSS based empirical dividing line \citep{Kauffmann2003} on the `Baldwin, Phillips \& Terlevich' (BPT) diagram \citep{Baldwin1981}, we thus conclude that the galaxy is not the host of an AGN and has the properties of a starforming galaxy.

    \begin{figure*}
    \includegraphics[width=12.8cm]{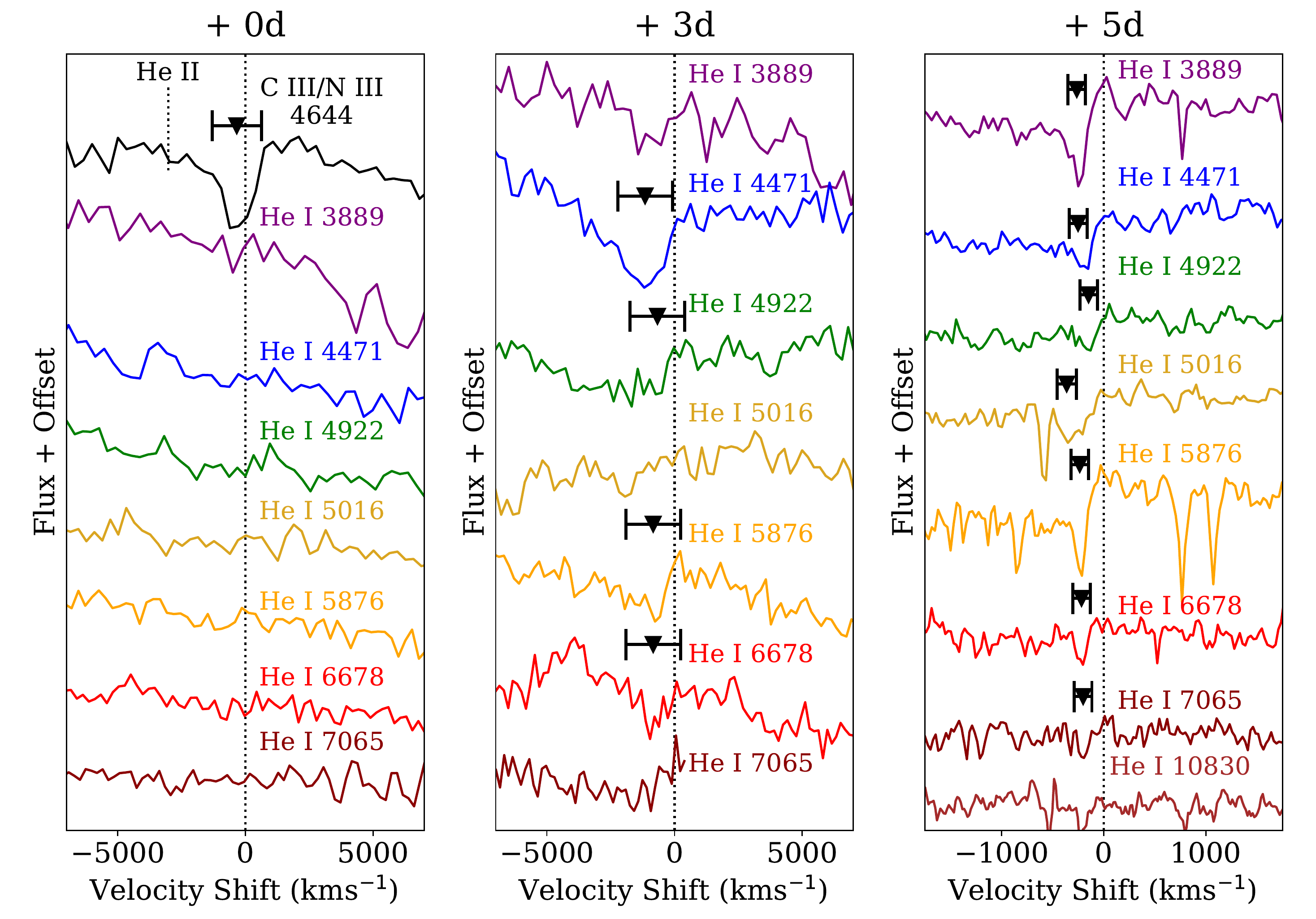}
    \caption{Spectral regions containing the strongest identified narrow \Hei\ features from 0 to $+$5~d after maximum light (left to right). Each feature displayed in velocity space relative to the rest location of the corresponding \Hei\ line. The absorption minima of detections are marked as solid downward facing triangles. The velocity of the blended \Ciii\ + \Niii\ feature in the initial spectrum if treated as \Heii\ is also indicated as a short vertical dotted line.}
    \label{He_Velocities_Evolution}
    \end{figure*}

  \begin{figure}
    \includegraphics[width=8cm]{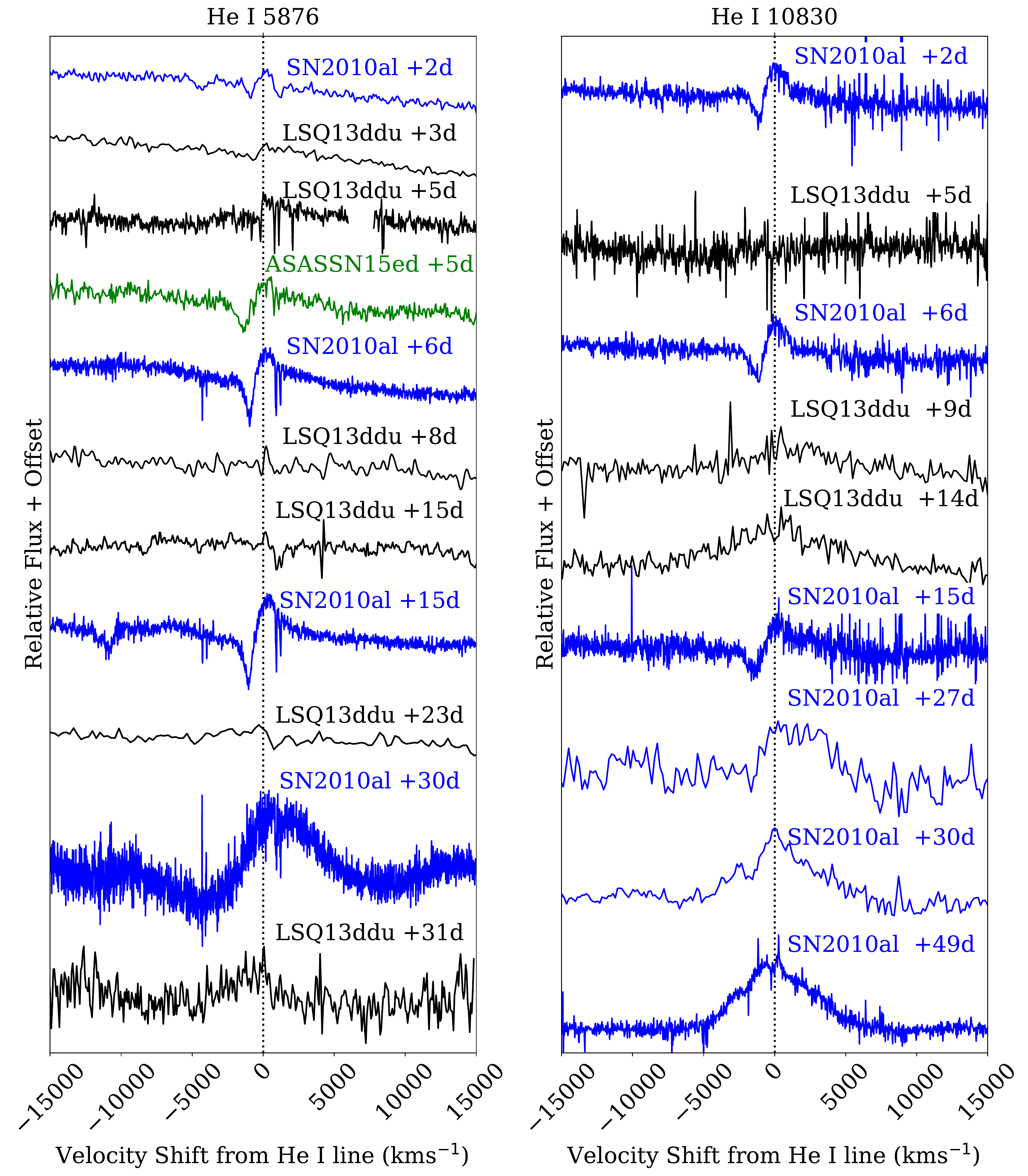}
    \caption{Comparison between the \Hei\ 5876 and 10830 \AA\ lines of LSQ13ddu (black), SN~2010al (blue) and ASSSN-15ed (green). Phases are from maximum light in the observer frame.}
    \label{ComparisonHeI}
    \end{figure}  

\subsubsection{Presence of narrow P-Cygni features} 
\label{Subsection:vel_he}
We investigated the presence of the weak and narrow He-related features in the spectra of LSQ13ddu. Fig.~\ref{He_Velocities_Evolution} shows the time evolution from 0 to $+$5 d of the narrow P-Cygni features that are visible in the early optical spectra. A P-Cygni feature initially linked to $\Heii$~4686~\AA\ but as discussed below, more likely a blend of C and N, was clearly detected in the earliest spectrum of LSQ13ddu at $+$0.1~d (left panel). This is the only clear feature identified in the maximum light spectrum and is no longer visible at $+$3.3~d. Features similar to this have previously been identified in SN~2010al \citep{Pastorello2015e} and tentatively in SN~2014av \citep{Pastorello2016}.
In particular the feature in the earliest two spectra obtained of SN~2010al initially presented as a double-peaked emission feature before transitioning to two distinct P-Cygni profiled absorption lines by 6 days after maximum, before disappearing by 15 days after maximum light. This feature was identified as a combination of the \Heii\ 4686~\AA\ line, responsible for the redder peak, and a blended \Ciii\ 4648~\AA\ + \Niii\ 4640~\AA\ feature producing the bluer emission peak. The velocities of these features measured based on their absorption minima was found to be $\sim$~1,000~\kms\ consistent with measurements of \Hei\ P-Cygni absorption minima in the same spectrum at 1,000 -- 1,100~\kms. The identification of \Heii\ in SN~2010al was confirmed through the detection of the \Heii\ 5411 and 8236~\AA\, features that are not observed in LSQ13ddu.
 
The velocity of the absorption minimum of the feature in the first spectrum of LSQ13ddu was measured to be 2950~\kms\ if treated as \Heii\ 4686~\AA\, 600~\kms\ if treated as \Ciii\ 4648~\AA, and 80~\kms\ if treated as \Niii, giving a mean \Ciii\ + \Niii\ (i.e. 4644~\AA) blend velocity of 340~\kms\ . We discuss the most likely identification of this feature below in the context of the \Hei\ line velocities.

In common with SNe Ibn, narrow \Hei\ features were clearly observed in multiple spectra of LSQ13ddu starting at $+$3~d after maximum light and continuing until $\sim$11~d with a decrease in strength with time, with these features appearing largely in absorption, with faint emission components observed in the higher resolution +5~d XShooter spectrum. Figure~\ref{ComparisonHeI} is a comparison between the behaviour of the \Hei\ lines in the spectra of LSQ13ddu and those in the spectra of SN~2010al and ASSASSN-15ed. The strongest (although still weak) features seen in LSQ13ddu are those of \Hei\ 3889, 4471, 5876 and 6678~\AA). In the higher resolution XShooter and WiFeS spectra obtained at $+$5 and $+$6~d, respectively, weaker narrow \Hei\ features at 3889, 4922, 5016, and 7065~\AA\ can also be identified. Marginal detections of \Hei\ 4471, 5016 and 5876 \AA\ features are seen in the first spectrum of LSQ13ddu at velocities comparable to those seen in later spectra. However, as they are of low significance, they are excluded from further analysis. 

\begin{figure}
    \includegraphics[width = 8cm]{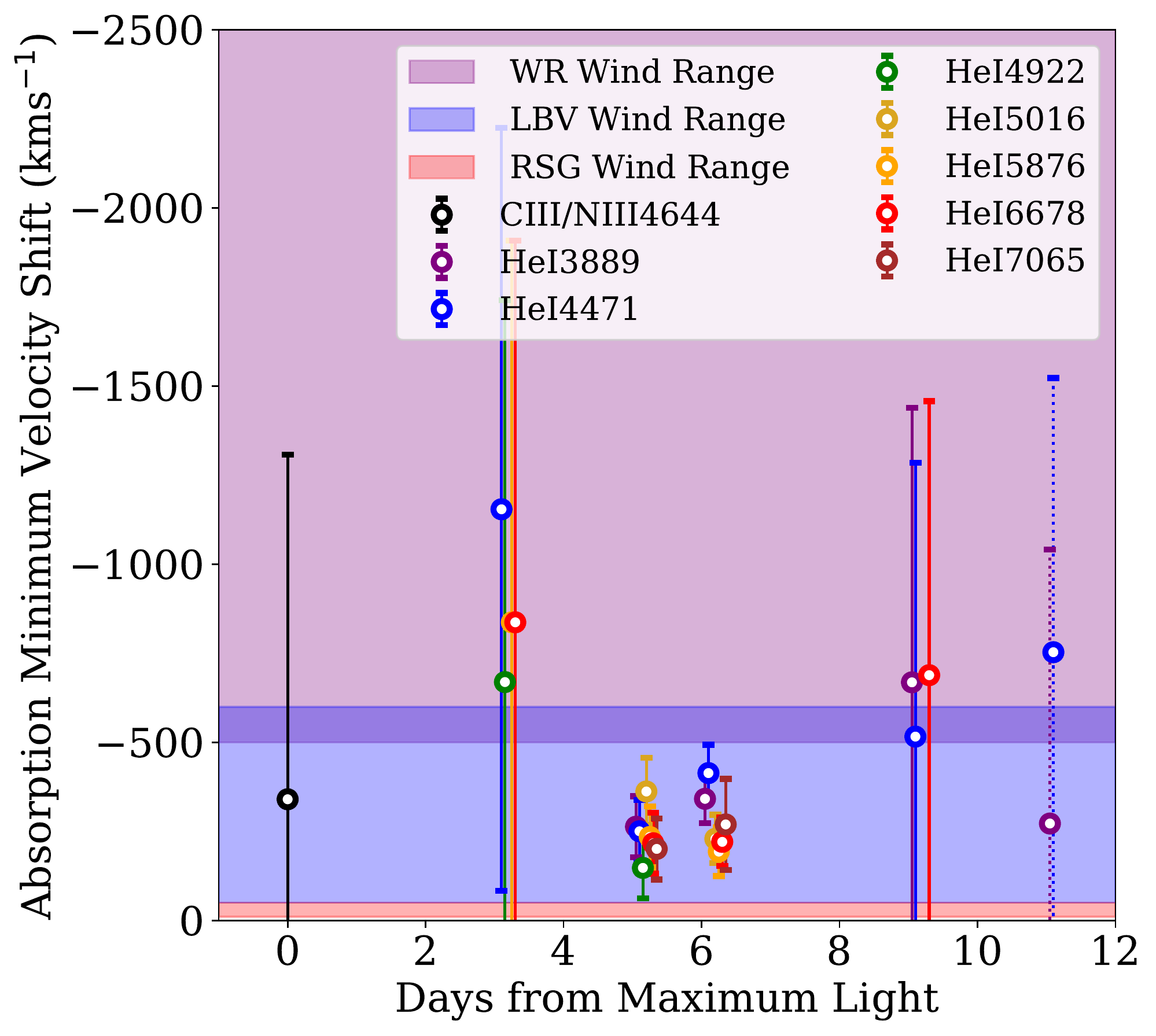}
    \caption{Velocity measurements of \Hei\ features observed in the spectra of LSQ13ddu along with the \Ciii + \Niii\ blend observed in the first spectrum. Points with dashed error bars indicate features with uncertain identification due to lower signal-to-noise ratios. The expected range of WR wind speeds (500--3200 \kms) \citep{Crowther2007} is shown as the purple shaded region, the expected wind speeds of LBV stars (50--600 \kms) is shown in blue \citep{Smith2014}  and the expected wind speed range of RSG stars (10--50 \kms) shown in red \citep{vanLoon2005}) is shown as the red shaded region. The data for each He line are offset by $+$0.05~d in phase for clarity. Errors are largely dominated by the resolution of the instruments.}
    \label{He_Velocities}
    \end{figure}

The blueshifted absorption minimum velocities were estimated for the identified \Hei\ features and found to be in the range $\sim$90 -- 1120~\kms\ (Fig.~\ref{He_Velocities}). The uncertainties on the velocities were estimated by taking the uncertainties on the fits added in quadrature to the uncertainties due to the spectral resolution. The highest resolution spectrum is the XShooter one at $+$5~d with a resolution of $\sim$35 \kms, while the lowest resolution obtained with EFOSC2 is $\sim$600~\kms, with the higher resolution spectra giving significantly lower absorption velocities.

To determine if differing spectral resolution is the main source of the variation in the measured velocities, we rebinned the XShooter spectrum to 3~\AA\ to match our lower resolution spectra and remeasured the velocities of the \Hei\ features. When measured from the rebinned spectrum, the lines were found to be broadened significantly, the absorption velocities of the 4471 and 6678~\AA\ lines increased from 250$\pm$90 and 220$\pm$90~\kms\ to 860$\pm$200 and 850$\pm$180~\kms, respectively. Additionally the weaker feature at 7065~\AA\ was no longer distinguishable in the spectrum. Therefore, we conclude the \Hei\ velocities are constant within the uncertainties in range of $+$3 to $+$11~d and using the highest resolution XShooter and WiFeS spectra, we obtain a weighted mean \Hei\ absorption minimum velocity of 250$\pm$20 \kms, which we use hereafter as our preferred value. The velocity of \Hei\ features in SNe Ibn has been shown to display little evolution with time \citep{Pastorello2016} and are typically higher than the values obtained for LSQ13ddu but this may be a consequence of the higher than normal resolution of our LSQ13ddu spectra compared to the literature sample.

If we assume that the material producing the \Hei\ features is also responsible for producing the features observed in the first spectrum of LSQ13ddu, then to obtain a consistent velocity, the first spectrum feature is most likely a blend of \Ciii\ 4648~\AA\ and \Niii\ 4640~\AA\ at $\sim$340 \kms, and not \Heii\ 4686~\AA. This is corroborated by the lack of detection of other \Heii\ lines, such as those at 5411 and 8236~\AA. Although there is no detection of a narrow P-Cygni \Hei\ 10830~\AA\ feature in the $+$5~d spectrum of LSQ13ddu, a broad \Hei\ 10830~\AA\ emission feature is observed in the $+$9 and $+$14~d spectra and is discussed further in the context of the other broad spectral features in Section \ref{subsec:Broadfeatures}. 

    \begin{figure}
    \includegraphics[width=8cm]{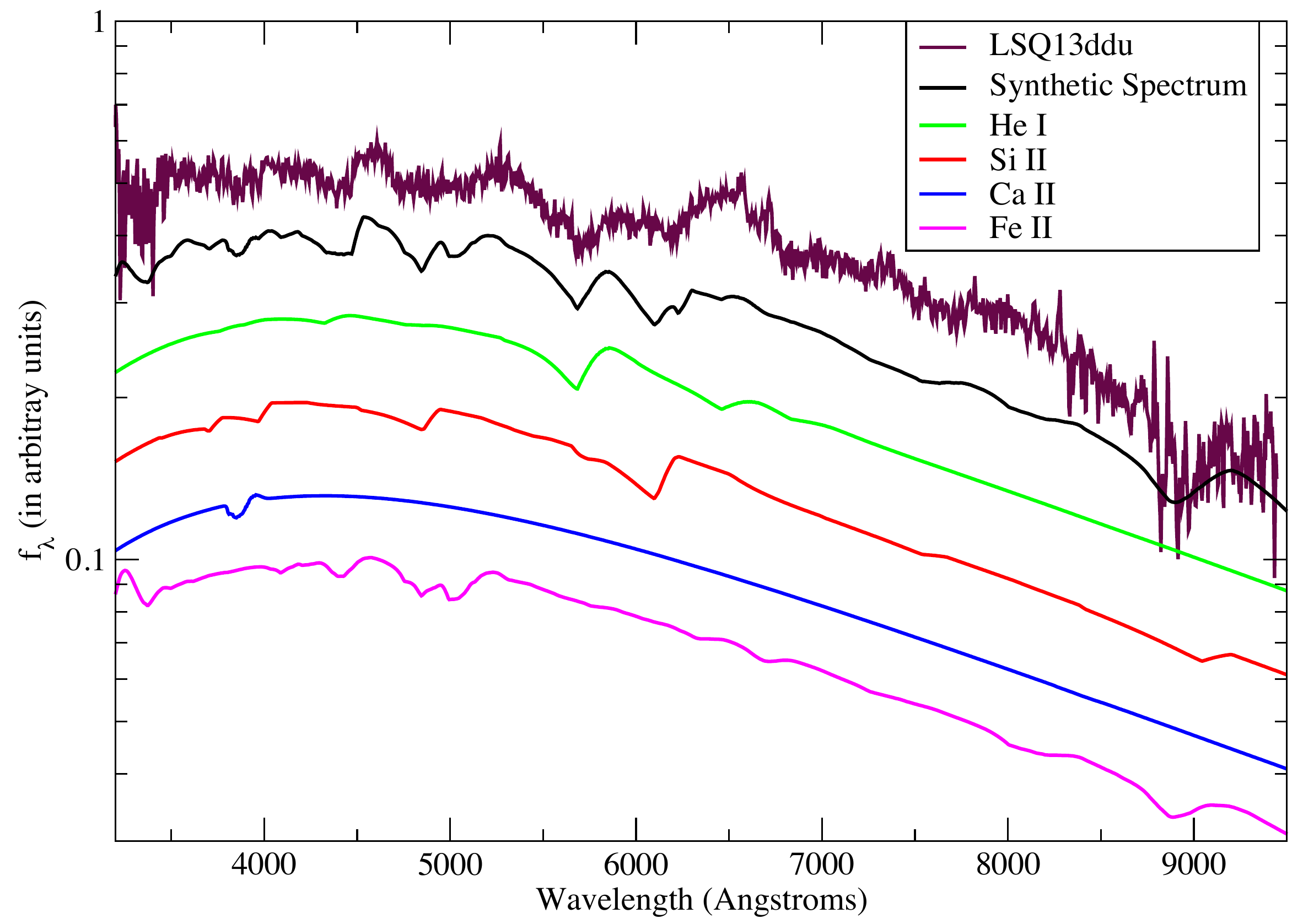}
    \caption{The $+$9~d spectrum of LSQ13ddu (purple) along with the best \textsc{synapps} fit (black). The individual line identifications for the unambiguously determined ionic species of \Hei\ (green), \Sii\ (red), \Caii\ (blue), and \Feii\ (magenta) are plotted underneath. All have a photospheric velocity of 10200 \kms. Additional, weaker lines from \Oi, \Coii\ may also be present, but do not significantly contribute to the spectral fit.}
    \label{synapps_v1}
    \end{figure}

\subsubsection{Spectral evolution of broad features and line identification}
\label{subsec:Broadfeatures}
In addition to the narrow features seen in LSQ13ddu, a broad emission feature is visible in the range $\sim$3700--3900 \AA\ in the first $+$0.1~d spectrum of LSQ13ddu (see Fig.~\ref{LSQ13dduSpectralSequenceVIS}). The identification of the elements producing these features is uncertain but this region typically contains \Feii\ lines \citep{Pastorello2016}. If this broad feature is taken to be a single feature, it displays a FWHM velocity of 15700$\pm$2600 \kms, which is typical of early-phase SN ejecta. A similar feature was identified in a maximum-light spectrum of SN~2014av (and attributed to a \Caii\ H\&K and \Hei\ line blend) but its peak was blueshifted by $\sim$6000 \kms\ with respect to that of LSQ13ddu. By the time of the second spectrum at $+$3~d, the broad feature at 3700--3900 \AA\ has disappeared. New broad features have developed at 3350--3750 \AA, 4440--4700 \AA\, and 4850--5300 \AA, which are likely due to \Feii\ emission \citep{Pastorello2016}.

At later times ($>+$9~d), the spectra have evolved to become dominated by these broad features. We have used the spectral fitting code \textsc{synapps} described in \cite{Thomas_Synapps} to the $+$9~d spectrum, to determine the elements contributing to the broad SN-like features that develop with time. \textsc{synapps} is based on the \textsc{synow} fitting code \citep{Fisher2000} but instead of being interactive, it optimises the input parameters to produce a best fit. \textsc{synapps} has the same assumptions and limitations of \textsc{synow} and is best for line identification and rough estimates of the spectral velocities since it does not provide measures of the ionic species abundances and therefore, cannot be used to infer the masses of the elements present in the ejecta. 

Fig.~\ref{synapps_v1} shows the \textsc{synapps} fits to the $+$9~d spectrum, along with the key features contributing to each line. The ions used in the fits are: \Cii, \Caii, \Coii, \Feii, \Hei, \Niii, \Oi, and \Siii. In particular, underlying broad \Hei\ is essential to fit the $\sim$5200--6000 and $\sim$6300--6600 \AA\ regions. The velocities of these ions are 10200 \kms. The presence of \Heii\ and H were checked for but neither provided a significant improvement to the fit and thus we can not unambiguously identify their presence in the spectrum. 

A broad \Hei\ 10830 \AA\ emission feature is observed in the $+$9 and $+$14 d spectra, with the strength of this feature seen to increase with time (Fig.~\ref{LSQ13dduSpectralSequenceNIR}). Its measured FWHM velocity is 9400$\pm$700 and 10200$\pm$700~\kms\ at $+$9 and $+$14~d past maximum light, respectively, which are consistent within the uncertainties. The 10830~\AA\ feature at $+$9~d appears to be redshifted by $\sim$1200~\kms\ with respect to its rest wavelength but the $+$14~d feature is consistent with no velocity offset. This could suggest a residual though unresolved narrow \Hei\ feature is causing the offset in the first spectrum and that is gone by the time of the later spectrum.

The NIR \Caii\ triplet emission feature at 8200--9000 \AA\ (which is likely blended with an \Oi\ line) first appears at $+$9~d and grows stronger with time. This evolution is clearly seen in the NIR spectra shown in Fig.~\ref{LSQ13dduSpectralSequenceNIR}. The strength of the potential Fe-group emission, at 9000--9550 \AA\ and 11300--11900 \AA\ and features due to \Oi\ and \Mgii\ emission (7650--8100 \AA) are also seen to grow with time. These broad features are similar to those seen in stripped-envelope SNe (SE-SNe) and are compared to a sample of them in Section \ref{subsection:Hybrid}.

\subsection{Powering the light curve of LSQ13ddu}
\label{section:LightcurvePowering}

The rapid rise of LSQ13ddu to such a bright peak luminosity followed by a decay over a similar timescale is difficult to achieve with a purely \nick\ power source, with significant tension between the large \nick\ mass required to produce such a bright event and its rapid photometric evolution (see Section \ref{mosfitnick}).
Studies of SNe Ibn have suggested that their light curves are likely powered by the decay of radioactive \nick\ combined with an additional component from CSM interaction, i.e.~as if they are CSM-enshrouded SE-SNe \citep[e.g.][]{Chugai2009}. Although the He-interaction features in LSQ13ddu are much weaker than in the SN Ibn sample studied to date, an additional (or alternative) powering source to \nick\ decay is needed to explain its light-curve behaviour. A combined \nick\ and CSM interaction model agrees qualitatively with the observed spectral evolution of LSQ13ddu, where narrow He features suggestive of CSM interaction are seen in the early spectra but fade away and become undetectable within a few weeks of maximum light. 

\begin{figure*}
     \includegraphics[width=15cm]{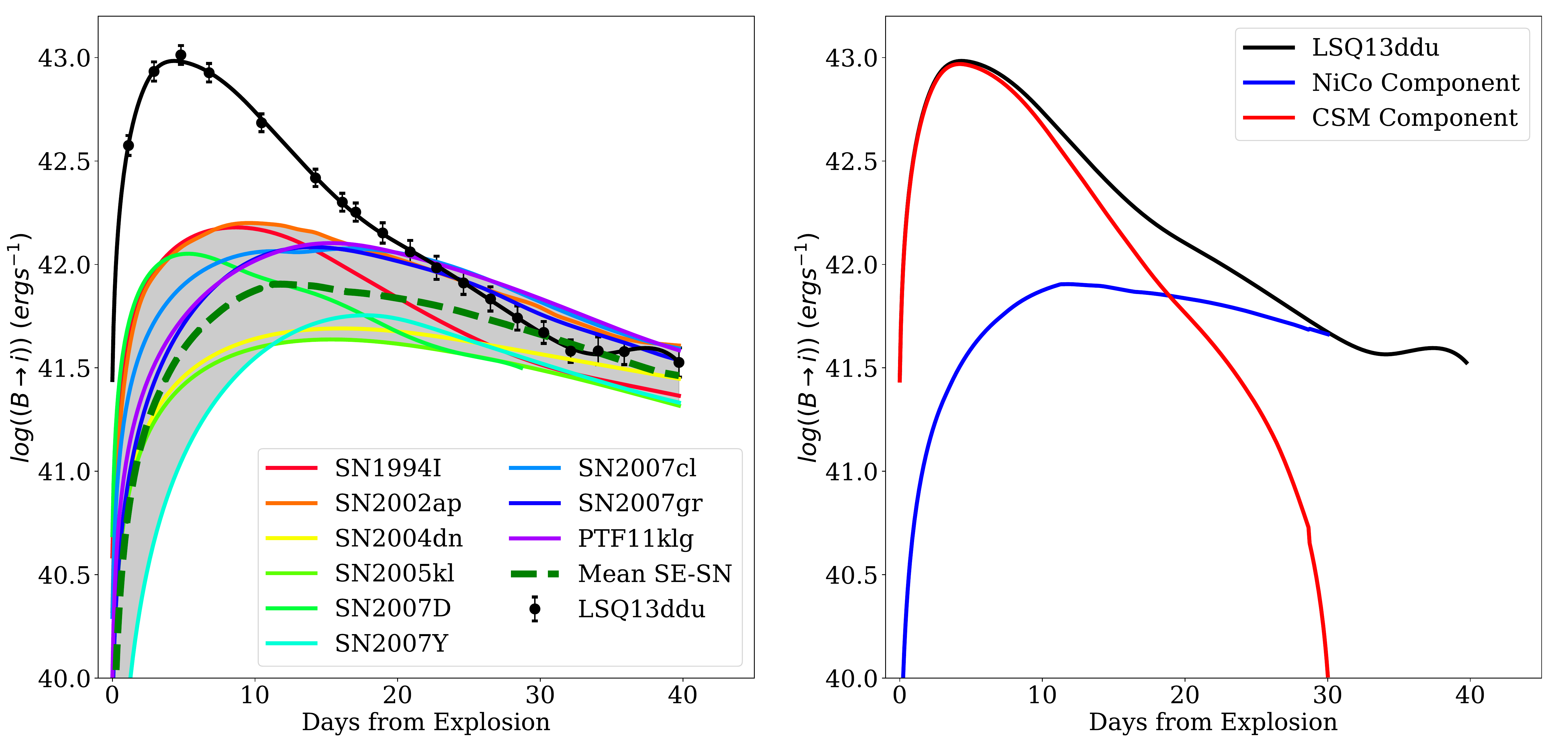}
    \caption{Left panel: The bolometric light curve (black data points with black solid line) of LSQ13ddu compared to a subset of SE-SNe from \citep{Prentice2016} showing similar late time photometric behaviour. The individual events are shown as coloured solid lines and the grey shaded region highlights the range within this sample. A SE-SN template was calculated as the mean value of this range, shown as a green dashed line. Right panel: A comparison showing the additional flux required above that provided by the mean SE-SN template assuming SE-SNe are powered by radioactive \nick\ decay (NiCo component, blue solid line). The additional flux needed to power the light curve of LSQ13ddu (shown in black) is shown to arise from CSM interaction (shown as a red solid line).}
    \label{PrenticeIcComparisonPlot}
    \end{figure*}

To determine if LSQ13ddu could be powered by a combination of early-time CSM interaction and the \nick\ decay of an underlying SE-SN light curve dominating at later times, we compared the bolometric light curve of LSQ13ddu with a light curve sample of SE-SNe from \cite{Prentice2016}. We identified a subset of objects with similar late-time photometric evolution to LSQ13ddu (left panel of Fig.~\ref{PrenticeIcComparisonPlot}) and used them to determine an average SE-SN light curve template. The sample used to produce this mean light curve contains one SN Ib (SN~2007Y), two broad-lined Ic SNe (Ic-BL, SNe~2002ap and 2007D) with the remaining six objects classified as SNe Ic. This template light curve was then subtracted from the LSQ13ddu light curve to obtain the additional luminosity component required to power the light curve. We find at least qualitatively using this comparison that the light curve of LSQ13ddu could be powered by an underlying SE-SN with a contribution at early times from CSM interaction (right panel of Fig.~\ref{PrenticeIcComparisonPlot}). This is supported by the weakening of the narrow CSM linked \Hei\ features which are not detected in the spectra of LSQ13ddu at phases greater than $+$11~d post maximum. The late-time spectra much more closely resembling those of a SE-SN. We also explored the plausibility of a light curve with a CSM contribution through photometric modelling of LSQ13ddu (see Section \ref{mostfitanalysis}).

\subsection{Photometric Modelling}
\label{mostfitanalysis}

A number of photometric modelling codes using semi-analytic models to provide fits to the observed light curves of SNe and other transients have become available in recent years, such as \textsc{MOSFiT} \citep{Guillochon2018} and \textsc{TigerFit} \citep{Tigerfit}. In this section we discuss the use of \textsc{MOSFiT}\footnote{\textsc{MOSFiT} version 1.1.1 was used throughout this work} to produce model fits to the light curve of LSQ13ddu. We chose to use \textsc{MOSFiT} due to its configurable nature, both in terms of the model to apply and the parameter inputs which can be constrained based on available observational data. Additionally, \textsc{MOSFiT} is a Monte-Carlo code and as such provides more robust statistical uncertainties on its output results than some other codes.

\textsc{MOSFiT} has previously been used to model the photometric properties of SLSNe \citep{Nicholl2017} and in the wider exploration of the duration-luminosity phase space occupied by astrophysical transients \citep{Villar2017}. One of the main limitations of \textsc{MOSFiT} is the assumption of single (or multiple) blackbodies to describe the SED, which is not appropriate for objects whose spectra deviate significantly from a blackbody due to the presence of strong absorption or emission features. 

For LSQ13ddu, an assumption of a blackbody SED is appropriate around maximum light but less so at later times when its spectra are dominated by broad emission and absorption features. \textsc{MOSFIT} has also not been previously applied to such a rapidly evolving transient that is modelled with both CSM and nickel-powered components. As such, the results of these semi-analytic models should be treated as indicative of the applicability of a powering mechanism rather than quantitative constraints on the specific configuration of the CSM or the ejecta mass required. For further limitations of modelling with \textsc{MOSFiT} see \cite{Nicholl2017} and \cite{Villar2017}.

Based on our earlier analysis of the photometric and spectroscopic evolution of LSQ13ddu, we compared the light curves of LSQ13ddu to three potential powering models, i) radioactive \nick\ decay, ii) CSM-interaction driven, and iii) magnetar spin down. \textsc{MOSFiT} has a combined \nick\ + CSM model that has been used to model high mass, superluminous SNe \cite[SLSNe][]{Nicholl2017}. The CSM aspect of this model is based on the work of \cite{Chatzopoulos2013a}, and makes several assumptions that, while appropriate for use in modelling high-mass SLSNe, limit its applicability to fast-evolving events with low CSM masses. In particular, the diffusion time within the model is significantly overestimated and as such when presented with a fast-evolving light curve, the model favours a very low CSM and ejecta mass with an increased ejecta velocity. This leads to the model drifting to a region where other underlying assumptions (e.g.~the diffusion of the products of \nick\ decay through both the ejecta and optically thick region of the CSM) become less applicable. As such we do not employ this model in this analysis. 

Instead to measure a potential CSM contribution to the light curve with \textsc{MOSFiT}, we fitted the two components obtained in Section~\ref{section:LightcurvePowering} separately, with this choice ensuring the CSM assumptions (and their associated uncertainties) only affect the CSM contribution and not the standard \nick\ model. The ejecta velocity of the CSM component was allowed to reach values much higher than was observed to compensate for the overestimated diffusion time discussed above.

As inputs, \textsc{MOSFiT} took a number of parameters constrained from the photometric and spectroscopic observations of LSQ13ddu, such as the inferred explosion date, minimum blackbody temperature, and host galaxy extinction. Where no strict constraints were available, broad priors were placed on these parameters. Following previous work, we assumed that the optical opacity, $\kappa$, is in the range, 0.05--0.1~cm$^{2}$ g$^{-1}$ and the gamma-ray opacity, $\kappa_{\gamma}$, is in the range, 0.01--0.1 cm$^{2}$ g$^{-1}$ \citep[e.g.][]{Swartz1995,Wheeler2015,Wang2017b}. The photosphere for the models makes use of a temperature floor with a single value taken from an allowed range of 4000--7000 K, which is constrained based on the derived temperature measurements (see Section~\ref{sec:BBandColours}). The variables and their input values or ranges used for each of the models are described in Table \ref{MOSFiT_Parameters}. The fits were made using a Markov chain Monte Carlo routine implemented in \textsc{MOSFiT} utilising the \textsc{emcee} python package \citep{EMCEE}. A description of the main features of the two input models is provided below.

    \begin{figure*}
    \includegraphics[width=14cm]{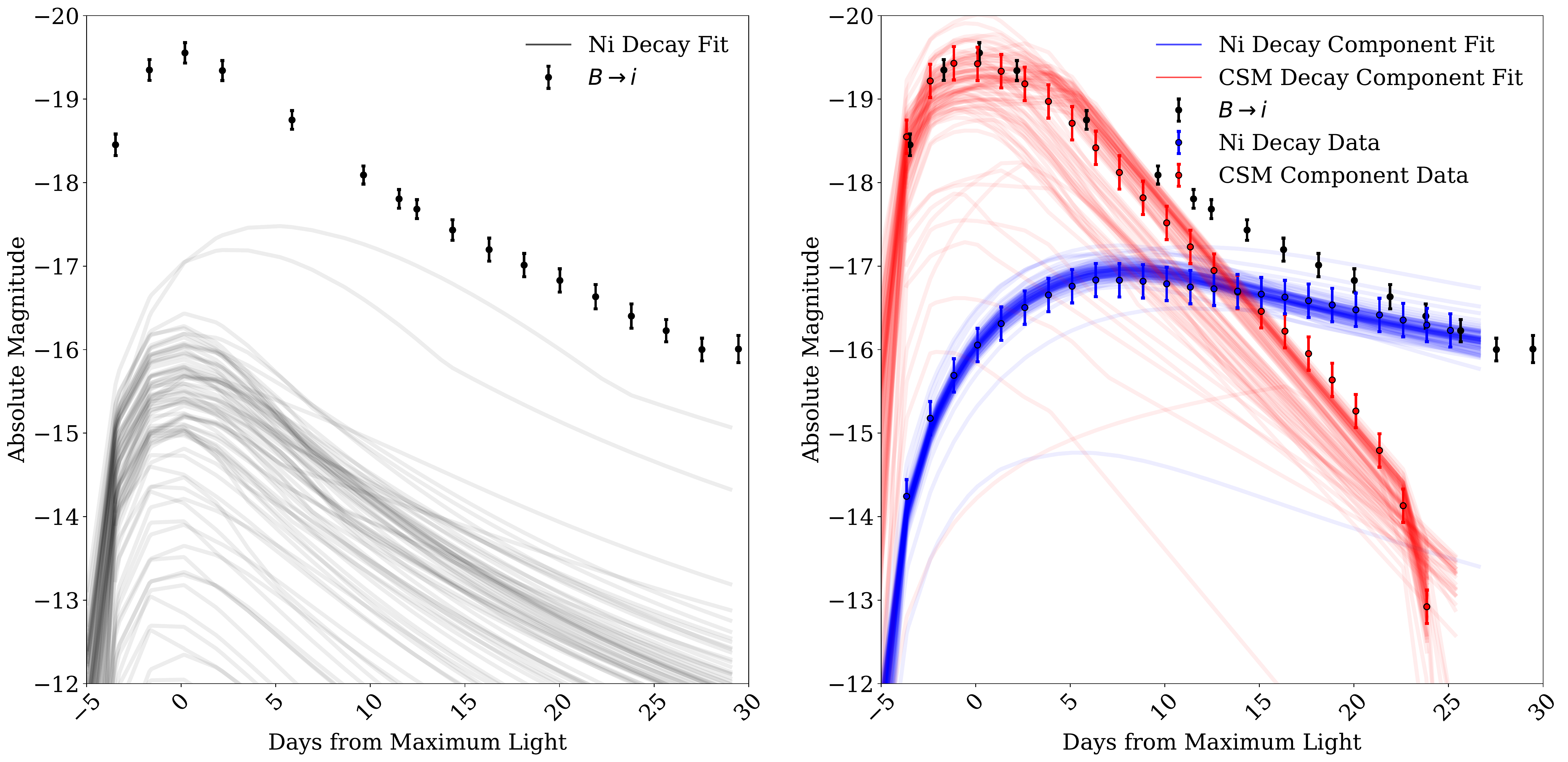}
    \caption{Left panel: Comparison of the light curves produced using \textsc{MOSFiT} for a \nick\ powered light curve with a limit on the \nick\ to total ejecta mass fraction of $<$0.6. The optical bolometric light curve points are in black and the fitted light curves with the top 25\% of scores are shown in grey. Right panel: The  \nick\ model outputs compared to the SE-SN light curve component identified in Section~\ref{section:LightcurvePowering}. The optical bolometric light curve points are in black and the fitted light curves with the top 25\% of scores of the \nick\ fit are shown in blue with the corresponding fitting of the CSM component shown in red.}
    \label{MOSFiTLCComp}
    \end{figure*}
    
    \begin{figure}
    \includegraphics[width=8cm]{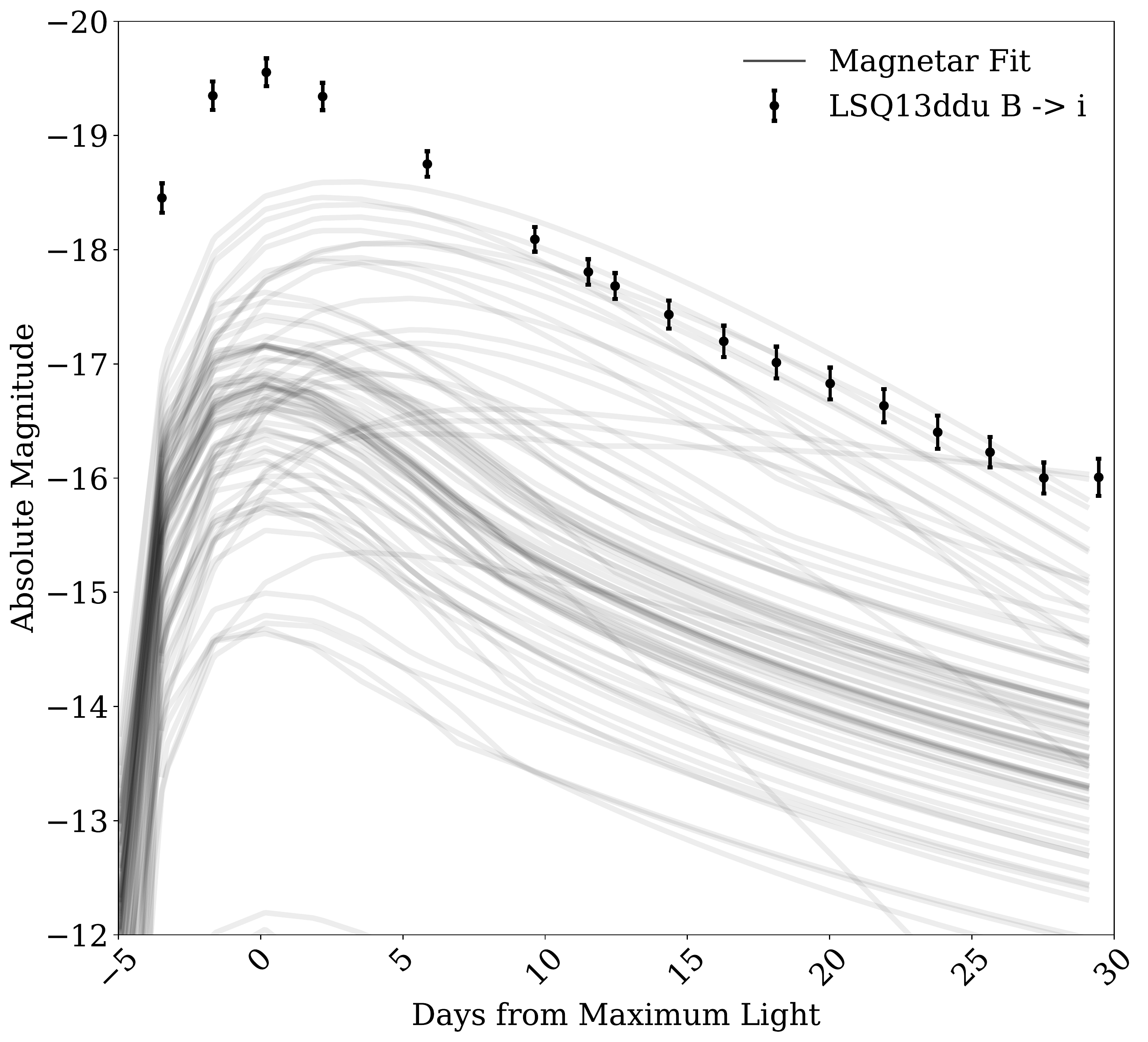}
    \caption{Comparison of the light curves produced using \textsc{MOSFiT} for a magnetar-powered light curve with parameters outlined in Table~\ref{MOSFiT_Parameters}. The optical bolometric light curve points are in black and the fitted light curves with the top 25\% of scores are shown in grey.}
    \label{MOSFIT_Mag_Fig}
    \end{figure}   

\subsubsection{Radioactive \nick\ decay}
\label{mosfitnick}
The bulk of SE-SN light curves can be explained by the radioactive decay of \nick\  \citep[e.g][]{Woosley_1993J}. Due to the high peak luminosity and rapid rise of LSQ13ddu, we used \textsc{MOSFiT} to explore if a physically realistic configuration of \nick\ decay based on the model of \cite{Nadyozhin1994} could generate the observed light curve. The \nick\ mass fraction has been constrained to be $<$0.6 of the ejecta mass, which includes the range expected of CC-SNe \citep[$\sim$0.2;][]{Umeda2008} extending to the higher values observed in SNe Ia. This model is fitted to the photometric data available from explosion to 29~d past maximum. The exclusion of the data at later phases is to reduce the effect of a non-constant opacity at later times on the validity of the underlying model assumptions. We also fit the underlying \nick\ component determined through the photometric analysis outlined in Section \ref{section:LightcurvePowering} to test if this component is consistent with that of a \nick\ decay-powered event. 

\begin{table}
\setlength{\extrarowheight}{4pt}
\caption{Input parameters and their allowed ranges for the \textsc{MOSFiT} models explored in this work: a \nick\ only model fit to the optical pseudo-bolometric light curve, a \nick\ model fit to the underlying SE-SN light curve `SE-SN component', a `CSM Component' for the corresponding CSM interaction contribution, and the magnetar model.}
\label{MOSFiT_Parameters}
\begin{tabular}{lccccc}
\hline
Parameter         & Input range  & Output value    \\
\hline
\textbf{\nick\ only model:}\\
Ejecta mass (\msol)           & 0.01 -- 10       &  $0.07^{+0.05}_{-0.03}$  \\
Ejecta velocity (\kms)          & 9,000 -- 11,000$^a$ & $10,000^{+700}_{-700}$ \\
\nick\ fraction               & 0.001 -- 0.6 & $0.13^{+0.25}_{-0.12}$ \\
\nick\ mass (\msol) &$10^{-5}$-- 6 & $0.01^{+0.04}_{-0.01}$\\ 
\\[-3ex]
\textbf{SE-SN Component:}\\
Ejecta mass (\msol)           & 0.01 -- 10       &  $1.30^{+0.43}_{-0.34}$  \\
Ejecta velocity (\kms)          &  9,000 -- 11,000$^a$ & $10,100^{+600}_{-700}$ \\
\nick\ fraction               & 0.001 -- 0.6 & $0.06^{+0.02}_{-0.01}$ \\
\nick\ mass (\msol) &$10^{-5}$-- 6 & $0.08^{+0.06}_{-0.03}$\\ 
\textbf{CSM Component:$^b$}\\
Ejecta mass (\msol)           & 0.01 -- 10       &  $0.08^{+0.07}_{-0.04}$  \\
CSM mass (\msol)           & 0.01 -- 10       &  $0.59^{+0.19}_{-0.14}$  \\
log$_{10}$ (CSM density) (g cm$^{-3}$)           & -16 -- -10       &  $-10.99^{+0.58}_{-0.69}$  \\
Progenitor Radius (AU)           & 0.01 -- 5       &  $1.58^{+1.94}_{-0.80}$  \\
Ejecta velocity (\kms)          & 9,000 -- 50,000$^c$ & $34,600^{+8,200}_{-7,100}$ \\
\\[-3ex]
\textbf{Magnetar:}\\
Ejecta mass (\msol)           & 0.01 -- 10  & $0.02^{+0.02}_{-0.00}$\\
Ejecta velocity (\kms)          & 9,000 -- 11,000$^a$ &$10,100^{+700}_{-800}$\\
Spin Period (\ms)           &1--10&$7.34^{+1.89}_{-2.70}$\\
B Field (10$^{14}$\G)          &0.1--10&$0.30^{+0.93}_{-0.16}$\\
\\[-3ex]
\hline
\end{tabular}
\begin{flushleft}
$^a$From \textsc{synapps} fitting and \Hei\ 10830~\AA\ line measurements.\\
$^b$For an isotropic, wind-like CSM with a density profile s=2.\\
$^c$Expanded range to compensate for the model's diffusion time overestimation.
\end{flushleft}
\end{table}

\subsubsection{CSM-powered light curve contribution}
\label{sec:CSM_Powering}
To determine if the CSM component identified by the light curve analysis in Section \ref{section:LightcurvePowering} is compatible with physically realistic parameter values, we have fit it with the \textsc{MOSFiT} CSM interaction model. As discussed above, this CSM model is based on \cite{Chatzopoulos2013a} and is subject to the same limitations.

In \cite{Chatzopoulos2013a} a parameter study was conducted to explore the robustness of the parameter outputs of the CSM model on which \textsc{MOSFiT} is based. Specifically, they tested changing the CSM density profile from that of a shell (s=0) to being wind-like (s=2) and found that there is a factor of 2-3 uncertainty in the majority of the parameters, with the exception of the ejecta mass and progenitor radius which differed by up to an order of magnitude, despite not producing significantly different light curves.

The CSM configuration of LSQ13ddu and other such events is uncertain, both as there is still debate as to their progenitor systems which likely plays a significant role, and the difficultly in determining the configuration from available observations. We have chosen to model the CSM with a wind-like profile but the true configuration is unknown.

\subsubsection{\nick\ Decay and CSM modelling results}
We found that the rapid rise and bright peak luminosity of LSQ13ddu can not be fit with the purely \nick-powered model. The output parameters for this fully \nick-powered model are given in Table \ref{MOSFiT_Parameters} and shown on the left panel of Fig.~\ref{MOSFiTLCComp}. Therefore, we explored if the light curve could be explained by a combination of a \nick\ and a CSM component by fitting each component separately.  
We found that \textsc{MOSFiT} was able to produce a good fit to the underlying \nick\ component with a \nick\ mass of $\sim$0.08 \msun\ and an ejecta mass of $\sim$1.30 \msun. The early flux excess in LSQ13ddu as seen in Fig.~\ref{PrenticeIcComparisonPlot} was found to be well fit by the \textsc{MOSFiT} CSM interaction model with a CSM mass of $\sim$0.6 \msun, a density of $\sim$1x10$^{-11}$ g cm$^{-3}$, and initial/progenitor radius of $\sim$1.6 AU. As noted above, by \cite{Chatzopoulos2013a} the ejecta mass and progenitor radius are the most weakly constrained parameters and may have systematic uncertainties of up to an order of magnitude depending on the CSM input parameters.

The output parameters are outlined in Table~\ref{MOSFiT_Parameters} with the two component fits shown in the right panel of Fig.~\ref{MOSFiTLCComp}. Several parameters are shared between the \nick\ and CSM models (opacity, minimum temperature floor, explosion epoch, host extinction contribution, ejecta mass and ejecta velocity) with good agreement between the majority. However, there is significant tension between the values of two parameters: ejecta velocity and ejecta mass. The tension in the fitted values of ejecta velocity was previously described as a consequence of the overestimated diffusion time of the CSM model for fast evolving events like LSQ13ddu. The tension in the values of ejecta mass ($1.30^{+0.43}_{-0.34}$ \msol\ for the \nick\ component and $0.08^{+0.07}_{-0.04}$ \msol\ for the CSM component) is most likely due to the propagation of the CSM density profile uncertainties \citep{Chatzopoulos2013a}, as well as the manner in which the model treats diffusion timescales and the resulting increased ejecta velocity. Despite these caveats, the results of this modelling show that the use of two distinct power sources (combined $^{56}$Ni and CSM) fits the light curve of LSQ13ddu well.

\subsubsection{Photometric modelling of magnetar powering}
\label{section:mag_powering}
For some luminous and rapidly evolving transients such as iPTF16asu \citep{Whitesides2017a}, and AT~2018cow \citep{Prentice2018_cow}, powering through the spin down of a magnetar has been investigated. Since LSQ13ddu also displays a fast rise and high peak luminosity, we have tested if LSQ13ddu could be fit by a magnetar model. We find that the best fitting models, while able to reproduce the shape of the light curve has several output parameters that appear physically unlikely. For models both with and without a constraint on the expansion velocity to match the observed velocities of 9,000 -- 11,000 \kms, the ejecta mass is extremely low ($\sim$0.02--0.04\msol). When the model velocities are constrained to realistic values, it is also unable to replicate the observed light curve shape, with the peak luminosity underestimated as shown in Figure~\ref{MOSFIT_Mag_Fig}. Very low ejecta masses have previously been seen in purely magnetar fits to iPTF16asu \citep{Whitesides2017a} and a range of fast transients in \cite{Arcavi2016} with these being disfavoured as powering sources as the massive star progenitors for such objects would require excessive envelope stripping to produce explosions with such low ejecta masses. We disfavour the magnetar spindown model for LSQ13ddu for the same reason.

\section{Discussion}
\label{section:Discussion}

\begin{figure*}
\includegraphics[width=16cm]{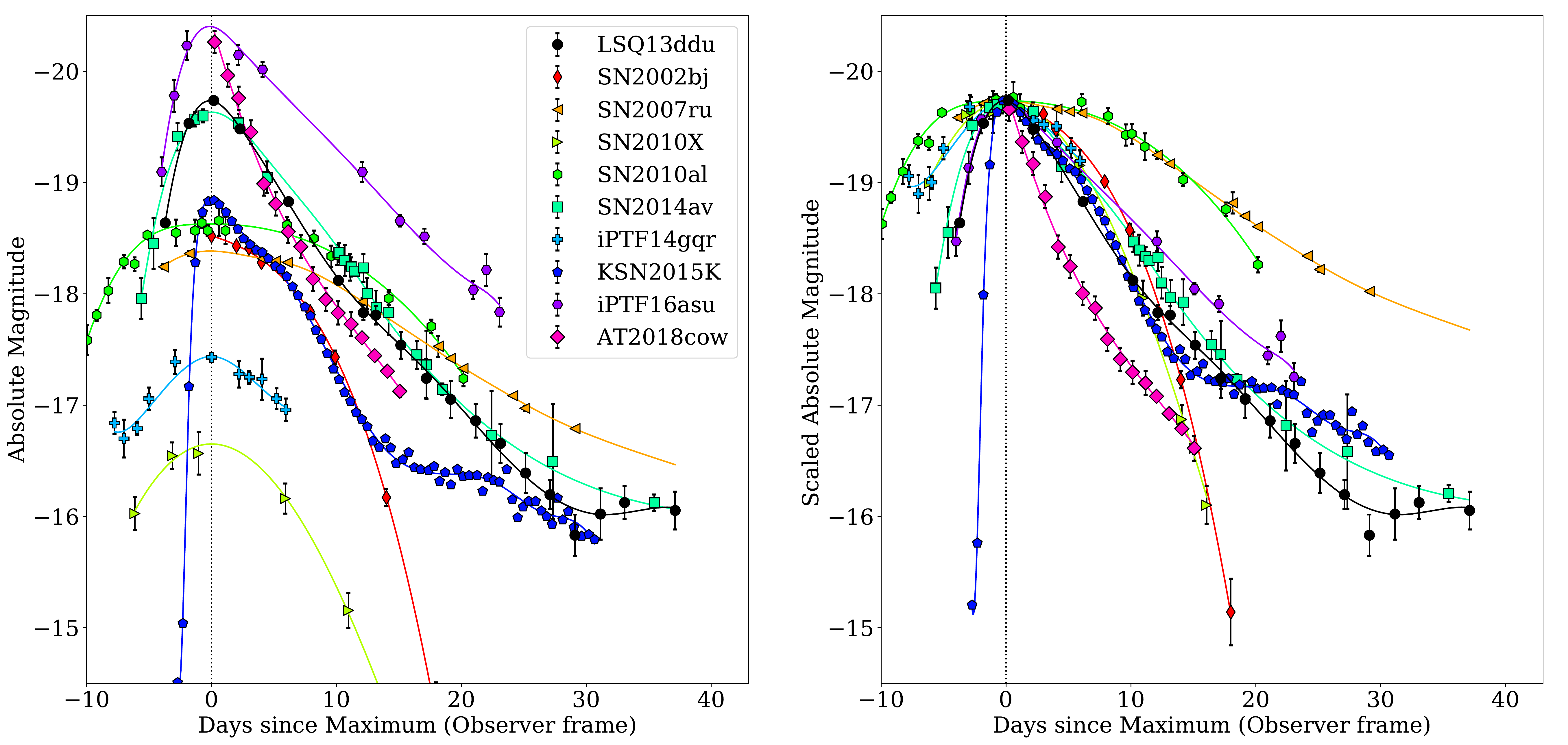}
\caption{Left panel: Absolute magnitude \textit{R}-band light curves of a sample of literature objects compared to the \textit{LSQgr}-band light curve of LSQ13ddu. Right panel: The light curves of the comparison objects have been scaled to match the peak of LSQ13ddu. Solid lines in both panels display the smoothed cubic spline fit to the observations. Details of the comparison objects are given in Table~\ref{LC_ComparisonSummary_Table}.}
\label{ComparativeLC}
\end{figure*}

New classes of rapidly evolving transients have been discovered in recent years and attempts have been made to divide these into sub-classes arising from distinct progenitor pathways. LSQ13ddu is one such rapidly evolving transient that challenges existing progenitor scenarios. It displayed a rapid photometric evolution, early spectroscopic features similar to, but significantly weaker than those seen in SNe Ibn, before developing broad spectral features more similar to SE-SNe. In this section, we compare the photometric and spectroscopic properties of LSQ13ddu with those of both common and rare transient sub-classes to attempt to place it in the context of previous events. We also discuss the evidence for this being a hybrid SN Ibn and the implications this has for its progenitor scenario.

\subsection{Light curve comparisons}
\label{section:GeneralTransientLightCurveComparisons}

\begin{figure*}
  \includegraphics[width = 17cm]{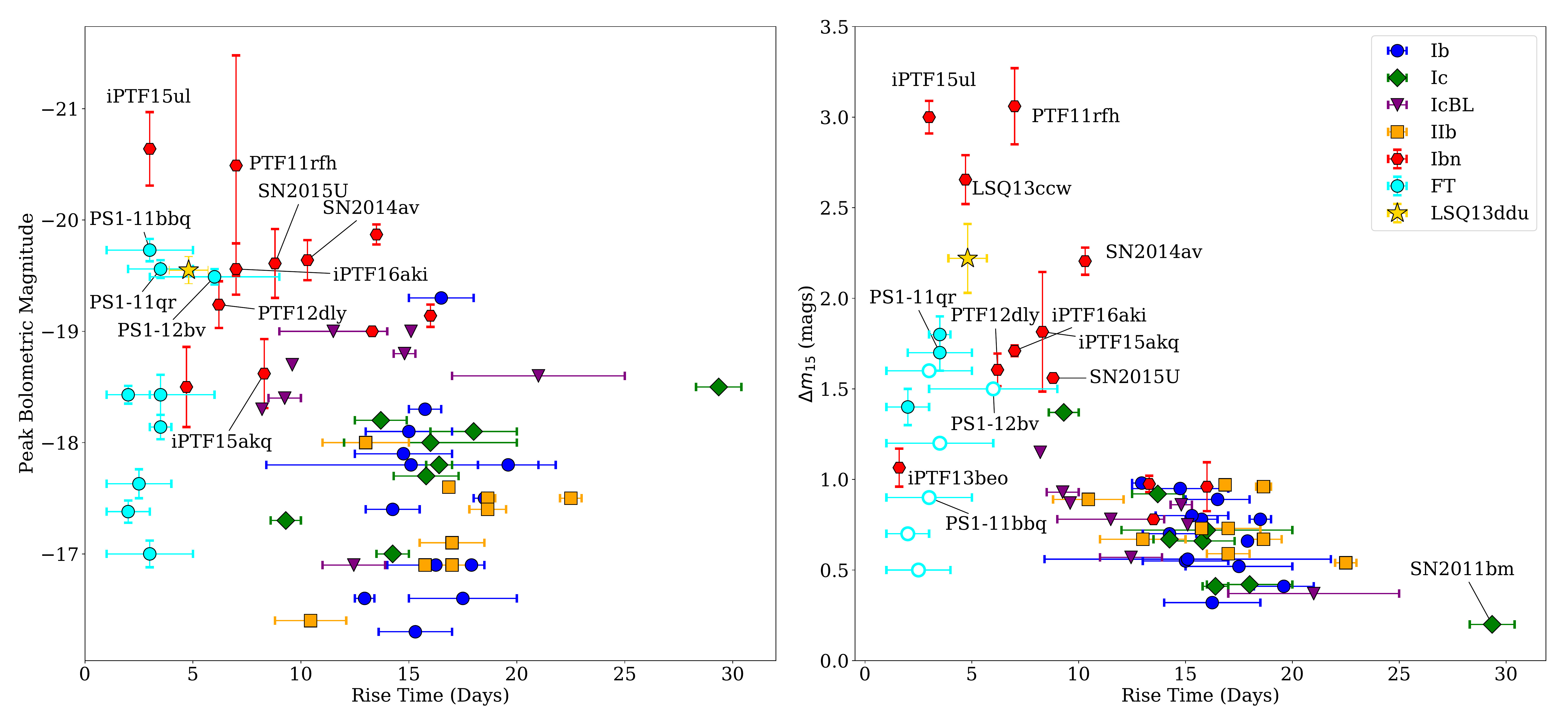}
   \caption{Left Panel: Peak bolometric magnitudes of a sample of SE-SNe from \protect\cite{Lyman2016} are plotted against their bolometric rise times to peak. For the Ibn sample of \protect\cite{Hosseinzadeh2017} the values shown are those of the \textit{R}, \textit{r} or \textit{g} band data, with the `fast' transients from \protect\cite{Drout2014} shown using \textit{PS1-r} data. Right panel: The decline rates in the first 15~d post maximum light are shown against their bolometric rise times. Open markers indicate where the $\Delta m_{15}$ value is a lower limit. The different classes of objects are colour-coded as detailed in the legend, with LSQ13ddu shown as a yellow star.}
   \label{BalerionComparisonDecayRise}
  \end{figure*}
  
In Fig.~\ref{ComparativeLC} we compare the \textit{LSQgr} light curve of LSQ13ddu with those of a sample of fast-evolving SE-SNe and unusual transient events (see Table~\ref{LC_ComparisonSummary_Table} for details of the comparison objects). Each light curve has been fitted using a smoothed cubic spline with their observed time of maximum light taken from the source work where possible or otherwise obtained from the spline fit. Each light curve has also been scaled to match the peak of LSQ13ddu to better study any differences in relative evolution.

Overall, as in the comparison of the (optical) pseudo-bolometric light curves, the scaled light curve behaviour of LSQ13ddu is most similar to that of the Ibn SN~2014av \citep{Pastorello2016}, with both showing a nearly identical rise time and initial decay rate (0.15$\pm$0.01 and 0.13$\pm$0.01 mag~d$^{-1}$, respectively, during the first 15~d post maximum). The decline rate of LSQ13ddu slows to 0.09$\pm$0.02 mag~d$^{-1}$ between 20 and 35~d after maximum, which is similar to that of the Ic-BL SN~2007ru, which declines at a rate of 0.09$\pm$0.01 mag~d$^{-1}$ at a similar phase.

In contrast, the behaviours of SNe~2010X \citep{Kasliwal2010} and 2002bj \citep{Poznanski2009} differ significantly from LSQ13ddu, with both showing slower initial relative decline rates that increase over time so that their overall decline is more rapid than that of LSQ13ddu. These events are significantly less luminous with peak absolute magnitudes of $-$16.6 and $-$18.8 mag for SNe~2010X and 2002bj, respectively, compared to the $-$19.6 mag of LSQ13ddu. The origin of these events remains unclear but they could be He-shell detonations on the surface of a white dwarf (`.Ia SN') or ultra-stripped SNe (USSNe).

USSNe are a proposed class linked to rapidly fading transients that are thought to be produced through the interaction of evolved massive stars, which have lost the majority of their envelopes via stripping by a neutron star in a close binary orbit \citep{Tauris2015}. iPTF14gqr \citep{De2018}, and perhaps SN~2010X \citep{Moriya2017a}, have been suggested to be members of this class. USSNe reach a \textit{V}-band peak of approximately $-$16 mag, whilst LSQ13ddu is brighter by more than a factor of 100 at a peak of $-$19.7 mag. This drastic difference in peak luminosity leads us to rule this class out as a physical explanation for LSQ13ddu.

\begin{table}
\caption{Summary of the comparison objects used in Fig.~\ref{ComparativeLC}, showing the object name, its predicted type, the peak \textit{R}-band magnitude, $R_\mathrm{max}$, and $\Delta m_{15}$.}
\label{LC_ComparisonSummary_Table}
\begin{tabular}{lcccc}
\hline
Object & Classification & $R_\mathrm{max}$  & $\Delta m_{15}$ & Ref. \\ 
&&(mag) &  (mag) \\
\hline
LSQ13ddu &--  & $-$19.7$^a$ & 2.2&- \\
SN~2002bj &.Ia or SE-SN$^b$  & $-$18.5 & 2.8 & 1 \\
SN~2007ru & Ic-BL & $-$18.4 & 0.7 & 2 \\
SN~2010X &.Ia or USSN$^b$ & $-$16.7 & 2.7 & 3\\
SN~2010al & hybrid Ibn & $-$18.6 & 0.7 &4  \\
SN~2014av & Ibn & $-$19.6 & 2.0 & 5\\
iPTF14gqr & USSN$^b$  & $-$17.4 & 0.5 &  6\\
KSN2015K &  SB$^b$ & $-$18.8 & 2.3 & 7\\
iPTF16asu & Magnetar and/or SB$^b$ & $-$20.4$^c$ & 1.7 & 8\\
AT~2018cow & ?  & $-$20.2 & 3.2 & 9\\
\hline
\end{tabular}
\begin{flushleft}
$^a$LSQ13ddu value is given using \textit{LSQgr}-band data.\\
$^b$USSN = Ultra-stripped SN. SB = shock-breakout into CSM. SE-SN = Stripped envelope SN\\
$^c$iPTF16asu value is given using \textit{g}-band data.\\
References: (1)~\cite{Poznanski2009}, (2)~\cite{Sahu2008}, (3)~\cite{Kasliwal2010}, (4)~\cite{Pastorello2015e}, (5)~\cite{Pastorello2016}, (6)~\cite{De2018}, (7)~\cite{Rest2018}, (8)~\cite{Whitesides2017a}, (9)~\cite{Prentice2018_cow}  \\
\end{flushleft}
\end{table}

In Fig.~\ref{BalerionComparisonDecayRise}, the optical bolometric light curve peak magnitude, the rise time, and the decline rate over the first 15~d post peak ($\Delta m_{15}$) are compared to a sample of SE-SN full bolometric light curves from \cite{Lyman2016}, supplemented with SN Ibn data from \cite{Hosseinzadeh2017} and `fast' transients from \cite{Drout2014}. The magnitudes quoted for the Ibn sample are not bolometric but are \textit{R}, \textit{r} or \textit{g} band values, with the objects from \cite{Drout2014} given in \textit{PS1-r} \citep{Tonry2012}. The offset between the \textit{R} and the optical bolometric magnitude for LSQ13ddu is 0.84 mag, while the offset between the optical and full bolometric magnitude of LSQ13ddu is much smaller at 0.24 mag, with a similar 0.16 mag value determined for the \textit{g}-band, suggesting that the use of the \textit{R}/\textit{r} or \textit{g}-band magnitudes for the Ibn sample is valid.

The light curve properties of LSQ13ddu are located with the parameter space occupied by the SNe Ibn class, with brighter peak magnitudes and a faster photometric evolution (rise and decay) that those of `normal' SE-SNe. In comparison to the `fast' transients of \cite{Drout2014}, LSQ13ddu has a slower rise and larger $\Delta m_{15}$ value than the bulk of the sample, which can reach their peak luminosity in as little as day and decay more slowly after peak. While the available spectra of these objects are limited, those that are available do not provide good matches to those of LSQ13ddu and we conclude that they are more likely to represent a distinct class of objects.

   \begin{figure}
    \includegraphics[width=7.6cm]{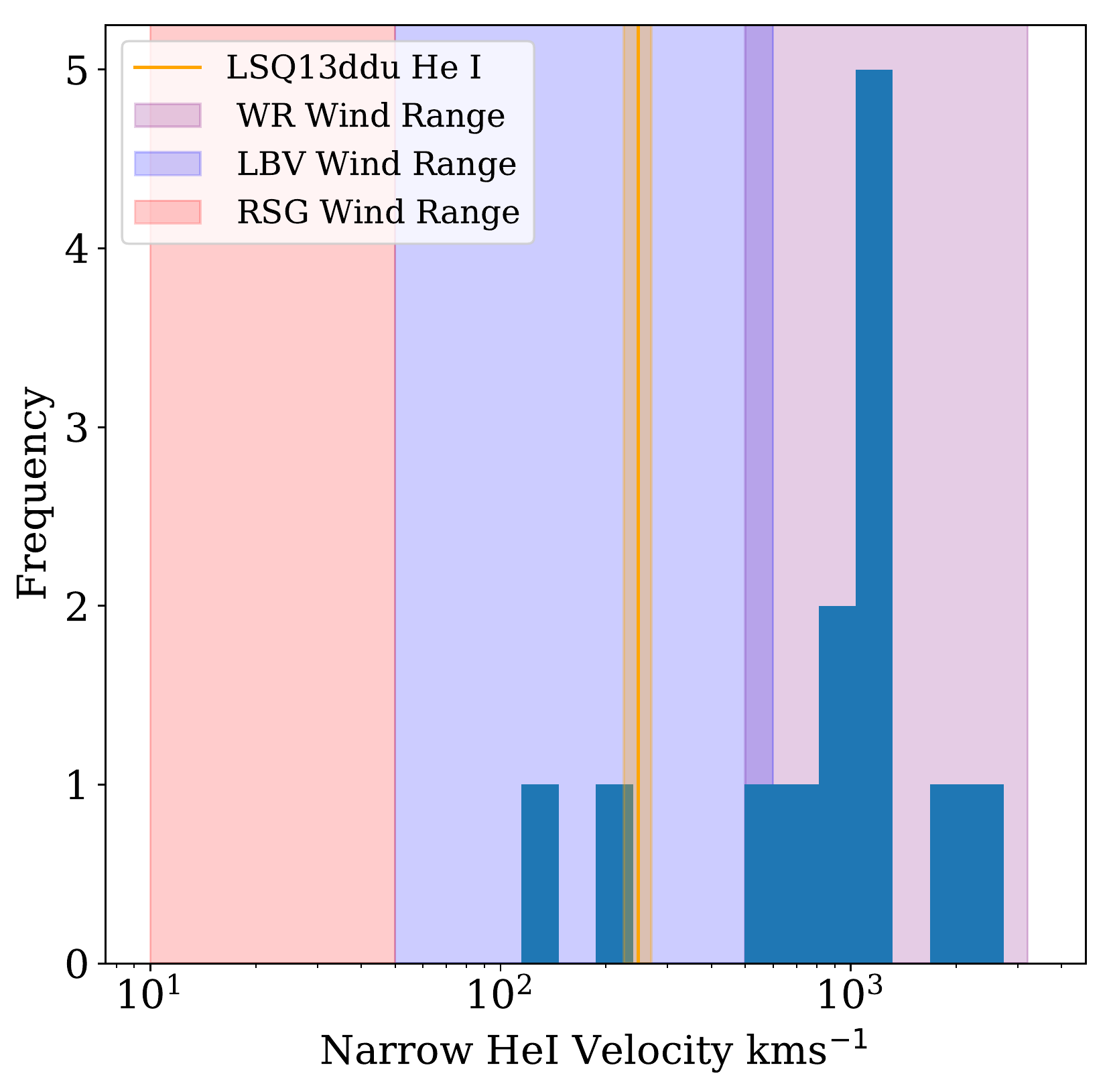}
    \caption{The mean narrow \Hei\ velocity of LSQ13ddu as a weighted mean of the \Hei\ features visible in the $+$5 and $+$6~d spectra (250$\pm$20~\kms, shown in yellow with its uncertainty shown by shading), compared to a literature sample of SNe Ibn obtained from \protect\cite{Pastorello2016} (blue histogram). The wind speed ranges for WR, LBV and RSG stars are indicated by the purple, blue and red shaded regions, respectively. The three SNe Ibn in the LBV wind range other than LSQ13ddu are PS1-12sk, SNe~2011he and 2005la in order of increasing velocity.}
    \label{NarrowHeIVelocities_Hist}
    \end{figure}

\subsection{He velocities and the connection to pre-explosion mass loss}
\label{subsection:NarrowFeatures}

Fig.~\ref{NarrowHeIVelocities_Hist} shows the absorption minimum velocities of the \Hei\ lines in LSQ13ddu, compared to a sample of SNe Ibn with velocity measurements within 20~d of maximum light obtained from \cite{Pastorello2016}. As discussed in Section~\ref{Subsection:vel_he},  the \Hei\ velocity of LSQ13ddu of 250$\pm$20~\kms\ is the weighted mean of the measurements in the two highest resolution spectra. As can be seen in Fig.~\ref{NarrowHeIVelocities_Hist}, this value is lower than the majority of SNe Ibn. The three objects with the most similar \Hei\ velocities to those of LSQ13ddu are PS1-12sk \citep{Sanders2013}, SN~2011hw \citep{Smith2012, Pastorello2015e} and SN~2005la \citep{Pastorello2008b} with velocities of 130, 200--250, and 500 \kms, respectively \citep{Pastorello2016}. These objects, along with LSQ13ddu, fall in the range of wind velocities normally seen for LBV and outside the range typically seen for WR stars of 500--3200~\kms\ \citep{Crowther2007}. Red supergiants have winds that are much slower, with values of a few tens of \kms\ \citep{vanLoon2005}. 

As discussed in Section \ref{Subsection:vel_he}, without the availability of higher resolution ($\sim$35 \kms) spectra for LSQ13ddu, we would have measured a significantly higher \Hei\ velocity of up to $\sim$850 \kms\ that would be more in agreement with the majority of SNe Ibn. To determine the role that spectral resolution plays in the obtained velocities for narrow features in SNe Ibn, we investigated the SN Ibn sample of \cite{Pastorello2016}. We found that the SNe Ibn with the lowest velocities, such as PS1-12sk and SN2011hw, have spectra with higher than normal resolution suggesting that there may be other objects that have similarly low \Hei\ velocities ($<$500 \kms) but did not have spectral measurements at high enough resolution. However, there are a number of SNe Ibn with high \Hei\ velocities ($>$1000 \kms) that have sufficiently high spectral resolution where lower velocities could have been resolved e.g. SN~2010al. Therefore, we conclude that there is an intrinsically wide range in \Hei\ velocities in SNe Ibn but that the percentage with low \Hei\ velocities may be underestimated due to the resolution of the spectra. 

SNe Ibn have been associated with explosions of WR stars because of the similar velocities of their winds and the \Hei\ features in SNe Ibn, as well as the H-free nature of WR atmospheres that matches with the H-free spectra of SNe Ibn \citep{Pastorello2008a}. The objects that have \Hei\ velocities most similar to LSQ13ddu and fall in the LBV wind range all displayed some degree of peculiarity. PS1-12sk was found in an old stellar population and was suggested to be associated with a white dwarf transient or a rare massive star event in a low star-formation region \citep{Sanders2013,Hosseinzadeh2019}. SN~2011hw and SN~2005la both displayed narrow H$\alpha$ lines in their spectra suggesting they are transitional between Type Ibn and Type IIn events \citep{Pastorello2008b} and perhaps consistent with stars exploding during the LBV to WR transition \citep{Pastorello2016}. As discussed in Section \ref{sec:HostGalaxy}, based on the line widths, the SN-related contribution to the H$\alpha$ flux is consistent with zero, though a minority contribution cannot be ruled out. This is at odds with the H-rich composition expected in LBVs. Therefore, it appears difficult to match LSQ13ddu to a particular progenitor scenario given its combination of observed properties, but we discuss potential systems in Section \ref{ProgentiorSystemOfIcCSM}.

\begin{figure}
\includegraphics[width=7.9cm]{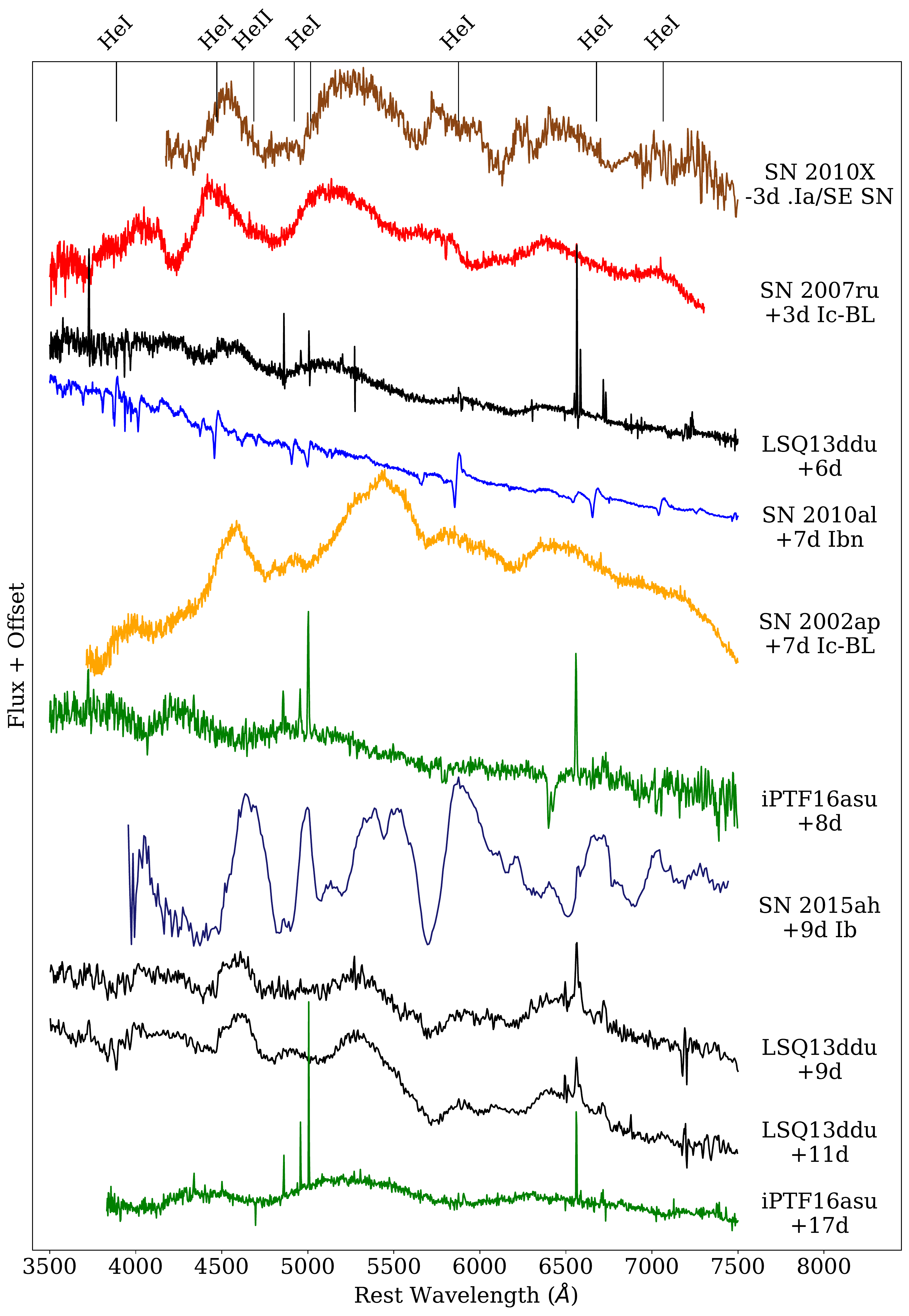}
\caption{Spectral comparison between LSQ13ddu (black), the Ic-BL SNe 2007ru (red) \citep{Sahu2008} and 2002ap (orange) \citep{Modjaz2014}, the Type Ib SN~2015ah (light blue) \citep{Prentice2019_SESN}, the Ibn SN~2010al (dark blue) \citep{Pastorello2015e} iPTF16asu (green) \citep{Whitesides2017a}, and the unusual SN~2010X (brown) \citep{Kasliwal2010}. Vertical ticks mark the location of the He lines.}
\label{SpecComp}
\end{figure}

\subsection{Is LSQ13ddu a hybrid SNe Ibn?}
\label{subsection:Hybrid}

Some SNe Ibn, such as SN~2010al \citep{Pastorello2015e} and ASASSN-15ed \citep{Pastorello_15ed}, show hybrid properties between those of `normal' Ibn and Ib SNe, with narrow \Hei\ CSM features in their early spectra along with broad He features coming from He-rich ejecta at late times. LSQ13ddu displays a number of spectral similarities to SN~2010al, with both displaying narrow WR wind-related features in their earliest spectra. SN~2010al displayed early-time spectral features indicative of material related to CNO-cycle enrichment in the wind of the progenitor star but these are not seen in LSQ13ddu. Transitional Ibn to Ib SNe show a broad diversity in their light curves and do not form a separate subclass based on their light curves alone.

A spectral comparison between the $+$6--11~d spectra of LSQ13ddu with a sample of SE-SNe and Ibn events (including the hybrid Ibn to Ib SN~2010al) is shown in Fig.~\ref{SpecComp}. The broad spectral features in LSQ13ddu most closely resemble those of SNe Ic-BL, such as SN~2007ru \citep{Sahu2008} and SN~2002ap \citep{Modjaz2015}, particularly in the 4450--4750, 5250--5800 and 6250--6700 \AA\ regions, though the continuum of the latter is redder than that of LSQ13ddu. As discussed in Section~\ref{Subsection:vel_he},  SN~2010al displays strong narrow \Hei\ features that are not seen in LSQ13ddu at the same epoch, suggesting a faster evolution for LSQ13ddu. \textsc{synapps} fitting (Fig.~\ref{synapps_v1}) showed that the broad optical features seen at $+9$ d in LSQ13ddu can be explained by lines of \Feii, \Caii, \Oi, \Ciii, along with \Hei\ features at the same photospheric velocity (10200 \kms). A broad \Hei\ 10830 \AA\ feature is seen in the $+$9 and $+$14~d of LSQ13ddu with a FWHM velocity of $\sim$9500 \kms, suggesting underlying ejecta with He present. 

\begin{figure}
    \includegraphics[width=7.5cm]{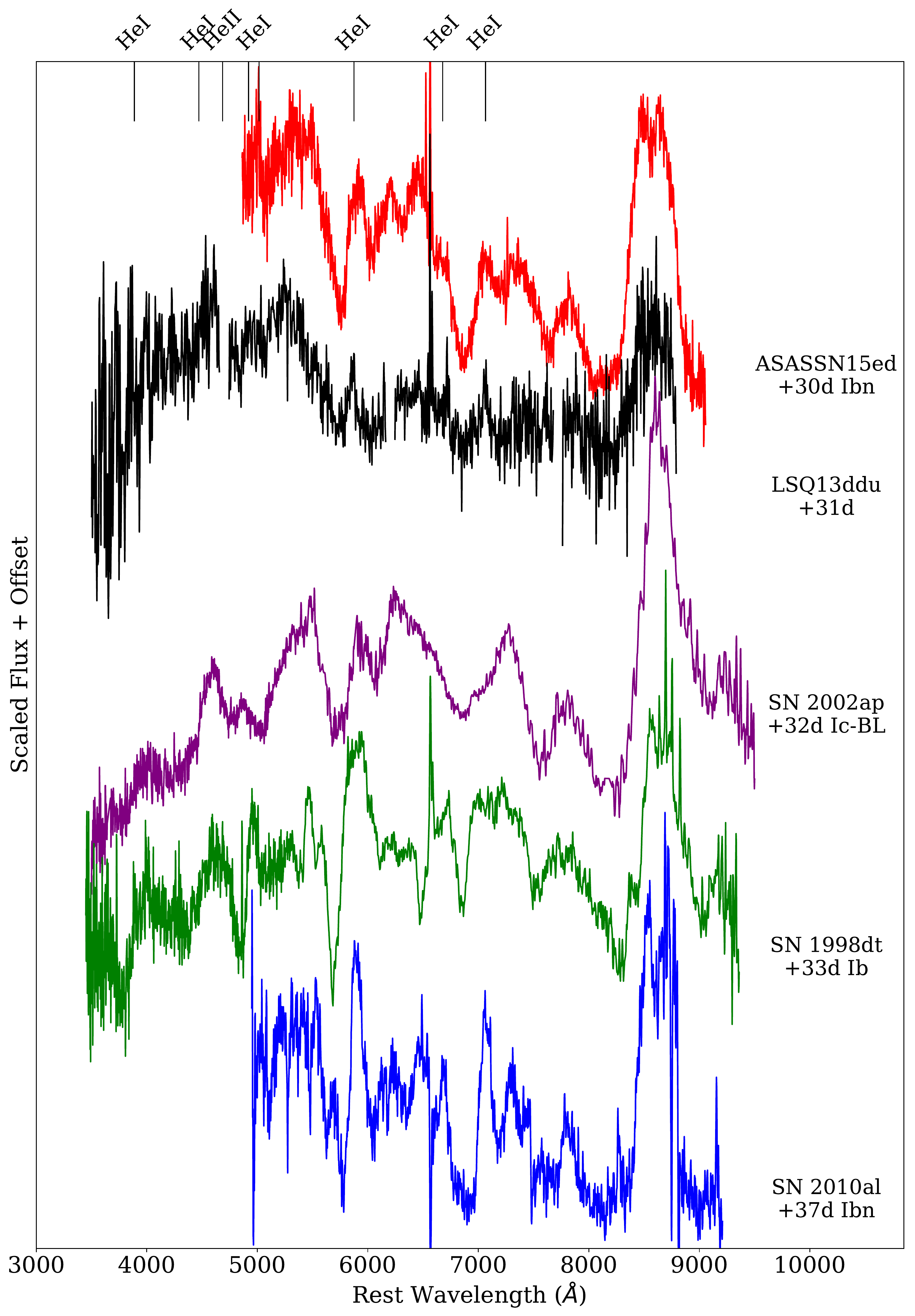}
    \caption{A spectral comparison of the broad spectral features in LSQ13ddu with a sample of SE-SNe and Ibn to Ib hybrid event. Vertical lines mark the location of the He lines for clarity.}
    \label{ASASSN15e_2010al_comp}
    \end{figure}

The late-time ($+31$ d) spectrum of LSQ13ddu is compared in Fig.~\ref{ASASSN15e_2010al_comp} to those of the transitional Ibn to Ib events, SN~2010al and ASASSN-15ed, as well as Ib SN~1998dt and the Ic-BL SN~2002ap \citep{Modjaz2014}. The spectrum of LSQ13ddu is most similar to the transitional Ibn to Ib SNe that display broad \Hei\ features (FWHM of 5000--6000 \kms) but it has significantly weaker \Hei\ features. This suggests that LSQ13ddu may be a transitional Ibn to Ic SN with a small amount of residual He present within the ejecta. Such residual He has previously been seen in other SNe Ic e.g.~the Ic-BL SN~2016coi, which displayed traces of \Hei\ within its ejecta \citep{Prentice2018_coi}.

The light curve of LSQ13ddu is well fit with the early peak brightness being powered by CSM interaction before an underlying $^{56}$Ni decay component becomes dominant at $\sim$10 d post peak brightness. Hybrid Ibn such as SN~2010al and ASASSN-15ed are suggested to be powered at early times by CSM interaction and although they have stronger \Hei\ interaction features at early times compared to LSQ13ddu, the differences in line strength may represent a continuum of CSM properties (e.g. mass and density profiles). The next weakest \Hei\ features were those of ASASSN-15ed, which was also a hybrid Ibn to Ib event. A full understanding of the connection between the strength of the narrow He features and properties of the underlying CSM is difficult and would require detailed knowledge of the CSM, as well as consideration of non-thermal effects \citep{Boyle2017}.

No intermediate and broad components due to shocked CSM are seen in LSQ13ddu, which is also the case in $\sim$50 per cent of Ibn SNe \citep{Hosseinzadeh2017}. The preferred scenario to explain the lack of intermediate-width features is the presence of an optically thick outer CSM reprocessing the shocked material that masks these intermediate-width features \citep{Pastorello2015, Hosseinzadeh2017, Chevalier2011, Ginzburg2012}.The geometric configuration of the system along with effects of viewing angles can also mask the appearance of ejecta-CSM interaction signatures from the observed spectra of some objects e.g. iPTF14hls \citep{Andrews2018}. In this case, it has been proposed that fast SN ejecta may obscure CSM ongoing ejecta - CSM interaction that is predominately occurring in an equatorial disk. Given the previously described uncertainty in the CSM configuration, along with supporting observational and theoretical studies, the absence of intermediate width lines does not exclude a significant contribution from CSM interaction to the luminosity of LSQ13ddu.

\subsection{Search for a GRB companion}
\label{subsubsection:SearchforaGRBCompanion}
A subset of Ic-BL SNe have been found to have accompanying long-duration Gamma Ray Bursts (GRBs) \citep{Woosley1999,Woosley2006}. Since the later time spectra of LSQ13ddu have some similar broad features to those of Ic-BL SNe, we conducted a search in the Fermi GBM Burst Catalogue \citep{Paciesas_2012,Gruber_2014,Bhat_2016} for events coincident with LSQ13ddu. Specifically, we searched for triggers within 15 deg of the spatial location of LSQ13ddu within the time period covering one week prior to its first optical detection. One such event was identified meeting these criteria -- GRB131123A. 
GRB131123A was detected on 2013 November 23.5 (MJD 56619.5), which is 2.8~d prior to the first optical detection of LSQ13ddu and 0.8~d before the last \textit{LSQgr} non-detection at a limiting magnitude of $>$21.84 mag. Samples of Ic-BL SNe with associated GRBs do not show significant lag times between the GRB and the optical transient \citep[e.g.][]{Woosley2006, Cano2017}. GRB131123A occurred at R.A. 53.240 deg, Decl. $-$20.880 deg (J2000) with a statistical uncertainty on the position of 8.34 deg \citep{FermiGRBTeam2013}. The location of LSQ13ddu is outside this error region at a separation of 10.7 deg from the GRB position. Furthermore, its T\textsubscript{90} duration was 3.14$\pm$0.72 s, placing it in the intermediate region between short- and long-GRBs \citep{Paciesas1999}. Given the offset in the locations, the delay between the GRB and the first visible detection of the SN, and the relatively short GRB duration compared to typical Ic-BL-GRB SNe, we conclude that GRB131123A is an unrelated event. 

\subsection{The potential progenitor system of LSQ13ddu}
\label{ProgentiorSystemOfIcCSM}

The progenitors of SNe Ibn have been proposed as evolved WR stars that explode before the residual CSM from prior mass loss events (either winds or outbursts) have dissipated. This poses a problem for stellar evolution models as WR stars are expected to survive for a sufficient period of time prior to a SN explosion for such material to have dissipated \citep[e.g.][]{Meynet2004}. 
Alternatively, interacting binary systems have also been proposed as possible progenitors for SNe Ibn \citep{Pastorello2008a}. The prototype for the Ibn class, SN~2006jc, was observed to undergo a large eruptive mass loss event approximately two years prior to its terminal explosion \citep{Pastorello2007}. Such events are problematic for single-star progenitor scenarios as no WR star has yet been observed to undergo significant mass-loss events. This may be alleviated if the SN~2006jc system consisted of an LBV and a WR star, with the LBV responsible for the outburst before its companion WR exploded in the years following this explosive mass-loss episode. This scenario, however, is somewhat fine-tuned and is unlikely to explain the full class of Ibn objects.

If SNe Ibn are produced though the explosion of WR stars or stars in a transition phase from LBV to WR, the velocity of the CSM should be consistent with measurements of these classes of objects. The bulk of observed SNe Ibn have CSM expansion velocities consistent with those of WR stars. However, a subset of Ibn, as well as LSQ13ddu, have lower wind velocities (250$\pm$20 \kms), more consistent with those of LBV stars (see Fig.~\ref{NarrowHeIVelocities_Hist}). The narrow \Hei\ features of LSQ13ddu are also weaker compared to the general SNe Ibn population at similar phases. Due to a lack of higher resolution spectra for some SNe Ibn, we have found that a larger percentage of SNe Ibn than previously thought may have lower \Hei\ velocities that are more consistent with LBV- instead of WR-like winds, although as previously discussed, the composition of these winds does not match what is expected of a H-rich LBV star.

Similarly to the hybrid Ibn/Ib SNe 2010al and ASASSN-15ed, LSQ13ddu evolves from an early blue continuum to show broad features consistent with He-containing ejecta. The amount of He present in the ejecta (not the CSM component) appears to be lower in LSQ13ddu compared to both Ib SNe and these hybrid events, with LSQ13ddu having weaker broad \Hei\ features at late-times than is typically seen, with its late-time spectral and photometric evolution more similar to those of SNe Ic.

The modelling of the CSM light curve component with \textsc{MOSFiT}, that whilst subject to several assumptions can inform this discussion. With the required CSM mass of $0.59^{+0.19}_{-0.14}$ \msol\ easily producible through either LBV eruptive mass loss events, which have been observed to ejecta up to a few solar masses of material \citep{Smith2017}, though in most cases the stars seen to produce these large outbursts have high masses in tension with the low values of ejecta mass produced by the modelling ($\sim$0.08--1.30\msol). Alternatively, if the CSM is produced largely via winds at the expected mass-loss rates of either LBV or WR stars ($\sim$ 1x10$^{-5}$ -- 1x10$^{-3}$ \msol\ yr$^{-1}$; \citealt{Moriya2016a, Smith2017}) the material could accumulate within a few thousand years, though at the relative speeds of ejection the ejecta-wind interaction would be expected to occur over several years rather than the short duration observed in LSQ13ddu. The determined progenitor radius from the modelling of $1.58^{+1.94}_{-0.80}$ AU is more consistent with an extended LBV-like progenitor than WR star \citep{Sholukhova2015}, with LBV stars observed with radii 0.33--0.93~AU compared to the much more compact WR stars with radii as small as a few \rsun\ ($\sim$0.001~AU) \citep{Moffat1996,Petrovic2006}. However, as described in Section~\ref{sec:CSM_Powering}, the radius value outputted by the model is highly uncertain and may have a systematic uncertainty of up to an order of magnitude depending on the input CSM configuration. Further modelling, therefore, is necessary to conclusively constrain the progenitor based on its estimated radius.

For the initial rapid rise and peak of LSQ13ddu to be powered largely through CSM interaction, the CSM must have been located very close to the progenitor star at the time of explosion. The CSM signatures also disappeared quickly on the timescale of less than two weeks after maximum. This fast appearance and disappearance could suggest a very recent pre-explosion outburst (or unusually rapid  mass-loss period) so that the CSM material that we see as early interaction is constrained to quite a small radius around the star. For SN~2006jc, the eruptive pre-explosion event was suggested to be caused by the coincidental eruption of an LBV companion star very shortly before the terminal explosion of the WR star. For LSQ13ddu, a binary companion in an earlier stage of evolution where mass loss is still removing He-rich material could provide CSM of required mass and composition to generate the detected interaction signatures, through a combination of winds and/or eruptions. Though the low \Hei\ velocities observed for LSQ13ddu that are coincident with LBV wind velocities or eruption velocities are at odds with the non-detection of H emission in its spectra although a small amount at the $\sim$10\% level cannot be excluded.

Some rapidly evolving transients such as iPTF16asu \citep{Whitesides2017a} have been investigated to see if their light curve could be powered by magnetar formation but it was found that the derived ejecta masses were too low \citep[e.g.][]{Arcavi2016}. We found a similar result for LSQ13ddu, where the very low ejecta mass from the magnetar model implies an unreasonable amount of envelope stripping to produce such a low ejecta mass.

\section{Conclusion}
\label{section:Conclusion}

LSQ13ddu is likely a hybrid event that transitions from showing weak, narrow P-Cygni \Hei\ features on a blue continuum to showing spectra and light curves more consistent with Ic SNe with some residual He remaining in its ejecta. The rapid 4.8~d rise from explosion to a bright absolute magnitude of $-$19.71$\pm$0.02 mag, along with the detection of narrow P-Cygni profiled He features, are properties common to SNe Ibn. Analysis of its light curve suggests that the photometric behaviour of LSQ13ddu is well described by a combination of sources: CSM interaction to power the early light curve peak and enable the rapid rise in luminosity and a component powered by the radioactive \nick\ decay to explain the late-time evolution. The presence of the weak, narrow \Hei\ CSM signatures also disappeared after a few weeks, consistent with the photometric analysis showing the underlying radioactive \nick\ component becoming more dominant over time. The weakness of the narrow \Hei\ features even at early times suggests the presence of less He in the surrounding environment than seen in SNe Ibn. The similarity of the later time spectra of LSQ13ddu to more normal SE-SNe indicates that the underlying SN is likely to have been produced by a heavily stripped star with the source of the CSM arising from an unseen binary companion or from a recent phase of increased mass loss. 

The population of hybrid He-interacting events has been shown to span from events with residual H within the CSM, to more stripped events lacking H but with significant He within their progenitors at the time of explosion (SN~2010al, ASASSN-15ed) to even more stripped events such as iPTF16asu \citep{Whitesides2017a}. iPTF16asu displayed an early blue continuum similar to those of SNe Ibn but without \Hei\ or \Heii\ lines and then evolved to be most similar to a Ic-BL SN. LSQ13ddu falls on this continuum, with early time spectra and photometric behaviour similar to those of a Ibn SN with weak \Hei\ lines before evolving to more closely match the behaviour of a Ic SN with some residual He within its ejecta at later times.

As on-going high-cadence surveys continue to discover more fast evolving and transitional events interacting with He-dominated CSM, it will be essential that new theoretical models of massive stars, their mass loss, and the probability of explosion during as LBV or WR stars, are explored that can hopefully explain the diversity of these classes of events and unveil their most likely progenitor scenarios.

\section*{Acknowledgements}
\label{sec:Acknowledgements}
	
PC acknowledges funding from DfE.
KM acknowledges support from the UK STFC through an Ernest Rutherford Fellowship (ST/M005348/1) and from EU/H2020/ERC grant no.~758638. GH and DAH are supported by NSF grant AST-1313484.
EYH, MS, and CA acknowledge support from the National Science Foundation under grant no.~AST-1613472.
T-WC acknowledges funding from the Alexander von Humboldt Foundation.
MF is supported by a Royal Society - Science Foundation Ireland University Research Fellowship. 
GP acknowledges support by the Ministry of Economy, Development, and Tourism's Millennium Science Initiative through grant IC120009, awarded to The Millennium Institute of Astrophysics, MAS. MS is supported in part by a grant (13261) from VILLUM FONDEN and a project grant from the Independent Research Fund Denmark. MS acknowledges support from EU/FP7-ERC grant no.~615929.  
Based on observations collected at the European Organisation for Astronomical Research in the Southern Hemisphere, Chile as part of PESSTO, (the Public ESO Spectroscopic Survey for Transient Objects Survey) ESO program ID 188.D-3003.
The CSP-II has been funded by the USA's NSF under grants AST-0306969, AST-0607438, AST-1008343, AST-1613426, AST-1613455, and AST-1613472, and in part by  a Sapere Aude Level 2 grant funded by the Danish Agency for Science and Technology and Innovation  (PI Stritzinger). 
This work makes use of the NASA/IPAC Extragalactic Database (NED), which is operated by the Jet Propulsion Laboratory, California Institute of Technology, under contract with the National Aeronautics and Space Administration. 
This work makes use of observations from the LCOGT Network.
The work made use of Swift/UVOT data reduced by P. J. Brown and released in the Swift Optical/Ultraviolet Supernova Archive (SOUSA). SOUSA is supported by NASA's Astrophysics Data Analysis Program through grant NNX13AF35G.
The Liverpool Telescope is operated on the island of La Palma by Liverpool John Moores University in the Spanish Observatorio del Roque de los Muchachos of the Instituto de Astrofisica de Canarias with financial support from the UK STFC (Project ID: PL13B10, PL13B16).




\bibliographystyle{mnras}
\bibliography{LSQ13ddu.bib} 

\begin{thebibliography}{}
\makeatletter
\relax
\def\mn@urlcharsother{\let\do\@makeother \do\$\do\&\do\#\do\^\do\_\do\%\do\~}
\def\mn@doi{\begingroup\mn@urlcharsother \@ifnextchar [ {\mn@doi@}
  {\mn@doi@[]}}
\def\mn@doi@[#1]#2{\def\@tempa{#1}\ifx\@tempa\@empty \href
  {http://dx.doi.org/#2} {doi:#2}\else \href {http://dx.doi.org/#2} {#1}\fi
  \endgroup}
\def\mn@eprint#1#2{\mn@eprint@#1:#2::\@nil}
\def\mn@eprint@arXiv#1{\href {http://arxiv.org/abs/#1} {{\tt arXiv:#1}}}
\def\mn@eprint@dblp#1{\href {http://dblp.uni-trier.de/rec/bibtex/#1.xml}
  {dblp:#1}}
\def\mn@eprint@#1:#2:#3:#4\@nil{\def\@tempa {#1}\def\@tempb {#2}\def\@tempc
  {#3}\ifx \@tempc \@empty \let \@tempc \@tempb \let \@tempb \@tempa \fi \ifx
  \@tempb \@empty \def\@tempb {arXiv}\fi \@ifundefined
  {mn@eprint@\@tempb}{\@tempb:\@tempc}{\expandafter \expandafter \csname
  mn@eprint@\@tempb\endcsname \expandafter{\@tempc}}}

\bibitem[\protect\citeauthoryear{Alard \& Lupton}{Alard \&
  Lupton}{1997}]{Alard1997}
Alard C.,  Lupton R.~H.,  1997, \mn@doi [ApJ] {10.1086/305984}, 503, 325

\bibitem[\protect\citeauthoryear{Andrews \& Smith}{Andrews \&
  Smith}{2018}]{Andrews2018}
Andrews J.~E.,  Smith N.,  2018, \mn@doi [MNRAS] {10.1093/mnras/sty584}, 477,
  74

\bibitem[\protect\citeauthoryear{Arcavi et~al.,}{Arcavi
  et~al.}{2016}]{Arcavi2016}
Arcavi I.,  et~al., 2016, \mn@doi [ApJ] {10.3847/0004-637X/819/1/35}, 819, 35

\bibitem[\protect\citeauthoryear{Asplund, Grevesse, Sauval  \& Scott}{Asplund
  et~al.}{2009}]{Asplund2009}
Asplund M.,  Grevesse N.,  Sauval A.~J.,   Scott P.,  2009, \mn@doi [ARA{\&}A]
  {10.1146/annurev.astro.46.060407.145222}, 47, 481

\bibitem[\protect\citeauthoryear{Baldwin, Phillips  \& Terlevich}{Baldwin
  et~al.}{1981}]{Baldwin1981}
Baldwin J.~A.,  Phillips M.~M.,   Terlevich R.,  1981, \mn@doi [PASP]
  {10.1086/130766}, 93, 5

\bibitem[\protect\citeauthoryear{Baltay et~al.,}{Baltay
  et~al.}{2013}]{Baltay2013_LaSillaQuest}
Baltay C.,  et~al., 2013, \mn@doi [PASP] {10.1086/671198}, 125, 683

\bibitem[\protect\citeauthoryear{Becker}{Becker}{2015}]{Becker2015}
Becker A.,  2015, Astrophysics Source Code Library, ascl:1504.004

\bibitem[\protect\citeauthoryear{Bellm et~al.,}{Bellm et~al.}{2019}]{ZTF_2019}
Bellm E.~C.,  et~al., 2019, \mn@doi [PASP] {10.1088/1538-3873/aaecbe}, 131,
  018002

\bibitem[\protect\citeauthoryear{Bhat et~al.,}{Bhat et~al.}{2016}]{Bhat_2016}
Bhat P.~N.,  et~al., 2016, \mn@doi [ApJS] {10.3847/0067-0049/223/2/28}, 223, 28

\bibitem[\protect\citeauthoryear{Boyle, Sim, Hachinger  \& Kerzendorf}{Boyle
  et~al.}{2017}]{Boyle2017}
Boyle A.,  Sim S.~A.,  Hachinger S.,   Kerzendorf W.,  2017, \mn@doi [Astron.
  Astrophys.] {10.1051/0004-6361/201629712}, 599, 1

\bibitem[\protect\citeauthoryear{Brown et~al.,}{Brown et~al.}{2013}]{Brown2013}
Brown T.~M.,  et~al., 2013, \mn@doi [PASP] {10.1086/673168}, 125, 1031

\bibitem[\protect\citeauthoryear{Brown, Breeveld, Holland, Kuin  \&
  Pritchard}{Brown et~al.}{2014}]{Brown2014}
Brown P.~J.,  Breeveld A.~A.,  Holland S.,  Kuin P.,   Pritchard T.,  2014,
  \mn@doi [Astrophys. Space Sci.] {10.1007/s10509-014-2059-8}, 354, 89

\bibitem[\protect\citeauthoryear{Buzzoni et~al.,}{Buzzoni
  et~al.}{1984}]{Buzzoni1984}
Buzzoni B.,  et~al., 1984, ESO Messenger (ISSN 0722-6691), pp 9--13

\bibitem[\protect\citeauthoryear{Cano, Wang, Dai  \& Wu}{Cano
  et~al.}{2017}]{Cano2017}
Cano Z.,  Wang S.-Q.,  Dai Z.-G.,   Wu X.-F.,  2017, \mn@doi [Adv. Astronon.]
  {10.1155/2017/8929054}, 2017, 1

\bibitem[\protect\citeauthoryear{Chatzopoulos}{Chatzopoulos}{2018}]{Tigerfit}
Chatzopoulos M.,  2018, {TigerFit}, \url
  {https://github.com/manolis07gr/TigerFit}

\bibitem[\protect\citeauthoryear{Chatzopoulos, Wheeler, Vinko, Horvath  \&
  Nagy}{Chatzopoulos et~al.}{2013}]{Chatzopoulos2013a}
Chatzopoulos E.,  Wheeler J.~C.,  Vinko J.,  Horvath Z.~L.,   Nagy A.,  2013,
  \mn@doi [AJ] {10.1088/0004-637X/773/1/76}, 773, 76

\bibitem[\protect\citeauthoryear{Chevalier \& Irwin}{Chevalier \&
  Irwin}{2011}]{Chevalier2011}
Chevalier R.~A.,  Irwin C.~M.,  2011, \mn@doi [ApJ]
  {10.1088/2041-8205/729/1/L6}, 729, L6

\bibitem[\protect\citeauthoryear{Childress et~al.,}{Childress
  et~al.}{2016}]{Childress2016}
Childress M.~J.,  et~al., 2016, \mn@doi [PASA] {10.1017/pasa.2016.47}, 33

\bibitem[\protect\citeauthoryear{Chugai}{Chugai}{2009}]{Chugai2009}
Chugai N.~N.,  2009, \mn@doi [MNRAS] {10.1111/j.1365-2966.2009.15506.x}, 400,
  866

\bibitem[\protect\citeauthoryear{Crowther}{Crowther}{2007}]{Crowther2007}
Crowther P.~A.,  2007, \mn@doi [ARA{\&}A]
  {10.1146/annurev.astro.45.051806.110615}, 45, 177

\bibitem[\protect\citeauthoryear{De et~al.,}{De et~al.}{2018}]{De2018}
De K.,  et~al., 2018, \mn@doi [Science] {10.1126/science.aas8693}, 362, 201

\bibitem[\protect\citeauthoryear{Dopita, Hart, McGregor, Oates, Bloxham  \&
  Jones}{Dopita et~al.}{2007}]{Dopita2007}
Dopita M.,  Hart J.,  McGregor P.,  Oates P.,  Bloxham G.,   Jones D.,  2007,
  \mn@doi [Astrophys. Space Sci.] {10.1007/s10509-007-9510-z}, 310, 255

\bibitem[\protect\citeauthoryear{Dressler et~al.,}{Dressler
  et~al.}{2011}]{Dressler2011}
Dressler A.,  et~al., 2011, \mn@doi [PASP] {10.1086/658908}, 123, 288

\bibitem[\protect\citeauthoryear{Drout et~al.,}{Drout et~al.}{2014}]{Drout2014}
Drout M.~R.,  et~al., 2014, \mn@doi [ApJ] {10.1088/0004-637X/794/1/23}, 794, 23

\bibitem[\protect\citeauthoryear{{Fermi GRB Team}}{{Fermi GRB
  Team}}{2013}]{FermiGRBTeam2013}
{Fermi GRB Team} 2013, {GCN/Fermi Notice: GRB 131123A}, \url
  {https://gcn.gsfc.nasa.gov/other/406904521.fermi}

\bibitem[\protect\citeauthoryear{Fisher}{Fisher}{2000}]{Fisher2000}
Fisher A.~K.,  2000, PhD thesis, Univ. Oklahoma

\bibitem[\protect\citeauthoryear{Fitzpatrick}{Fitzpatrick}{1999}]{Fitzpatrick1998}
Fitzpatrick E.~L.,  1999, \mn@doi [PASP] {10.1086/316293}, 111, 63

\bibitem[\protect\citeauthoryear{Foley, Smith, Ganeshalingam, Li, Chornock  \&
  Filippenko}{Foley et~al.}{2007}]{Foley2007}
Foley R.~J.,  Smith N.,  Ganeshalingam M.,  Li W.,  Chornock R.,   Filippenko
  A.~V.,  2007, \mn@doi [ApJ] {10.1086/513145}, 657, L105

\bibitem[\protect\citeauthoryear{Foreman-Mackey, Hogg, Lang  \&
  Goodman}{Foreman-Mackey et~al.}{2013}]{EMCEE}
Foreman-Mackey D.,  Hogg D.~W.,  Lang D.,   Goodman J.,  2013, \mn@doi [PASP]
  {10.1086/670067}, 125, 306

\bibitem[\protect\citeauthoryear{Freudling, Romaniello, Bramich, Ballester,
  Forchi, Garc{\'{i}}a-Dabl{\'{o}}, Moehler  \& Neeser}{Freudling
  et~al.}{2013}]{Freudling2013}
Freudling W.,  Romaniello M.,  Bramich D.~M.,  Ballester P.,  Forchi V.,
  Garc{\'{i}}a-Dabl{\'{o}} C.~E.,  Moehler S.,   Neeser M.~J.,  2013, \mn@doi
  [A{\&}A] {10.1051/0004-6361/201322494}, 559, A96

\bibitem[\protect\citeauthoryear{Ginzburg \& Balberg}{Ginzburg \&
  Balberg}{2012}]{Ginzburg2012}
Ginzburg S.,  Balberg S.,  2012, \mn@doi [ApJ] {10.1088/0004-637X/757/2/178},
  757, 178

\bibitem[\protect\citeauthoryear{Gruber et~al.,}{Gruber
  et~al.}{2014}]{Gruber_2014}
Gruber D.,  et~al., 2014, \mn@doi [ApJS] {10.1088/0067-0049/211/1/12}, 211

\bibitem[\protect\citeauthoryear{Guillochon, Nicholl, Villar, Mockler, Narayan,
  Mandel, Berger  \& Williams}{Guillochon et~al.}{2018}]{Guillochon2018}
Guillochon J.,  Nicholl M.,  Villar V.~A.,  Mockler B.,  Narayan G.,  Mandel
  K.~S.,  Berger E.,   Williams P. K.~G.,  2018, \mn@doi [ApJS]
  {10.3847/1538-4365/aab761}, 236, 6

\bibitem[\protect\citeauthoryear{Hamuy et~al.,}{Hamuy et~al.}{2006}]{CSP1_Lowz}
Hamuy M.,  et~al., 2006, \mn@doi [PASP] {10.1086/500228}, 118, 2

\bibitem[\protect\citeauthoryear{{Henden}, {Welch}, {Terrell}  \&
  {Levine}}{{Henden} et~al.}{2009}]{Henden2009}
{Henden} A.~A.,  {Welch} D.~L.,  {Terrell} D.,   {Levine} S.~E.,  2009, in AAS
  Meeting Abstracts \#214. p. 407.02

\bibitem[\protect\citeauthoryear{Hosseinzadeh et~al.,}{Hosseinzadeh
  et~al.}{2017}]{Hosseinzadeh2017}
Hosseinzadeh G.,  et~al., 2017, \mn@doi [ApJ] {10.3847/1538-4357/836/2/158},
  836, 158

\bibitem[\protect\citeauthoryear{Hosseinzadeh, McCully, Zabludoff, Arcavi,
  French, Howell, Berger  \& Hiramatsu}{Hosseinzadeh
  et~al.}{2019}]{Hosseinzadeh2019}
Hosseinzadeh G.,  McCully C.,  Zabludoff A.~I.,  Arcavi I.,  French K.~D.,
  Howell D.~A.,  Berger E.,   Hiramatsu D.,  2019, \mn@doi [ApJ]
  {10.3847/2041-8213/aafc61}, 871, L9

\bibitem[\protect\citeauthoryear{Hsiao et~al.,}{Hsiao et~al.}{2019}]{Hsiao2019}
Hsiao E.~Y.,  et~al., 2019, \mn@doi [PASP] {10.1088/1538-3873/aae961}, 131,
  014002

\bibitem[\protect\citeauthoryear{Hunter et~al.,}{Hunter
  et~al.}{2009}]{Hunter2009}
Hunter D.~J.,  et~al., 2009, \mn@doi [A{\&}A] {10.1051/0004-6361/200912896},
  508, 371

\bibitem[\protect\citeauthoryear{Jenness \& Economou}{Jenness \&
  Economou}{2015}]{Jenness2015}
Jenness T.,  Economou F.,  2015, \mn@doi [Astron. {\&} Comput.]
  {10.1016/j.ascom.2014.10.005}, 9, 40

\bibitem[\protect\citeauthoryear{Jones, Oliphant, Peterson  \& Others}{Jones
  et~al.}{2001}]{SciPy}
Jones E.,  Oliphant T.,  Peterson P.,   Others 2001, {SciPy: Open source
  scientific tools for Python}, \url {http://www.scipy.org/}

\bibitem[\protect\citeauthoryear{Jones et~al.,}{Jones et~al.}{2009}]{Jones2009}
Jones D.~H.,  et~al., 2009, \mn@doi [MNRAS] {10.1111/j.1365-2966.2009.15338.x},
  399, 683

\bibitem[\protect\citeauthoryear{Karamehmetoglu et~al.,}{Karamehmetoglu
  et~al.}{2017}]{Karamehmetoglu2017}
Karamehmetoglu E.,  et~al., 2017, \mn@doi [A{\&}A]
  {10.1051/0004-6361/201629619}, 602, A93

\bibitem[\protect\citeauthoryear{Kasliwal et~al.,}{Kasliwal
  et~al.}{2010}]{Kasliwal2010}
Kasliwal M.~M.,  et~al., 2010, \mn@doi [ApJ] {10.1088/2041-8205/723/1/L98},
  723, L98

\bibitem[\protect\citeauthoryear{Kauffmann et~al.,}{Kauffmann
  et~al.}{2003}]{Kauffmann2003}
Kauffmann G.,  et~al., 2003, \mn@doi [MNRAS]
  {10.1111/j.1365-2966.2003.07154.x}, 346, 1055

\bibitem[\protect\citeauthoryear{Kausch et~al.,}{Kausch
  et~al.}{2015}]{Kausch2015}
Kausch W.,  et~al., 2015, \mn@doi [A{\&}A] {10.1051/0004-6361/201423909}, 576,
  A78

\bibitem[\protect\citeauthoryear{Kewley, Dopita, Sutherland, Heisler  \&
  Trevena}{Kewley et~al.}{2001}]{Kewley2001}
Kewley L.~J.,  Dopita M.~A.,  Sutherland R.~S.,  Heisler C.~A.,   Trevena J.,
  2001, \mn@doi [ApJ] {10.1086/321545}, 556, 121

\bibitem[\protect\citeauthoryear{Lyman, Bersier, James, Mazzali, Eldridge,
  Fraser  \& Pian}{Lyman et~al.}{2016}]{Lyman2016}
Lyman J.~D.,  Bersier D.,  James P.~A.,  Mazzali P.~A.,  Eldridge J.~J.,
  Fraser M.,   Pian E.,  2016, \mn@doi [MNRAS] {10.1093/mnras/stv2983}, 457,
  328

\bibitem[\protect\citeauthoryear{Makarov, Prugniel, Terekhova, Courtois  \&
  Vauglin}{Makarov et~al.}{2014}]{Makarov2014}
Makarov D.,  Prugniel P.,  Terekhova N.,  Courtois H.,   Vauglin I.,  2014,
  \mn@doi [A{\&}A] {10.1051/0004-6361/201423496}, 570, A13

\bibitem[\protect\citeauthoryear{McCully et~al.,}{McCully
  et~al.}{2018}]{McCully2018}
McCully C.,  et~al., 2018, \mn@doi [Zenodo] {10.5281/zenodo.1257560}

\bibitem[\protect\citeauthoryear{Meynet \& Maeder}{Meynet \&
  Maeder}{2005}]{Meynet2004}
Meynet G.,  Maeder A.,  2005, \mn@doi [A{\&}A] {10.1051/0004-6361:20047106},
  429, 581

\bibitem[\protect\citeauthoryear{Modjaz et~al.,}{Modjaz
  et~al.}{2014}]{Modjaz2014}
Modjaz M.,  et~al., 2014, \mn@doi [AJ] {10.1088/0004-6256/147/5/99}, 147, 99

\bibitem[\protect\citeauthoryear{Modjaz, Liu, Bianco  \& Graur}{Modjaz
  et~al.}{2016}]{Modjaz2015}
Modjaz M.,  Liu Y.~Q.,  Bianco F.~B.,   Graur O.,  2016, \mn@doi [ApJ]
  {10.3847/0004-637X/832/2/108}, 832, 108

\bibitem[\protect\citeauthoryear{Moffat \& Marchenko}{Moffat \&
  Marchenko}{1996}]{Moffat1996}
Moffat A. F.~J.,  Marchenko S.~V.,  1996, Astron. Astrophys.

\bibitem[\protect\citeauthoryear{Moorwood, Cuby  \& Lidman}{Moorwood
  et~al.}{1998}]{Moorwood1998}
Moorwood A.,  Cuby J.,   Lidman C.,  1998, ESO Messenger, 53, 1

\bibitem[\protect\citeauthoryear{Moriya \& Maeda}{Moriya \&
  Maeda}{2016}]{Moriya2016a}
Moriya T.~J.,  Maeda K.,  2016, \mn@doi [ApJ] {10.3847/0004-637X/824/2/100},
  824, 100

\bibitem[\protect\citeauthoryear{Moriya et~al.,}{Moriya
  et~al.}{2017}]{Moriya2017a}
Moriya T.~J.,  et~al., 2017, \mn@doi [MNRAS] {10.1093/mnras/stw3225}, 466, 2085

\bibitem[\protect\citeauthoryear{Nadyozhin}{Nadyozhin}{1994}]{Nadyozhin1994}
Nadyozhin D.~K.,  1994, \mn@doi [ApJS] {10.1086/192008}, 92, 527

\bibitem[\protect\citeauthoryear{Nicholl}{Nicholl}{2018}]{Nicholl2018}
Nicholl M.,  2018, \mn@doi [RNAAS] {10.3847/2515-5172/aaf799}, 2, 230

\bibitem[\protect\citeauthoryear{Nicholl, Guillochon  \& Berger}{Nicholl
  et~al.}{2017}]{Nicholl2017}
Nicholl M.,  Guillochon J.,   Berger E.,  2017, \mn@doi [ApJ]
  {10.3847/1538-4357/aa9334}, 850, 55

\bibitem[\protect\citeauthoryear{Oemler, Clardy, Kelson, Walth  \&
  Villanueva}{Oemler et~al.}{2017}]{Oemler2017}
Oemler A.,  Clardy K.,  Kelson D.,  Walth G.,   Villanueva E.,  2017,
  Astrophysics Source Code Library, record ascl:1705.001

\bibitem[\protect\citeauthoryear{Paciesas et~al.,}{Paciesas
  et~al.}{1999}]{Paciesas1999}
Paciesas W.~S.,  et~al., 1999, \mn@doi [ApJS] {10.1086/313224}, 122, 465

\bibitem[\protect\citeauthoryear{Paciesas et~al.,}{Paciesas
  et~al.}{2012}]{Paciesas_2012}
Paciesas W.~S.,  et~al., 2012, \mn@doi [ApJS] {10.1088/0067-0049/199/1/18}, 199

\bibitem[\protect\citeauthoryear{Pastorello et~al.,}{Pastorello
  et~al.}{2007}]{Pastorello2007}
Pastorello A.,  et~al., 2007, \mn@doi [Nature] {10.1038/nature05825}, 447, 829

\bibitem[\protect\citeauthoryear{Pastorello et~al.,}{Pastorello
  et~al.}{2008a}]{Pastorello2008a}
Pastorello A.,  et~al., 2008a, \mn@doi [MNRAS]
  {10.1111/j.1365-2966.2008.13602.x}, 389, 113

\bibitem[\protect\citeauthoryear{Pastorello et~al.,}{Pastorello
  et~al.}{2008b}]{Pastorello2008b}
Pastorello A.,  et~al., 2008b, \mn@doi [MNRAS]
  {10.1111/j.1365-2966.2008.13603.x}, 389, 131

\bibitem[\protect\citeauthoryear{Pastorello et~al.,}{Pastorello
  et~al.}{2015a}]{Pastorello2015e}
Pastorello A.,  et~al., 2015a, \mn@doi [MNRAS] {10.1093/mnras/stu2745}, 449,
  1921

\bibitem[\protect\citeauthoryear{Pastorello et~al.,}{Pastorello
  et~al.}{2015b}]{Pastorello2015c}
Pastorello A.,  et~al., 2015b, \mn@doi [MNRAS] {10.1093/mnras/stu2621}, 449,
  1941

\bibitem[\protect\citeauthoryear{Pastorello et~al.,}{Pastorello
  et~al.}{2015c}]{Pastorello2015}
Pastorello A.,  et~al., 2015c, \mn@doi [MNRAS] {10.1093/mnras/stv335}, 449,
  1954

\bibitem[\protect\citeauthoryear{Pastorello et~al.,}{Pastorello
  et~al.}{2015d}]{Pastorello_15ed}
Pastorello A.,  et~al., 2015d, \mn@doi [MNRAS] {10.1093/mnras/stv1812}, 453,
  3650

\bibitem[\protect\citeauthoryear{Pastorello et~al.,}{Pastorello
  et~al.}{2016}]{Pastorello2016}
Pastorello A.,  et~al., 2016, \mn@doi [MNRAS] {10.1093/mnras/stv2634}, 456, 853

\bibitem[\protect\citeauthoryear{Perez, Bagish, Bredthauer, Espoz, Jones  \&
  Pinto}{Perez et~al.}{2012}]{Perez_2012}
Perez F.,  Bagish A.,  Bredthauer G.,  Espoz J.,  Jones P.,   Pinto P.,  2012,
  in Proceedings Volume 8444, Ground-based and Airborne Telescopes IV. \url
  {https://doi.org/10.1117/12.927150}

\bibitem[\protect\citeauthoryear{Petrovic, Pols  \& Langer}{Petrovic
  et~al.}{2006}]{Petrovic2006}
Petrovic J.,  Pols O.,   Langer N.,  2006, \mn@doi [Astron. Astrophys.]
  {10.1051/0004-6361:20035837}, 450, 219

\bibitem[\protect\citeauthoryear{Pettini \& Pagel}{Pettini \&
  Pagel}{2004}]{Pettini2004}
Pettini M.,  Pagel B. E.~J.,  2004, \mn@doi [MNRAS]
  {10.1111/j.1365-2966.2004.07591.x}, 348, L59

\bibitem[\protect\citeauthoryear{Phillips et~al.,}{Phillips
  et~al.}{2019}]{Phillips2019}
Phillips M.~M.,  et~al., 2019, \mn@doi [PASP] {10.1088/1538-3873/aae8bd}, 131,
  014001

\bibitem[\protect\citeauthoryear{Poznanski et~al.,}{Poznanski
  et~al.}{2009}]{Poznanski2009}
Poznanski D.,  et~al., 2009, \mn@doi [Science] {10.1126/science.1181709}, 327,
  58

\bibitem[\protect\citeauthoryear{Prentice et~al.,}{Prentice
  et~al.}{2016}]{Prentice2016}
Prentice S.~J.,  et~al., 2016, \mn@doi [MNRAS] {10.1093/mnras/stw299}, 458,
  2973

\bibitem[\protect\citeauthoryear{Prentice et~al.,}{Prentice
  et~al.}{2018a}]{Prentice2018_coi}
Prentice S.~J.,  et~al., 2018a, \mn@doi [MNRAS] {10.1093/mnras/sty1223}, 478,
  4162

\bibitem[\protect\citeauthoryear{Prentice et~al.,}{Prentice
  et~al.}{2018b}]{Prentice2018_cow}
Prentice S.~J.,  et~al., 2018b, \mn@doi [ApJ] {10.3847/2041-8213/aadd90}, 865,
  L3

\bibitem[\protect\citeauthoryear{Prentice et~al.,}{Prentice
  et~al.}{2019}]{Prentice2019_SESN}
Prentice S.~J.,  et~al., 2019, \mn@doi [MNRAS] {10.1093/mnras/sty3399}, 485,
  1559

\bibitem[\protect\citeauthoryear{Pursiainen et~al.,}{Pursiainen
  et~al.}{2018}]{Pursiainen2018}
Pursiainen M.,  et~al., 2018, \mn@doi [MNRAS] {10.1093/mnras/sty2309}, 481, 894

\bibitem[\protect\citeauthoryear{Rest et~al.,}{Rest et~al.}{2018}]{Rest2018}
Rest A.,  et~al., 2018, \mn@doi [Nature Astron.] {10.1038/s41550-018-0423-2},
  2, 307

\bibitem[\protect\citeauthoryear{Rochowicz \& Niedzielski}{Rochowicz \&
  Niedzielski}{1999}]{Rochowicz1999}
Rochowicz K.,  Niedzielski A.,  1999, Acta Astronomica, 45, 307

\bibitem[\protect\citeauthoryear{Sahu, Tanaka, Anupama, Gurugubelli  \&
  Nomoto}{Sahu et~al.}{2009}]{Sahu2008}
Sahu D.~K.,  Tanaka M.,  Anupama G.~C.,  Gurugubelli U.~K.,   Nomoto K.,  2009,
  \mn@doi [ApJ] {10.1088/0004-637X/697/1/676}, 697, 676

\bibitem[\protect\citeauthoryear{Sanders et~al.,}{Sanders
  et~al.}{2013}]{Sanders2013}
Sanders N.~E.,  et~al., 2013, \mn@doi [ApJ] {10.1088/0004-637X/769/1/39}, 769,
  39

\bibitem[\protect\citeauthoryear{Scalzo et~al.,}{Scalzo
  et~al.}{2014}]{Scalzo2014a}
Scalzo R.~A.,  et~al., 2014, \mn@doi [MNRAS] {10.1093/mnras/stu1723}, 445, 30

\bibitem[\protect\citeauthoryear{Schlafly \& Finkbeiner}{Schlafly \&
  Finkbeiner}{2011}]{Schlafly2011}
Schlafly E.~F.,  Finkbeiner D.~P.,  2011, \mn@doi [ApJ]
  {10.1088/0004-637X/737/2/103}, 737

\bibitem[\protect\citeauthoryear{Schlegel}{Schlegel}{1990}]{Schlegel1990}
Schlegel E.~M.,  1990, MNRAS, 244, 269

\bibitem[\protect\citeauthoryear{Shappee et~al.,}{Shappee
  et~al.}{2014}]{Shappee2014}
Shappee B.~J.,  et~al., 2014, \mn@doi [ApJ] {10.1088/0004-637X/788/1/48}, 788,
  48

\bibitem[\protect\citeauthoryear{Sholukhova, Bizyaev, Fabrika, Sarkisyan,
  Malanushenko  \& Valeev}{Sholukhova et~al.}{2015}]{Sholukhova2015}
Sholukhova O.,  Bizyaev D.,  Fabrika S.,  Sarkisyan A.,  Malanushenko V.,
  Valeev A.,  2015, \mn@doi [MNRAS] {10.1093/mnras/stu2597}, 447, 2459

\bibitem[\protect\citeauthoryear{Simcoe et~al.,}{Simcoe
  et~al.}{2013}]{Simcoe2013}
Simcoe R.~A.,  et~al., 2013, \mn@doi [PASP] {10.1086/670241}, 125, 270

\bibitem[\protect\citeauthoryear{Smartt et~al.,}{Smartt
  et~al.}{2014}]{Smartt2014}
Smartt S.~J.,  et~al., 2014, \mn@doi [A{\&}A] {10.1051/0004-6361/201425237},
  579, A40

\bibitem[\protect\citeauthoryear{Smette et~al.,}{Smette
  et~al.}{2015}]{Smette2015}
Smette A.,  et~al., 2015, \mn@doi [A{\&}A] {10.1051/0004-6361/201423932}, 576,
  A77

\bibitem[\protect\citeauthoryear{Smith}{Smith}{2014}]{Smith2014}
Smith N.,  2014, \mn@doi [ARA{\&}A] {10.1146/annurev-astro-081913-040025}, 52,
  487

\bibitem[\protect\citeauthoryear{Smith}{Smith}{2017}]{Smith2017}
Smith N.,  2017, Interacting Supernovae: Types IIn and Ibn.
Springer International Publishing, Cham, pp 403--429, \url
  {https://doi.org/10.1007/978-3-319-21846-5_38}

\bibitem[\protect\citeauthoryear{Smith, Mauerhan, Silverman, Ganeshalingam,
  Filippenko, Cenko, Clubb  \& Kandrashoff}{Smith et~al.}{2012}]{Smith2012}
Smith N.,  Mauerhan J.~C.,  Silverman J.~M.,  Ganeshalingam M.,  Filippenko
  A.~V.,  Cenko S.~B.,  Clubb K.~I.,   Kandrashoff M.~T.,  2012, \mn@doi
  [MNRAS] {10.1111/j.1365-2966.2012.21849.x}, 426, 1905

\bibitem[\protect\citeauthoryear{Steele et~al.,}{Steele
  et~al.}{2004}]{Steele2004}
Steele I.~A.,  et~al., 2004, in Proceedings Volume 5489, Ground-based
  Telescopes. \url {https://doi.org/10.1117/12.551456}

\bibitem[\protect\citeauthoryear{Swartz, Sutherland  \& Harkness}{Swartz
  et~al.}{1995}]{Swartz1995}
Swartz D.~A.,  Sutherland P.~G.,   Harkness R.~P.,  1995, \mn@doi [ApJ]
  {10.1086/175834}, 446, 766

\bibitem[\protect\citeauthoryear{Taddia et~al.,}{Taddia
  et~al.}{2015}]{Taddia2015}
Taddia F.,  et~al., 2015, \mn@doi [A{\&}A] {10.1051/0004-6361/201525989}, 580,
  A131

\bibitem[\protect\citeauthoryear{Tauris, Langer  \& Podsiadlowski}{Tauris
  et~al.}{2015}]{Tauris2015}
Tauris T.~M.,  Langer N.,   Podsiadlowski P.,  2015, \mn@doi [MNRAS]
  {10.1093/mnras/stv990}, 451, 2123

\bibitem[\protect\citeauthoryear{Thomas, Nugent  \& Meza}{Thomas
  et~al.}{2011}]{Thomas_Synapps}
Thomas R.~C.,  Nugent P.~E.,   Meza J.~C.,  2011, \mn@doi [PASP]
  {10.1086/658673}, 123, 237

\bibitem[\protect\citeauthoryear{Tonry et~al.,}{Tonry et~al.}{2012}]{Tonry2012}
Tonry J.~L.,  et~al., 2012, \mn@doi [ApJ] {10.1088/0004-637X/750/2/99}, 750, 99

\bibitem[\protect\citeauthoryear{Tonry et~al.,}{Tonry
  et~al.}{2018}]{Tonry_2018}
Tonry J.~L.,  et~al., 2018, \mn@doi [PASP] {10.1088/1538-3873/aabadf}, 130,
  064505

\bibitem[\protect\citeauthoryear{Umeda \& Nomoto}{Umeda \&
  Nomoto}{2008}]{Umeda2008}
Umeda H.,  Nomoto K.,  2008, \mn@doi [ApJ] {10.1086/524767}, 673, 1014

\bibitem[\protect\citeauthoryear{Valenti et~al.,}{Valenti
  et~al.}{2016}]{Valenti2016}
Valenti S.,  et~al., 2016, \mn@doi [MNRAS] {10.1093/mnras/stw870}, 459, 3939

\bibitem[\protect\citeauthoryear{Vernet et~al.,}{Vernet
  et~al.}{2011}]{Vernet2011a}
Vernet J.,  et~al., 2011, \mn@doi [A{\&}A] {10.1051/0004-6361/201117752}, 536,
  A105

\bibitem[\protect\citeauthoryear{Villar, Berger, Metzger  \& Guillochon}{Villar
  et~al.}{2017}]{Villar2017}
Villar V.~A.,  Berger E.,  Metzger B.~D.,   Guillochon J.,  2017, \mn@doi [ApJ]
  {10.3847/1538-4357/aa8fcb}, 849, 70

\bibitem[\protect\citeauthoryear{Wang et~al.,}{Wang et~al.}{2017}]{Wang2017b}
Wang L.~J.,  et~al., 2017, \mn@doi [ApJ] {10.3847/1538-4357/aa9a38}, 851, 54

\bibitem[\protect\citeauthoryear{Wheeler, Johnson  \& Clocchiatti}{Wheeler
  et~al.}{2015}]{Wheeler2015}
Wheeler J.~C.,  Johnson V.,   Clocchiatti A.,  2015, \mn@doi [MNRAS]
  {10.1093/mnras/stv650}, 450, 1295

\bibitem[\protect\citeauthoryear{Whitesides et~al.,}{Whitesides
  et~al.}{2017}]{Whitesides2017a}
Whitesides L.,  et~al., 2017, \mn@doi [ApJ] {10.3847/1538-4357/aa99de}, 851,
  107

\bibitem[\protect\citeauthoryear{Woosley \& Bloom}{Woosley \&
  Bloom}{2006}]{Woosley2006}
Woosley S.,  Bloom J.,  2006, \mn@doi [ARA{\&}A]
  {10.1146/annurev.astro.43.072103.150558}, 44, 507

\bibitem[\protect\citeauthoryear{Woosley, Eastman, Weaver  \& Pinto}{Woosley
  et~al.}{1994}]{Woosley_1993J}
Woosley S.~E.,  Eastman R.~G.,  Weaver T.~A.,   Pinto P.~A.,  1994, \mn@doi
  [ApJ] {10.1086/174319}, 429, 300

\bibitem[\protect\citeauthoryear{Woosley, Eastman  \& Schmidt}{Woosley
  et~al.}{1999}]{Woosley1999}
Woosley S.~E.,  Eastman R.~G.,   Schmidt B.~P.,  1999, \mn@doi [ApJ]
  {10.1086/307131}, 516, 788

\bibitem[\protect\citeauthoryear{Yaron \& Gal-Yam}{Yaron \&
  Gal-Yam}{2012}]{Yaron2012a}
Yaron O.,  Gal-Yam A.,  2012, \mn@doi [PASP] {10.1086/666656}, 124, 668

\bibitem[\protect\citeauthoryear{van Loon}{van Loon}{2005}]{vanLoon2005}
van Loon J.~T.,  2005, ASP Conference Series, pp 1--13

\makeatother
\end{thebibliography}







\bsp	
\label{lastpage}

\end{document}